\documentclass[preprint, 10pt,a4paper,fleqn]{elsarticle}

\usepackage{custom_style}

\usepackage[utf8]{inputenc}
\usepackage[T1]{fontenc}
\usepackage{amsfonts}
\usepackage{amsmath,amssymb,setspace,amsthm,fancybox,mathtools}
\usepackage{graphicx}
\usepackage{subcaption}
\usepackage{graphbox}
\usepackage{bm}
\usepackage[english]{babel}
\usepackage{geometry}
\usepackage[dvipsnames]{xcolor}
\usepackage{todonotes}
\usepackage{float}
\usepackage{epstopdf}
\usepackage[hypertexnames=false,colorlinks=true]{hyperref}
\usepackage{placeins}
\usepackage[modulo]{lineno}
\usepackage{nth}
\usepackage{natbib}
\usepackage{bbm}
\usepackage[most]{tcolorbox}
\usepackage{array}
\usepackage{multirow}
\usepackage{booktabs}

\theoremstyle{definition}

\usepackage[ruled,lined,linesnumbered,resetcount]{algorithm2e} 
\SetKwComment{Comment}{$\triangleright$\ }{}
\SetCommentSty{}
\SetKwInOut{Given}{Given}
 
\colorlet{author}{magenta}
\colorlet{reviewer1}{blue}
\colorlet{reviewer2}{ForestGreen}
\colorlet{reviewer3}{red}

\graphicspath{{src/figs/}} 

\theoremstyle{plain}

\newcommand{\tns}[1]{\bm{\mathrm{#1}}}
\renewcommand{\vec}[1]{\bm{#1}}

\newcommand{\partialder}[2]{\frac{\partial #1}{\partial #2}}

\DeclareMathOperator{\cof}{cof}

\DeclareMathOperator{\tr}{tr}

\DeclareMathOperator{\atan}{atan}

\newcolumntype{C}[1]{>{\centering\arraybackslash}p{#1}}

\makeatletter
\def\ps@pprintTitle{%
    \let\@oddhead\@empty
    \let\@evenhead\@empty
    \def\@oddfoot{\footnotesize\itshape
         {Submitted preprint} \hfill}%
    \let\@evenfoot\@oddfoot
    }
\makeatother

\begin{document}
	
	\begin{frontmatter}

        \title{Constitutive Kolmogorov--Arnold Networks (CKANs): Combining Accuracy and Interpretability in Data-Driven Material Modeling}
        
        \author[1]{Kian P. Abdolazizi}
        \author[1,2]{Roland C. Aydin}
        \author[1,2]{Christian J. Cyron \corref{cor1}}
        \author[1]{Kevin Linka \corref{cor2}}

		\address[1]{Institute for Continuum and Material Mechanics, Hamburg University of Technology,\\ Ei\ss endorfer Stra\ss e 42, 21073 Hamburg, Germany}
        
        \address[2]{Institute of Material Systems Modeling, Helmholtz-Zentrum Hereon, \\Max-Planck-Straße 1, 21502 Geesthacht, Germany}

        \cortext[cor1]{Corresponding author: christian.cyron@hereon.de}
        \cortext[cor2]{Corresponding author: kevin.linka@tuhh.de}
		
		\begin{abstract}
           Hybrid constitutive modeling integrates two complementary approaches for describing and predicting a material's mechanical behavior: purely data-driven black-box methods and physically constrained, theory-based models. While black-box methods offer high accuracy, they often lack interpretability and extrapolability. Conversely, physics-based models provide theoretical insight and generalizability but may not capture complex behaviors with the same accuracy. Traditionally, hybrid modeling has required a trade-off between these aspects.
           In this paper, we show how recent advances in symbolic machine learning---specifically Kolmogorov--Arnold Networks (KANs)---help to overcome this limitation. We introduce Constitutive Kolmogorov--Arnold Networks (CKANs) as a new class of hybrid constitutive models. By incorporating a post-processing symbolification step, CKANs combine the predictive accuracy of data-driven models with the interpretability and extrapolation capabilities of symbolic expressions, bridging the gap between machine learning and physical modeling.
        \end{abstract}
		
		
	\end{frontmatter}


    \begin{figure}[h]
       \centering
       \includegraphics[width=1\linewidth]{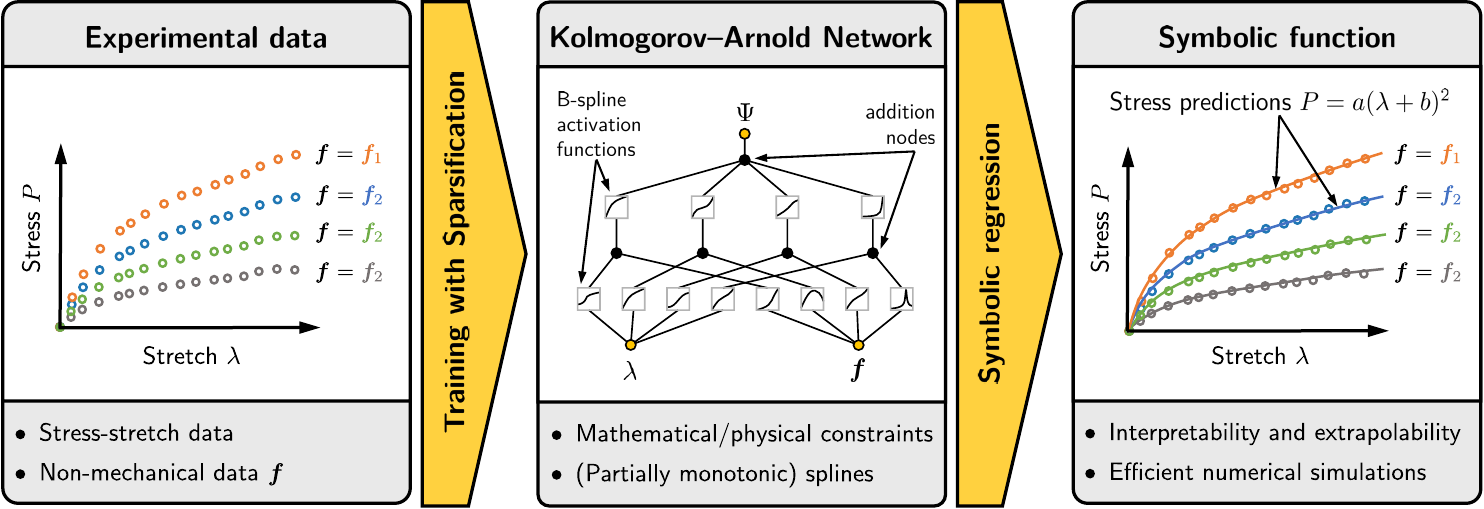}
       \caption{\textbf{Constitutive Kolmogorov-Arnold Networks (CKANs)}. Given experimental stretch $\lambda$ and stress $P$ data and potentially also non-mechanical data $\vec{f}$ (such as microstructural imaging data or information on materials processing), a CKAN learns the strain energy function $\Psi$ of a hyperelastic material. It inherently satisfies essential mathematical or physical requirements, such as objectivity, symmetry properties, or thermodynamic consistency. Its activation functions are univariate (partially monotonic) splines adapting during training. Combined with sparsification techniques, this leads to an accurate and efficient model. Finally, the activation functions are symbolized, yielding an interpretable and compact symbolic expression for the strain energy function $\Psi$.}
       \label{fig:graphical_abstract}
    \end{figure}
    
    \section{Introduction}\label{sec:intro}

    To describe the mechanical behavior of a material---specifically, the relationship between mechanical stress and strain---a constitutive model is required. Developing constitutive models that are accurate, predictive, and interpretable remains a persistent challenge. Historically, these models have primarily consisted of manually derived symbolic expressions that define functions or functionals mapping stress to strain or vice versa \cite{Treloar1943, Rivlin1951, Ogden1972, Fung1981, Holzapfel2000, Ehret2007}. However, this approach has three significant shortcomings. First, manually deriving symbolic expressions is both complex and time-consuming. Second, the limited complexity of manually tractable expressions often fails to capture the intricate behavior of materials. Third, while these symbolic expressions can describe known materials, they generally struggle to predict the behavior of new materials. 

    To overcome these shortcomings of traditional constitutive modeling, data-driven methods have attracted increasing attention over the last decade. The distance-minimizing method \cite{Kirchdoerfer2016, Carrara2020}, black-box neural networks \cite{Ghaboussi1991,Hashash2004} or methods based on spline interpolations~\cite{Sussman2009,Latorre2013,Latorre2014,Crespo2017,Dal2023,Wiesheier2024} were suggested. They overcome the need for manual derivation of symbolic equations. However, such purely data-driven approaches have great difficulties in extrapolating beyond known material data or predicting the behavior of new materials. Moreover, they are typically not directly interpretable. This lack of interpretability limits the applicability in particular in safety-critical areas where it is crucial to understand the precise nature of an applied model and to prove also certain bounds within which its output will remain under certain conditions to prevent situations where a model failure could cause major damage \cite{goldberg2024no}. 
    
    Therefore, hybrid constitutive models were recently introduced to bridge the gap between theory-based symbolic models and data-driven approaches \cite{fuhg2024review}. These models often embed domain knowledge into machine learning (ML) frameworks using physics-based loss functions or hard constraints. Examples include Physics-Augmented Neural Networks (PANNs), which integrate hyperelastic constraints into neural architectures \cite{Linden2023}, neural ordinary differential equations that enforce physical plausibility \cite{tacc2023data}, and models explicitly designed to ensure polyconvexity \cite{as2022mechanics}. Constitutive Artificial Neural Networks (CANNs) further expand the scope of data-driven constitutive modeling by incorporating mechanical and non-mechanical inputs \cite{Linka2021, Abdolazizi2023}.
    Hybrid constitutive models offer several key advantages. First, like all data-driven models, they eliminate the need for manual model derivation. Second, their flexible architectures allow them to accurately capture highly complex material behavior, given sufficient data is available. Third, integrating theoretical principles improves extrapolation capabilities significantly compared to classical black-box models. Fourth, certain hybrid models, such as CANNs \cite{Linka2022}, can leverage non-mechanical data---such as microstructural imaging or processing parameters---to predict the behavior of novel materials. Despite these strengths, hybrid constitutive models still face a critical limitation. While embedding physical knowledge makes them more transparent than purely data-driven black-box models, they often retain extensive, uninterpretable parts---such as deep neural networks. This residual opacity remains a significant drawback, particularly in high-stakes fields like biomedical or aerospace engineering, where reliability and traceability are crucial \cite{goldberg2024no}.

    To address this challenge, a complementary trend has focused on interpretable data-driven modeling, aiming to derive constitutive laws directly from experimental data. One approach combines large libraries of physics-inspired building blocks with sparse regression to identify relevant terms efficiently~\cite{Flaschel2021a}. Another strategy employs specialized neural networks that embed fundamental constitutive structures~\cite{Linka2022a, holthusen2024theory, pierre2023discovering, McCulloch2024, holthusen2025automated}, enabling the automatic selection of dominant terms. Additionally, symbolic regression, particularly through genetic programming, has gained traction for generating concise, human-readable equations~\cite{wang2019symbolic, Abdusalamov2022a}.
    While these methods enhance model interpretability, they come with drawbacks. Limited expressiveness can hinder their ability to capture complex material behavior, while high computational costs and sensitivity to data quality pose significant challenges.

    This paper aims to merge the strengths of hybrid constitutive models, such as Constitutive Artificial Neural Networks (CANNs), with the benefits of interpretable symbolic data-driven methods. To achieve this, we employ Kolmogorov–Arnold Networks (KANs), which bridge the gap between purely data-driven approaches and symbolic interpretation by enabling the extraction of interpretable features, modular structures, and closed-form expressions~\cite{Liu2024a}. KANs are particularly well-suited for discovering constitutive laws, as they provide functionally simple yet expressive representations of complex nonlinear behavior~\cite{Liu2024}.
    KANs leverage the Kolmogorov–Arnold representation theorem, which states that any continuous multivariate function can be represented as a composition of univariate functions and the addition operation. This property allows them to function as powerful universal approximators while maintaining interpretability. Building on these principles, we propose a novel framework that integrates the symbolic capabilities of KANs with the flexibility of CANNs, thereby overcoming the persistent trade-off between interpretability and performance.

    In this paper, we introduce Constitutive Kolmogorov--Arnold Networks (CKANs), a novel approach to data-driven constitutive modeling that leverages Kolmogorov--Arnold Networks (KANs) within a hybrid modeling framework. By systematically incorporating physically inspired constraints and pruning strategies, our method attains high predictive accuracy while generating transparent, symbolic constitutive laws. The overall workflow of the CKAN framework is illustrated in Figure \ref{fig:graphical_abstract}. 
    This paper is organized as follows: Section~\ref{sec:conti} introduces the necessary background from continuum mechanics. In Section~\ref{sec:methods}, we present the theoretical foundations of the CKAN framework, while Section~\ref{sec:sparse_symbol} details the sparsification and symbolification process necessary for deriving compact and interpretable symbolic expressions. We then demonstrate the effectiveness of our methodology through various numerical experiments and real-world data in Section~\ref{sec:results}. Finally, Section \ref{sec:discussion} summarizes the main findings and suggests possible future directions. To reduce the complexity of the discussion herein, we focus in this paper on isotropic hyperelastic materials. However, the ideas developed herein can easily be transferred to much broader classes of materials.

    \section{Continuum mechanics}\label{sec:conti}

    \subsection{Kinematics}
    Herein, we rely on the general framework of nonlinear continuum mechanics as outlined, for example, in \cite{Holzapfel2000a}. We consider a continuum body composed of material points $\tns{X}$ in the undeformed reference configuration. During a motion, the material points are mapped at time $t$ to the current position $\tns{x}(\tns{X},t)$ in the deformed current configuration. To quantify the deformation of the material in the body, one often uses the so-called deformation gradient $\tns{F}$ and its spectral decomposition:
    \begin{equation}\label{eq:def_grad}
        \tns{F} = \partialder{\tns{x}}{\tns{X}} = \sum_{\alpha=1}^3 \lambda_\alpha \, \vec{n}_\alpha \otimes \vec{N}_\alpha.
    \end{equation}
    Here $\lambda_\alpha>0$ are the principal stretches of the material, and $\vec{N}_\alpha$ and $\vec{n}_\alpha$ are the corresponding principal stretch directions in the reference and current configuration. The larger $\lambda_\alpha$, the more the material is stretched in the corresponding principal stretch direction. The Jacobian (determinant) $J=\det (\tns{F}) = \lambda_1\lambda_2\lambda_3$ characterizes local volume changes. For incompressible materials $J=1$. Based on the deformation gradient, one can define the so-called right Cauchy--Green deformation tensor $\tns{C}$ with its cofactor $\cof (\tns{C})$:
    \begin{equation}\label{eq:spec_decomp}
        \tns{C} =\tns{F}^\mathrm{T}\tns{F}= \sum_{\alpha=1}^3 \lambda_\alpha^2 \, \vec{N}_\alpha \otimes \vec{N}_\alpha, \qquad
        \cof (\tns{C}) = \det (\tns{C}) \tns{C}^{-\mathrm{T}} = \sum_{\alpha=1}^3 \nu_\alpha^2 \, \vec{N}_\alpha \otimes \vec{N}_\alpha,
    \end{equation}
    For an undeformed body, the deformation gradient and Cauchy-Green deformation tensor are identity tensors; $\nu_\alpha=J/\lambda_\alpha$ is the stretch of the area orthogonal to the $\alpha$-direction and can be computed as
    \begin{equation}\label{eq:area_stretch}
        \nu_1 = \lambda_2\lambda_3, \quad \nu_2 = \lambda_1\lambda_3, \quad \nu_3 = \lambda_1\lambda_2.
    \end{equation}    
    The principal invariants of $\tns{C}$, in terms of the principal stretches, are
    \begin{equation}\label{eq:invars}
        I_1 = \tr (\tns{C})=\lambda_1^2 + \lambda_2^2 +\lambda_3^2 , \qquad I_2 = \tr (\cof \tns{C}) = \nu_1^2 + \nu_2^2 + \nu_3^2, \qquad I_3 = \det \tns{C} = J^2 = \lambda_1^2 \lambda_2^2\lambda_3^2.
    \end{equation}
    To describe the behavior of rubber-like materials, it was found that using other sets of kinematic variables can help to bridge the gap between the kinematics on the micro- and macro-scale. For instance, the modified invariants
    \begin{equation}\label{eq:mod_invars}
        \iota_1 = \sqrt{\frac{\lambda_1^2 + \lambda_2^2 + \lambda_3^2}{3}} = \sqrt{\frac{I_1}{3}}, \qquad \iota_2 = \sqrt{\frac{\nu_1^2 + \nu_2^2 + \nu_3^2}{3}} = \sqrt[3]{\frac{I_2}{3}},
    \end{equation}
    can be interpreted as representing the rubber network's average stretch and tube contraction, respectively \cite{Dal2023, Kearsley1989, Arruda1993}. 

\begin{figure}
        \centering
        \includegraphics[width=1\linewidth]{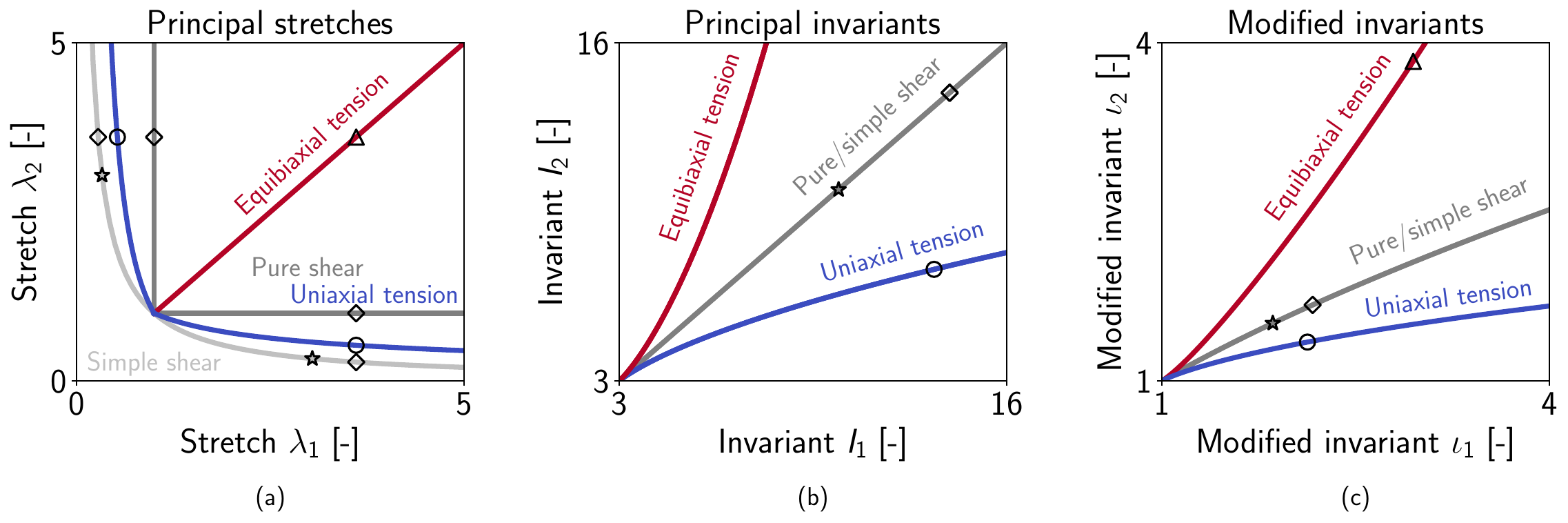}
        \caption{\textbf{Representation of special planar modes of deformation of a thin incompressible membrane using three different functional bases:} the complexity of the functional forms required to represent uniaxial tension, equibiaxial tension, and pure or simple shear vastly differs depending on whether one relies on representing them in terms of (a) principal stretches $\{\lambda_1, \lambda_2\}$, (b) principal invariants $\{I_1, I_2\}$, (c) modified invariants $\{\iota_1, \iota_2\}$. The markers indicate identical deformation states. Notably, not all the deformation states plotted in the left panel fit into the range plotted in the center panel so that there fewer symbols are visible.}
        \label{fig:input_space}
    \end{figure}

    Each of the above-introduced three sets of kinematic variables, i.e., $\{\lambda_1, \lambda_2, \lambda_3\}$, $\{I_1, I_2, I_3\}$, and $\{\iota_1, \iota_2, I_3\}$, provides a sufficient functional basis to characterize the local deformation of a solid body completely. However, depending on which functional basis one uses, the functions required to describe a specific physical phenomenon may take on a relatively complex or rather simple form. This is illustrated in Figure \ref{fig:input_space} by considering three specific deformation modes of a thin incompressible material sample, namely, uniaxial tension, equibiaxial tension, and pure shear (see Figure \ref{fig:experiments} for details). Using $\{\lambda_1, \lambda_2, \lambda_3\}$ as a functional basis, equibiaxial tension and pure shear can be expressed as linear functions, whereas uniaxial tension and simple shear require more complex functional forms. By contrast, using $\{I_1, I_2, I_3\}$, pure and simple shear can be represented by linear functions, whereas uniaxial and equibiaxial tension require more complex functional forms. 

    This article is devoted to the automated derivation of simple and yet accurate symbolic expressions to characterize the mechanical behavior of materials. Apparently, it is worth examining how the results of the framework developed herein depend on the functional basis chosen ($\{\lambda_1, \lambda_2, \lambda_3\}$, $\{I_1, I_2, I_3\}$ or $\{\iota_1, \iota_2, I_3\}$). This comparison thus forms an integral part of the discussion below. 

    \subsection{Isotropic hyperelasticity}
    The primary objective of this work is to illustrate how Kolmogorov--Arnold Networks can be applied to hybrid constitutive modeling of materials. For clarity, we focus on isotropic hyperelastic materials, exhibiting the same properties in all directions.

    \paragraph{Strain energy function and stress}
     Hyperelasticity postulates the existence of a strain energy function $\Psi=\Psi(\tns{F})$. The constitutive equation for the first Piola-Kirchhoff stress $\tns{P}$ of a hyperelastic material directly follows from the second law of thermodynamics generally as \cite{Truesdell1984}
    \begin{equation}\label{eq:1PK_F}
        \tns{P}=\partialder{\Psi(\tns{F})}{\tns{F}}.
    \end{equation}
    Objectivity of the strain energy function is inherently ensured by using the reduced formulation \cite{Ogden1985}
    \begin{equation}\label{eq:reduced_psi}
        \Psi=\Psi(\tns{F})=\Psi(\tns{C}),
    \end{equation}
    which is based on $\tns{C}$ and can thus be shown to be observer-independent (objective). The (symmetric) so-called second Piola-Kirchhoff stress tensors $\tns{S}$ is related to the first Piola-Kirchhoff stress tensor $\tns{P}$ and the strain energy through \cite{Holzapfel2000a}:
    \begin{equation}\label{eq:stress_hyper}
        \tns{P} = \tns{F}\tns{S}, \qquad \tns{S} = 2\partialder{\Psi(\tns{C})}{\tns{C}}.
    \end{equation}

    \paragraph{Functional bases} The strain energy function of an isotropic hyperelastic material must be a scalar-valued isotropic tensor function of $\tns{C}$ \cite{Truesdell1984}.
    The principal invariants \(I_i\), the principal stretches \(\lambda_i\), and the modified invariants \(\iota_i\) form so-called functional bases, in terms of which all scalar-valued isotropic tensor functions of $\tns{C}$ can be expressed \cite{Truesdell1984}. Thus, the strain energy function can be written as:
    \begin{equation}
        \Psi = \Psi(I_1, I_2, I_3) = \Psi(\lambda_1, \lambda_2, \lambda_3) = \Psi(\iota_1, \iota_2, I_3).
    \end{equation}
    
    The appropriate functional form of $\Psi$ for a specific material generally depends on material features, such as the composition, microstructure, or processing parameters. This information may be expressed in terms of some feature vector $\vec{f}$, leading to 
    \begin{equation}\label{eq:psi}
        \Psi = \Psi\bigl(I_1, I_2, I_3, \vec{f}\bigr) 
             = \Psi\bigl(\lambda_1, \lambda_2, \lambda_3, \vec{f}\bigr)
             = \Psi\bigl(\iota_1, \iota_2, I_3, \vec{f}\bigr).
    \end{equation}
    
    \paragraph{Incompressibility} Many important materials can be regarded as incompressible so that $J=1$. In this case, one has only two independent deformation variables, that is, ($\{\lambda_1, \lambda_2\}$, $\{I_1, I_2\}$ or $\{\iota_1, \iota_2\}$), while the respective third one is determined via the $J = I_3 = 1$ and thus $\lambda_3=(\lambda_1\lambda_2)^{-1}$. For incompressible isotropic hyperelastic materials, the strain energy functions can be postulated as
    \begin{equation}\label{eq:psi_iso}
        \Psi = \hat{\Psi}(I_1,I_2,\vec{
        f}) - \frac{p}{2}(I_3-1) = \hat{\Psi}(\lambda_1,\lambda_2,\lambda_3,\vec{
        f}) - p(J-1)= \hat{\Psi}(\iota_1,\iota_2,\vec{
        f}) - \frac{p}{2}(I_3-1),
    \end{equation}
    where $p$ denotes the hydrostatic pressure in the material determined from the boundary conditions. For the remainder of this work, the overset hat $\hat{(\cdot)}$ explicitly denotes strain energy functions for the incompressible case. In equation \eqref{eq:psi_iso}, we use the same symbol $p$ for the Lagrange multiplier ensuring incompressibility for all three different functional bases. However, the value that results for $p$ differs for these three cases even if the same boundary value problem is considered.     

    \paragraph{Separable principal stretch-based strain energy functions} Since the properties of isotropic materials do not depend on direction, principal stretch-based strain energy functions $\Psi(\lambda_1, \lambda_2, \lambda_3,\vec{f})$ must be symmetric with respect to the principal stretches. However, the space of all symmetric functions is vast, making it challenging to design suitable candidate functions. This complexity is significantly reduced by adopting a separable form, which is symmetric by definition and has proven effective in practice \cite{Xu2015}:
    \begin{equation}\label{eq:separable_comp}
        \Psi = \Psi(\lambda_1, \lambda_2, \lambda_3,\vec{f}) = \sum_{\alpha=1}^{3} \omega_1(\lambda_\alpha,\vec{f}) + \sum_{\alpha=1}^{3} \omega_{-1}(\nu_\alpha,\vec{f}) + \Omega(\lambda_1 \lambda_2 \lambda_3,\vec{f}),
    \end{equation} where $\omega_1$, $\omega_{-1}$, and $\Omega$ are univariate strain energy functions related to the changes in line, area, and volume elements, respectively. For incompressible materials, this simplifies further to:
    \begin{equation}\label{eq:separable_incomp}
        \Psi = \hat{\Psi}(\lambda_1, \lambda_2, \lambda_3,\vec{f}) - p(J-1) = \sum_{\alpha=1}^{3} \hat{\omega}_1(\lambda_\alpha,\vec{f}) + \sum_{\alpha=1}^{3} \hat{\omega}_{-1}(\lambda_\alpha^{-1},\vec{f}) - p(J-1).
    \end{equation}
    Often, even $\hat{\omega}_{-1}$ is discarded, reducing the strain energy function to a sum of a single scalar function $\hat{\omega}_{1}$ evaluated at the three principal stretches. This additive separability of the strain energy function is known as the Valanis--Landel hypothesis \cite{Valanis1967}, and has been widely adopted \cite{Ogden1972, Heinrich1997, Shariff2000, Attard2004}.
    
    \paragraph{Mixed invariant- and principal stretch-based strain energy functions} Any principal invariant-based strain energy function may be expressed in terms of principal stretches and vice versa \cite{Ehret2022}. That is, in principle, one has the freedom to represent strain energy either in terms of $\{\lambda_1, \lambda_2, \lambda_3\}$, $\{I_1, I_2, I_3\}$ or $\{\iota_1, \iota_2, I_3\}$. However, representations relying on just one of these functional bases may be complicated and sometimes lead to numerical instabilities \cite{Xu2015}. To overcome this problem, it can be helpful to rely on a fourth mixed functional basis $\{\lambda_1, \lambda_2, \lambda_3, I_1, I_2, I_3\}$. Using this basis, one often assumes an additive decomposition of the strain energy function \cite{Kaliske1999,Xiang2018,Davidson2013}:
    \begin{equation}\label{eq:fourthbasis}
        \Psi(I_1,I_2,I_3,\lambda_1,\lambda_2,\lambda_3,\vec{f}) = \Psi_I(I_1,I_2,I_3,\vec{f}) + \Psi_\lambda(\lambda_1,\lambda_2,\lambda_3,\vec{f}).
    \end{equation}
    While the first three functional bases we introduced are irreducible, the mixed fourth basis introduced in the context of \eqref{eq:fourthbasis} is reducible \cite{Itskov2015}, meaning it contains more elements than strictly necessary to span the space of interest.
    
    \section{Constitutive-Kolmogorov--Arnold Networks (CKANs)} \label{sec:methods} 
    
    This section summarizes the fundamentals of Kolmogorov–Arnold Networks (KANs) and introduces the Constitutive Kolmogorov–Arnold Network (CKAN) framework for interpretable and predictive material modeling. Standard types of neural networks like feedforward neural networks (FFNNs), recurrent neural networks (RNNs), convolutional neural networks (CNNs), and others are powerful approximators. However, their black-box nature limits interpretability and complicates adherence to physical constraints. KANs address these challenges by leveraging the Kolmogorov–Arnold representation theorem and representing the strain energy function using B-spline curves with learnable control points of local B-spline basis functions. This low-dimensional parameterization enables efficient training and direct incorporation of physical prior knowledge \cite{Liu2024a, Liu2024}.
        
    \subsection{Kolmogorov--Arnold Networks (KANs)}

\begin{figure}
        \centering
        \begin{subfigure}{0.46\linewidth}
        \centering
           \includegraphics[width=1\linewidth]{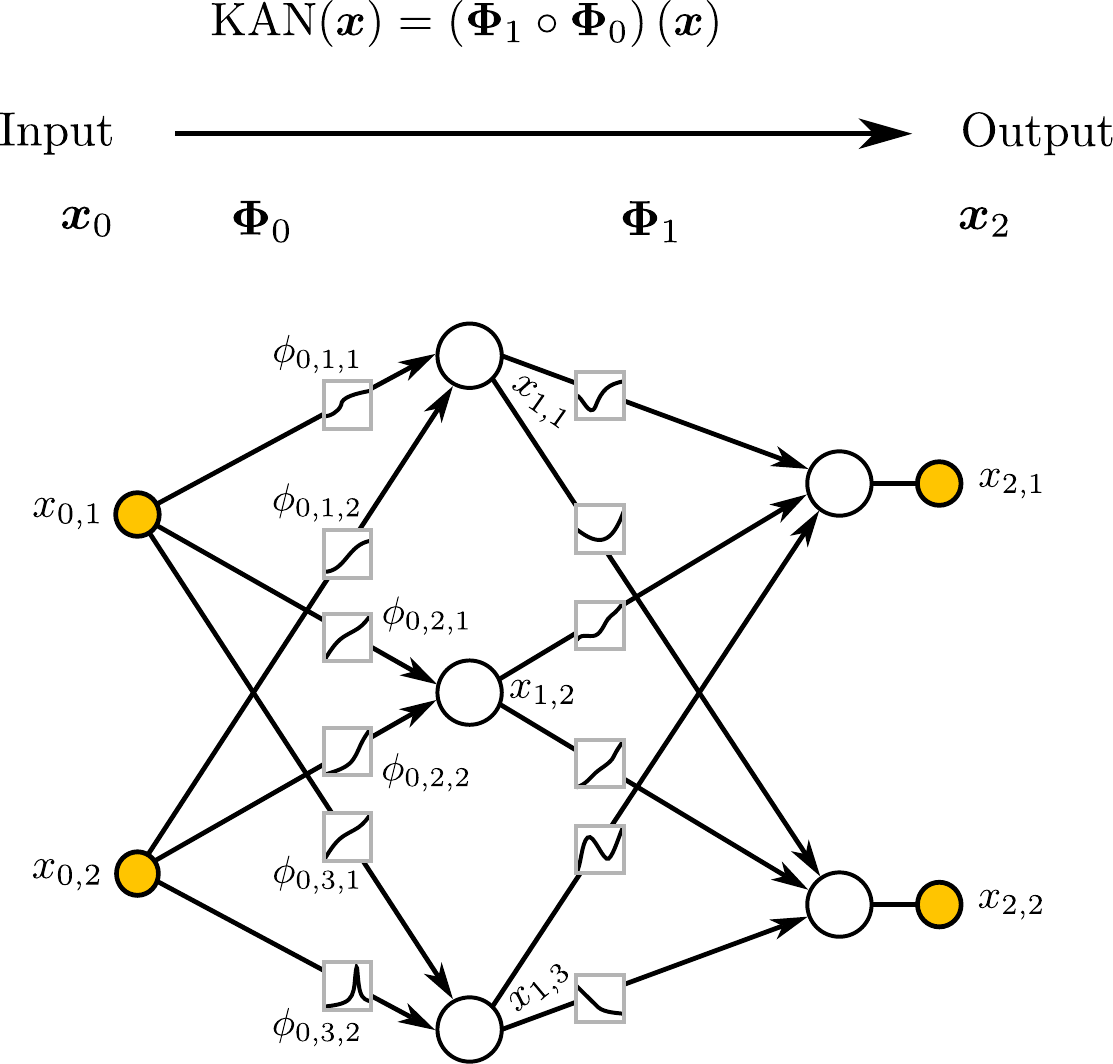}
        \caption{Kolmogorov--Arnold Network (KAN)}
        \label{fig:KAN} 
        \end{subfigure}
        \hfill
        \begin{subfigure}{0.46\linewidth}
        \centering
           \includegraphics[width=1\linewidth]{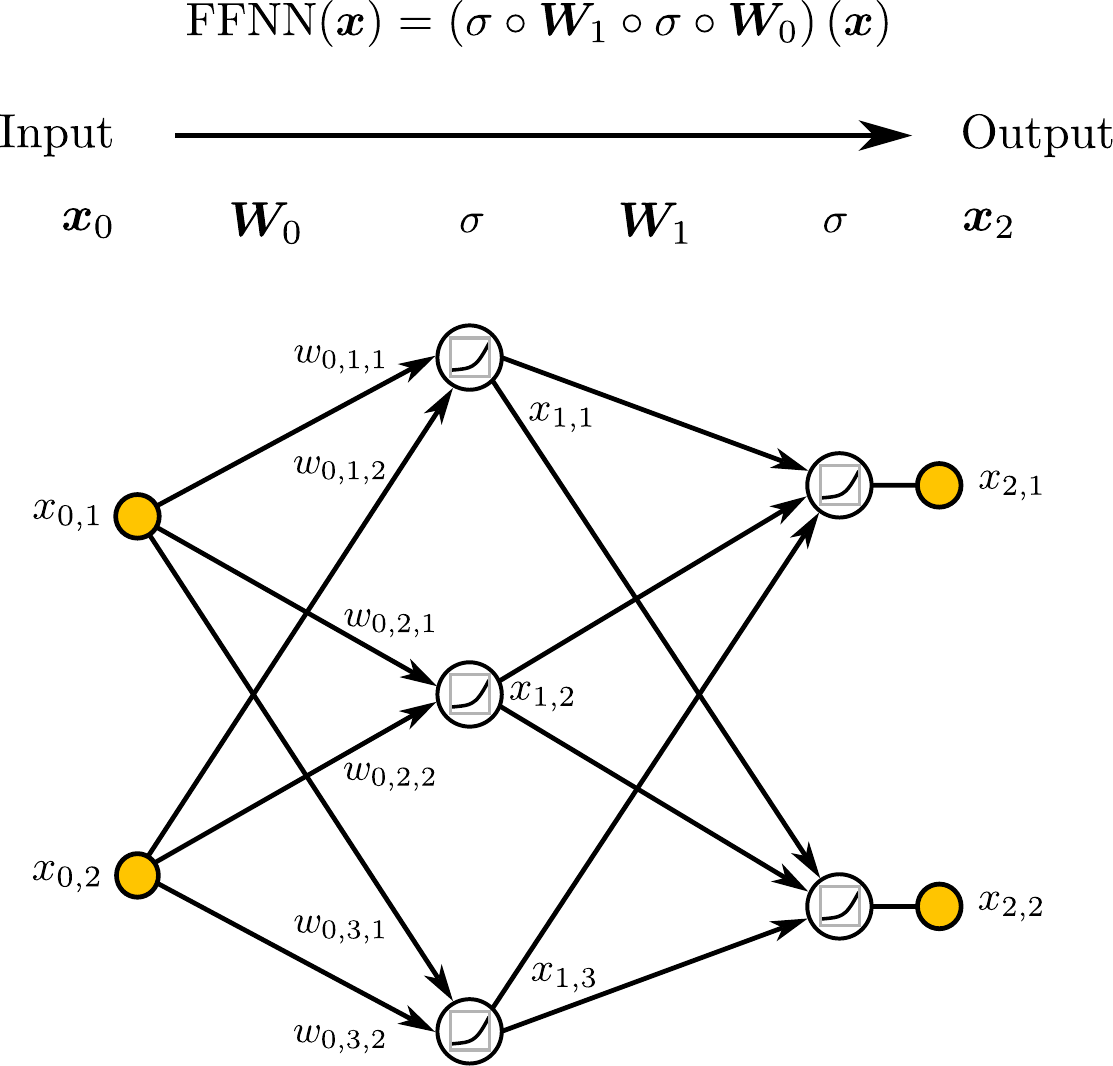}
            \caption{Feedforward neural network (FFNN)}
            \label{fig:FFNN} 
        \end{subfigure}       
        \caption{\textbf{Comparison of a Kolmogorov--Arnold Network (KAN) and a feedforward neural network (FFNN)}. A KAN (a) can be interpreted as a generalization of an FFNN (b) where the weights are no longer constants but nonlinear functions, typically spline functions of higher order, and where as a consequence the intra-neural (typically nonlinear) activation function $\sigma$ is no longer needed. While in an FFNN the weights $w_{l,j,i}$ are adjusted in a training process to learn a specific input-output relation, in a KAN the functions $\phi_{l,j,i}$ are. Orange nodes represent input and output nodes.}
        \label{fig:FFNN_KAN} 
    \end{figure}

    Kolmogorov--Arnold Networks (KANs) draw inspiration from the Kolmogorov--Arnold representation theorem (KART), which states that any continuous multivariate function $f:[0,1]^n\to\mathbb{R}$ can be represented as a finite composition of univariate continuous functions and addition operations:
    \begin{equation}\label{eq:KART}
        f(\vec{x}_0) = f(x_{0,1}, x_{0,2}, \dots, x_{0,n}) = \sum_{j=1}^{2n+1} \phi_{1,1,j} \left( \sum_{i=1}^{n} \phi_{0,j,i}(x_{0,i}) \right),
    \end{equation}
    where the $x_{0,1}$ are the elements of the vector $\vec{x}_0$, and the $\phi_{0,j,i}:[0,1]\to\mathbb{R}$ and $\phi_{1,1,j}:\mathbb{R}\to\mathbb{R}$ are univariate continuous functions.
    Equation \eqref{eq:KART} can be understood as a sort of neural network with $n$ input nodes, a middle layer with $2n+1$ nodes, and a single output node. The functions $\phi_{0,j,i}$ connect the input nodes to the nodes in the middle layer, and the functions $\phi_{1,j,i}$ connect the nodes in the middle layer to the output nodes, with a summation performed at each node over all incoming inputs.   
    Interpreting equation \eqref{eq:KART} as a neural network directly leads to the concept of Kolmogorov--Arnold Networks (KANs). Generally, a KAN may consist of $L$ layers, each containing $n_l$ nodes with $l=1,...,L$. Its topology can thus be characterized by the vector 
    \begin{equation}
        \bm{\mathcal{N}} = [n_0, n_1, \ldots, n_L].
    \end{equation}
    The output of layer $l$ is denoted by the vector $\vec{x}_l$. Specifically, its element $x_{l,i}$ is the output of node $i$ of layer $l$ and forms after passing through the function $\phi_{l,j,i}$ an input for node $j$ in layer $l+1$. The output of node $j$ in layer $l+1$ is the sum of all the inputs it receives from layer $l$, that is:
    \begin{equation}\label{eq:activation}
        x_{l+1,j} = \sum_{i=1}^{n_l} \phi_{l,j,i}(x_{l,i}), \qquad j = 1,2,\ldots, n_{l+1}.
    \end{equation}
    In the context of KANs, the functions   $\phi_{l,j,i}$ are usually referred to as activation functions.
    For the ease of notation, one can arrange them in a matrix $\vec{\Phi}_l$ so that
    \begin{equation}
        \vec{x}_{l+1} =
        \underbrace{
        \begin{bmatrix}
            \phi_{l,1,1}(\cdot) & \phi_{l,1,2}(\cdot) & \hdots & \phi_{l,1,n_l}(\cdot) \\
            \phi_{l,2,1}(\cdot) & \phi_{l,2,2}(\cdot) & \hdots & \phi_{l,2,n_l}(\cdot) \\
            \vdots              & \vdots              & \ddots & \vdots                \\
            \phi_{l,n_{l+1},1}(\cdot) & \phi_{l,n_{l+1},2}(\cdot) & \hdots & \phi_{l,n_{l+1},n_l}(\cdot) \\
        \end{bmatrix}}_{\vec{\Phi}_l \coloneqq}
        \vec{x}_{l},
    \end{equation}
   Generally, a KAN with $L$ layers then maps its input $\vec{x}_0$ to its output $\vec{x}_L$ via
    \begin{equation}\label{eq:KAN}
        \vec{x}_L = \text{KAN}(\vec{x}) = (\vec{\Phi}_{L-1} \circ \vec{\Phi}_{L-2} \circ \ldots \circ \vec{\Phi}_1 \circ \vec{\Phi}_0)\,(\vec{x}_0).
    \end{equation}
    Figure \ref{fig:FFNN_KAN} compares the architecture of a KAN to the one of a feedforward neural network (FFNN) with the same topology. Both have in common that they consist of layers of neurons, where, in general, each neuron in one layer may provide input to each neuron in the subsequent layer. However, while in the FFNN this input is simply multiplied with a scalar weight $w_{l,j,i}$, in the KAN, it passes through a function $\phi_{l,j,i}$. This equips KANs with a capability of approximating nonlinear relationships that allows them to eliminate the intra-neural activation function $\sigma$ required in FFNNs to this end.    
    The FFNN and KAN share a similar training process in which the $w_{l,j,i}$ and $\phi_{l,j,i}$, respectively, are adjusted such that, after training, the FFNN or KAN captures a certain input-output relation as well as possible. A scalar weight $w_{l,j,i}$ can be interpreted as a constant function, whereas in KANs the $\phi_{l,j,i}$ are usually chosen as spline functions of higher order.
    In this sense, a KAN can be viewed as a generalization of an FFNN, where the weights are no longer simple constant values (spline functions of order zero) but rather spline functions of higher order. This gives KANs much greater flexibility to approximate nonlinear relationships between subsequent layers. As a result, KANs can eliminate the need for intra-neural nonlinear activation functions $\sigma$, effectively choosing $\sigma$ as an identity function and removing it from the equations.
    Formally, the input-output relation of a KAN is given by equation \eqref{eq:KAN}, while for an FFNN with $L$ layers, it is described by the following equation:
    \begin{equation}
        \text{FFNN}(\vec{x}_L) = (\sigma \circ \vec{W}_{L-1} \circ \sigma \circ \vec{W}_{L-2} \circ \ldots \circ \sigma \circ \vec{W}_1 \circ \sigma \circ \vec{W}_0)\,(\vec{x}_0),
    \end{equation}
    where the matrices $\vec{W}_{l}$ assemble the scalar weights $w_{l,j,i}$, analogous to how the matrices $\vec{\Phi}_{l}$ in KANs assemble the functions $\phi_{l,j,i}$. 

    \subsection{Constitutive modeling with KANs} \label{sec:CANN-KAN}
\begin{figure}[h]
        \centering
        \vspace{0.2cm}
        \includegraphics[width=0.85\linewidth]{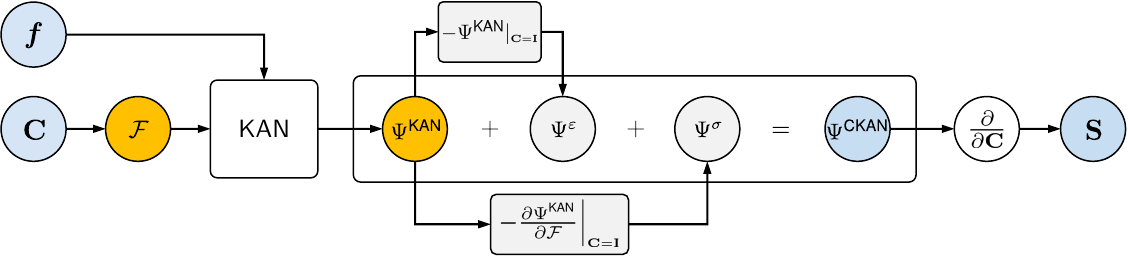}
        \caption{\textbf{Schematic overview of the Constitutive Kolmogorov--Arnold Network (CKAN) framework}: The right Cauchy Green deformation tensor $\tns{C}$ and the feature vector $\vec{f}$ serve as input to the CKAN. Depending on the chosen functional basis $\mathcal{F}$ (principal stretches, principal invariants, modified invariants, mixed basis) and the feature vector, the strain energy function $\Psi^\text{KAN}$ is represented by a KAN. For example, in a principal invariant-based CKAN, one of the KANs from Figure \ref{fig:cKAN_architectures} can be substituted in the box {\fboxsep=1pt\fbox{KAN}}. 
        The correction terms $\Psi^\sigma$ and $\Psi^\varepsilon$ ensure the normalization of the stress and strain energy function in the undeformed configuration $(\tns{C}=\tns{I})$, yielding the total strain energy function $\Psi^\text{CKAN}$. The correction terms depend on $\Psi^\text{KAN}$ and its derivatives with respect to the elements of the functional basis $\partial\Psi^\text{KAN}/\partial\mathcal{F}$, evaluated in the undeformed configuration. They iteratively adapt during training. Finally, the stress $\tns{S}$ is obtained as the gradient of $\Psi^\text{CKAN}$ with respect to $\tns{C}$.}
        \label{fig:overview}
    \end{figure}

    In this work, we aim to learn the strain energy function of an isotropic hyperelastic material from experimental data. To achieve this, we develop Constitutive Kolmogorov–Arnold Networks (CKANs), a machine learning architecture tailored to model the constitutive behavior of hyperelastic materials. Figure \ref{fig:overview} provides a schematic overview of the CKAN architecture.
    
    The first design choice in constructing CKANS is selecting an appropriate functional basis $\mathcal{F}$ to describe the deformation state as input. In this paper, we compare the four functional bases introduced in Section \ref{sec:conti}: $\{\lambda_1, \lambda_2, \lambda_3\}$, $\{I_1, I_2, I_3\}$, $\{\iota_1, \iota_2, I_3\}$, and $\{\lambda_1, \lambda_2, \lambda_3, I_1, I_2, I_3\}$. Further, we include any additional material features $\vec{f}$ into the input space. For example, $\vec{f}$ may contain information about the processing conditions or microstructure of a material, or about certain properties that are not measured in a classical strain-stress test, such as the material's hardness.    
    The output of the CKAN should be an approximation of the material's stress response derived from the strain energy function $\Psi^\text{CKAN}$, which completely characterizes the nonlinear hyperelasticity of the material. This strain energy function is defined by:
    \begin{equation}\label{eq:Psi_total}
        \Psi^\text{CKAN}(\mathcal{F},\vec{f}) \coloneqq \Psi^\text{KAN}(\mathcal{F},\vec{f}) + \Psi^\sigma(J,\vec{f}) + \Psi^\varepsilon(\vec{f}). 
    \end{equation}
    The strain energy contribution $\Psi^\text{KAN}$ is represented by a KAN and captures the basic constitutive behavior of the material. To ensure that both stress and strain energy vanish in the undeformed reference configuration, the correction terms $\Psi^\sigma$ and $\Psi^\varepsilon$ are introduced. These terms are not modeled by a KAN but instead depend on $\Psi^\text{KAN}$ and its derivatives with respect to the elements of the functional basis, evaluated in the undeformed state, see Figure \ref{fig:overview}. Details on each contribution to the total strain energy function $\Psi^\text{CKAN}$ are provided below.

    The CKAN framework with its resulting total strain energy function $\Psi^\text{CKAN}$ inherently ensures thermodynamic consistency, objectivity, material symmetry, symmetry of the second Piola–Kirchhoff stress tensor, as well as zero stress and zero strain energy in the undeformed state. 

    \paragraph{Training of CKANs} As mentioned above, only the contribution $\Psi^\text{KAN}$ is represented by a KAN. As a result, the only trainable parameters of the total strain energy function $\Psi^\text{CKAN}$ are the control points of the univariate spline activation functions associated with $\Psi^\text{KAN}$. Training the CKAN is thus reduced to optimizing these control points. In practice, direct measurements of the strain energy are typically unavailable. Instead, only stress data corresponding to specific tested strain states are accessible. As a result, the loss function for training the CKAN cannot be based on a direct comparison between $\Psi^\text{CKAN}$ and experimentally measured strain energies. Instead, it must rely on comparing their derivatives---i.e., the stresses. This requires differentiation of $\Psi^\text{CKAN}$, yielding the stress tensors via \eqref{eq:stress_hyper}, see Figure \ref{fig:overview}. Detailed expressions for these stress tensors, corresponding to the different functional bases, are provided in Appendix \ref{app:1PK_general}. Since all mathematical operations in a CKAN are fully differentiable, the equations from Appendix \ref{app:1PK_general} can be directly applied using automatic differentiation. The computed stresses are then compared with experimentally obtained stress values. Training is performed by minimizing the loss function $\mathcal{L}_\mathrm{data}$, which is defined in Appendix \ref{app:loss}.
    
\begin{figure}
        \centering
        \begin{subfigure}{0.31\linewidth}
        \centering
            \includegraphics[width=\linewidth]{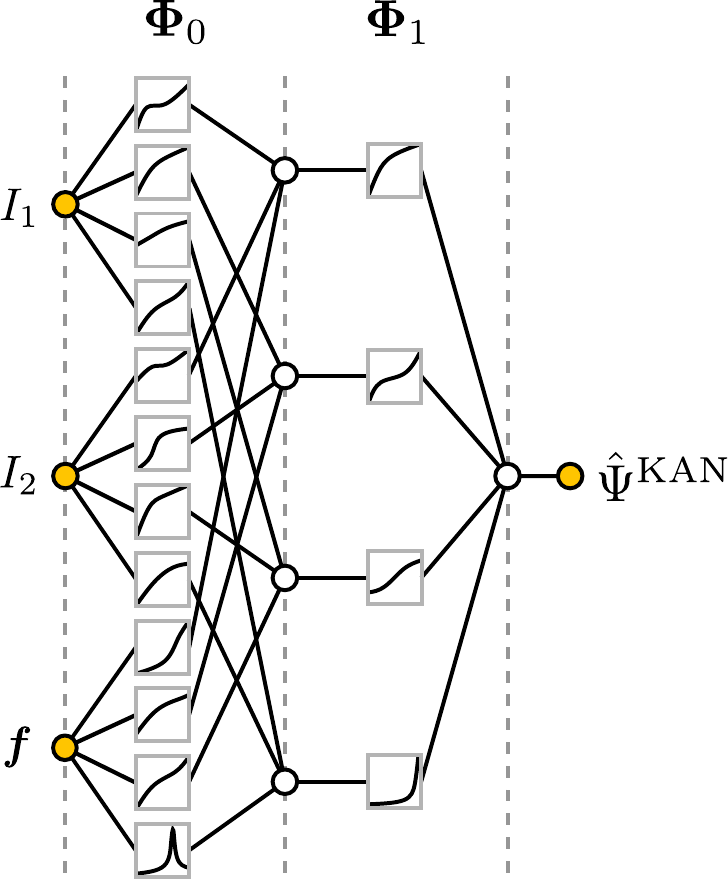}
            \caption{CKAN with topology $\bm{\mathcal{N}}=[3,4,1]$.}
            \label{fig:CKAN_intro}
        \end{subfigure}
        \hspace{2.cm}
        \begin{subfigure}{0.31\linewidth}
            \centering
            \includegraphics[width=0.9\linewidth]{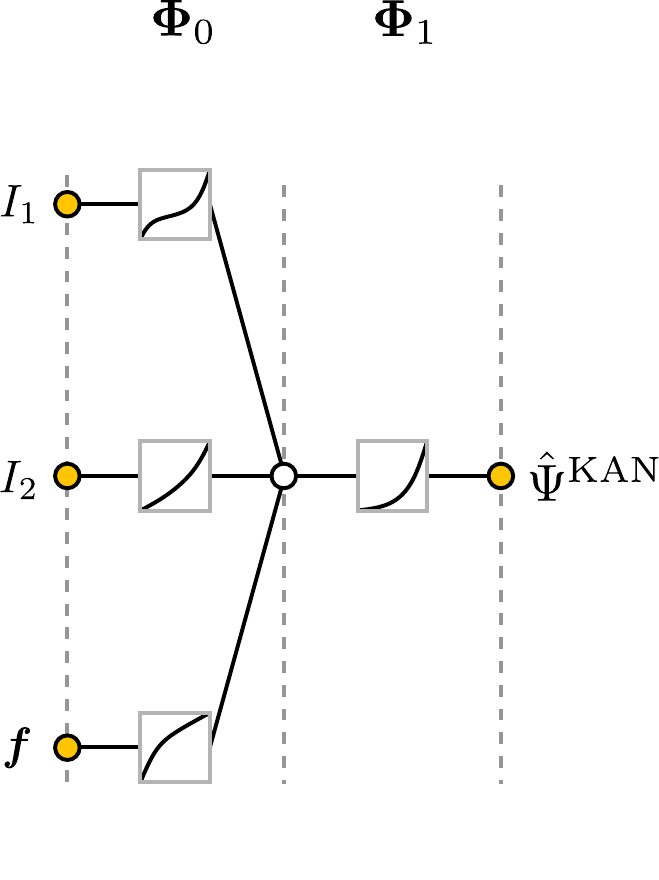}
            \caption{CKAN with topology $\bm{\mathcal{N}}=[3,1,1]$}
            \label{fig:CKAN_architecture_1}      
        \end{subfigure}
        \caption{\textbf{Examples of Kolmogorov--Arnold Networks (KANs)} representing the strain energy function $\Psi^\text{KAN}$, where the functional basis $\mathcal{F} = \{I_1, I_2, I_3\}$ was chosen and reduced to the setting of an incompressible material so that $I_3$ is no longer required as an explicit part of the input. Depending on the complexity of the constitutive behavior of the material of interest, the number of layers and neurons per layer can be adjusted.
        } 
        \label{fig:cKAN_architectures}
    \end{figure}

    \paragraph{Basic strain energy represented by KAN}
    The basic constitutive behavior of the material is captured by the strain energy contribution $\Psi^\text{KAN}$ represented by a KAN of arbitrary width and depth. The KAN takes as input the elements of the functional basis $\mathcal{F}$ and the feature vector $\vec{f}$. For example, Figure \ref{fig:cKAN_architectures} shows two possible KAN architectures, both using the principal invariants as the functional basis.
    For a strain energy function relying on the principal stretches as the functional basis, a more intricate architecture is required to ensure that $\Psi^\text{KAN}$ remains symmetric with respect to its arguments. This architecture is built upon the separable ansatz \eqref{eq:separable_comp} and \eqref{eq:separable_incomp}. Further details are provided in Appendix \ref{app:architecture_prin_stretch}. The only parameters optimized during training are the control points of the spline activation functions in the KAN representing $\Psi^\text{KAN}$.
    
    \paragraph{Normalization of stress} It can be expected that $\Psi^\text{KAN}$ will approximate stresses with reasonable quality. Minor inaccuracies are generally acceptable for most practical applications. However, there is an important exception: at zero strain, the stresses should be exactly zero to avoid numerical problems in computer simulations using CKAN-based strain energy functions. Therefore, from a practical point of view, it is advantageous to add a correction term $\Psi^\sigma$ to $\Psi^\text{KAN}$ to ensure that the stress at zero strain is also zero within numerical precision. To determine $\Psi^\sigma$, one can simply compute the stress that would result from $\Psi^\text{KAN}$ at zero strain, and then construct a potential $\Psi^\sigma$ such that the stresses resulting from its first derivatives exactly cancel those resulting from $\Psi^\text{KAN}$ at zero strain, cf. \cite{Dal2023, Linden2023}. Appendix \ref{app:normalization} demonstrates that $\Psi^\sigma$ generally takes on the form 
    \begin{equation}\label{eq:stress_offset}
        \Psi^\sigma(J, \vec{f}) \coloneqq -\Xi(\vec{f}) \, (J-1).
    \end{equation}
    Notably, $\Psi^\sigma(J, \vec{f})$ is zero in the undeformed state, where $J=1$. Furthermore, Appendix \ref{app:normalization} provides equations for computing $\Xi$ depending on the chosen functional basis. However, it is important to note that $\Xi$ depends on the derivative of $\Psi^\text{KAN}$ with respect to the functional basis, evaluated in the reference configuration. Consequently, $\Xi$ continuously adapts during training as $\Psi^\text{KAN}$ evolves and is \emph{not} computed in a post-processing step, but an integral part of the network architecture, cf. Figure \ref{fig:overview}. For isotropic incompressible materials, the stress normalization condition is satisfied through the hydrostatic pressure $p$ and its associated stress contribution, allowing $\Psi^\sigma$ to be set to zero.

    \paragraph{Normalization of strain energy}
    Since training the KAN relies on minimizing the model's stresses to experimentally observed values, and since stresses are related to first derivatives of $\Psi^\text{KAN}$, $\Psi^\text{KAN}$ may differ after the training from the real strain energy by an arbitrary constant. Since, by definition, $\Psi^\sigma$ vanishes in the undeformed configuration, applying $\Psi^\sigma$ ensures only zero stresses in the undeformed configuration but not zero strain energy. Therefore, one has to evaluate $\Psi^\text{KAN}$ at zero strain and add this constant $\Psi^\varepsilon$ multiplied by minus one to $\Psi^\text{KAN}$ to ensure zero strain energy in the undeformed state, i.e., 
    \begin{equation}\label{eq:energy_free}
        \Psi^\varepsilon(\vec{f})\coloneqq-\Psi^\text{KAN}|_{\tns{C}=\tns{I}}.
    \end{equation}

    \paragraph{Implementation} We implemented the CKAN framework using KAN~1.0 \cite{Liu2024a}, built on PyTorch. Following \cite{polo2024monokan}, the B-spline activation functions can be replaced with cubic Hermite splines, with constraints imposed on the control points to enforce partial monotonicity of the KAN. In this context, partial monotonicity allows the user to specify, for each input individually, whether the KAN's output should be monotonic (either non-decreasing or non-increasing) or remain unrestricted. This modification was adopted in the CKAN framework.
    For example, enforcing non-decreasing monotonicity of the strain energy function with respect to the strain inputs can improve robustness and extrapolation. Further technical details on the implementation are provided in Appendix~\ref{sec:implement}.
        
    \section{Symbolic Constitutive Modeling} \label{sec:sparse_symbol}

    Deriving an interpretable symbolic constitutive model requires two additional steps. First, the dense CKAN must be sparsified during training by iteratively removing redundant activation functions, resulting in a more compact network that reduces overfitting and enhances generalizability. Second, the remaining activation functions must be expressed in symbolic form through regression.

    
    \subsection{Sparsificaiton}\label{sec:sparse}
    As depicted in Figure \ref{fig:cKAN_architectures}, KAN layers with varying widths can be stacked arbitrarily. In practice, the optimal topology $\bm{\mathcal{N}}$ of the CKAN architecture for a given dataset is often unknown unless the dataset is derived from a known symbolic expression, such as a predefined strain energy function. To address this problem, we propose starting with a large CKAN and employing sparsity regularization during training to determine its optimal topology automatically.
    In standard FFNNs, $L_1$ regularization is applied to weights to promote sparsity \cite{Abdolazizi2023, McCulloch2024}. In analogy, we regularize the $L_1$ norm of activation functions $\phi_{l,j,i}$. Passing an input batch with $n_s$ samples to a KAN results in $n_s$ inputs $x^{s}_{l,i}$ to each activation function $\phi_{l,j,i}$. Following \cite{Liu2024a}, the $L_1$ norm of an activation function $\phi_{l,j,i}$ is then the average magnitude over these $n_s$ samples:
    \begin{equation}\label{eq:L1_activation}
        |\phi_{l,j,i}|_1 \coloneqq \frac{1}{n_s} \sum_{s=1}^{n_s} \left| \phi_{l,j,i}(x^{s}_{l,i})\right|.
    \end{equation}
    Based on the $L_1$ norm of a single activation function, the $L_1$ norm of a KAN layer $\tns{\Phi}_l$ with $n_\text{in}$ inputs and $n_\text{out}$ outputs is given by:
    \begin{equation}\label{eq:L1_layer}
        |\tns{\Phi}_l|_1 \coloneqq \sum_{i=1}^{n_\text{in}}\sum_{j=1}^{n_\text{out}} |\phi_{l,j,i}|_1.
    \end{equation}
    Based on the findings that $L_1$ regularization alone is insufficient for sparsification, \cite{Liu2024a} introduced the entropy $S$ of KAN layer $\tns{\Phi}_l$ as:
    \begin{equation}\label{eq:entropy}
        S(\tns{\Phi}_l) = -\sum_{i=1}^{n_\text{in}}\sum_{j=1}^{n_\text{out}}\frac{|\phi_{l,j,i}|_1}{|\tns{\Phi}_l|_1} \log_2 \left( \frac{|\phi_{l,j,i}|_1}{|\tns{\Phi}_l|_1} \right).
    \end{equation}
    Then, the total loss function used for training the network is the sum of the data loss $\mathcal{L}_\text{data}$ (characterizing the difference between network output and training data) and the regularization loss
    \begin{equation}\label{eq:loss_total}
        \mathcal{L}_\text{total} = \mathcal{L}_\text{data} + \Lambda \left( \mu_1 \sum_{l=0}^{L-1} |\tns{\Phi}_l|_1 + \mu_2 \sum_{l=0}^{L-1} S(\tns{\Phi}_l)  \right),
    \end{equation}
    where the parameters $\mu_1$ and $\mu_2$ are usually set to $\mu_1 = \mu_2 =1$, and the hyperparameter $\Lambda$ controls the overall regularization magnitude.
    
    Evaluating the importance of edges using the $L_1$ norm of an activation function may be limiting, as it considers only local information. The most recent MultKAN paper \cite{Liu2024} addresses this limitation by introducing attribution scores for edges and nodes. These scores are computed iteratively, propagating importance backward from the output layer to the input layer. While we have successfully obtained sparse and very accurate CKANs using only $L_1$ and entropy regularization, we anticipate that incorporating these attribution scores into our framework in future research may further enhance the interpretability and performance of CKANs.

    \subsection{Pruning}  \label{sec:prune}   
    After training with the sparsification regularization \eqref{eq:loss_total}, the network may be pruned on the node level. Pruning of node $(l,i)$ is based on its incoming and outgoing scores $I_{l,i}$ and $O_{l,i}$, respectively, defined by:
    \begin{equation}
        I_{l,i}=\max_k \{ |\phi_{l-1,i,k}|_1 \}, \qquad O_{l,i}=\max_j \{ |\phi_{l,j,i}|_1 \}.
    \end{equation}
    If either $I_{l,i}$ or $O_{l,i}$ are below a certain threshold hyperparameter $\theta$, the corresponding node $(l,i)$ is pruned.    
    However, in our application, pruning was not beneficial, as it often removed nodes that had minimal impact on the absolute value of the strain energy function but substantially affected its derivatives, which are critical for accurate stress predictions. Additionally, it introduces the additional hyperparameter $\theta$.
    Therefore, we fully relied on sparsification regularization. If the spline-based activation function effectively approximated the zero function, we replaced it with a zero function during the symbolification process described in the following section.
     Nonetheless, exploring alternative pruning strategies in future research---such as those based on the non-local attribution scores proposed in \cite{Liu2024}---could be valuable.

    \subsection{Symbolification}\label{sec:symbolification}
    In principle, a user could manually inspect the spline activation functions $\phi_{l,j,i}$ and attempt to identify a symbolic counterpart. However, for large CKAN architectures, automatic identification is generally more practical. To determine a suitable symbolic replacement, the fitting performance of symbolic univariate candidate functions $f_{l,j,i}$ from a predefined function library $\mathcal{B}$ is evaluated for a given $\phi_{l,j,i}$. The function library includes common elementary functions such as monomials or the exponential function, e.g., $\mathcal{B}=\{x, x^2,\ldots,\ln(x),\exp(x),\cos(x), \ldots\}$. If the user has prior knowledge and expects a specific symbolic function, it can also be included in $\mathcal{B}$.
    
    Even if the graphs of the activation function $\phi_{l,j,i}$ and a candidate function $f_{l,j,i}$ appear similar, directly replacing $\phi_{l,j,i}$ with $f_{l,j,i}$ rarely yields a good fit. Rather, typically shifting and scaling of input and output is additionally required.  Thus, identifying a suitable symbolic expression translates to finding a suitable candidate function $f_{l,j,i}\in\mathcal{B}$ and corresponding affine parameters---$a_{l,j,i} \, , \, b_{l,j,i} \, , \, c_{l,j,i} \, , \, d_{l,j,i}\in\mathbb{R}$---such that
    \begin{equation}
        \phi_{l,j,i}(x_{l,i}) \approx y_{l,j,i}( x_{l,i}) \coloneqq c_{l,j,i}\, f_{l,j,i}(a_{l,j,i}\, x_{l,i} + b_{l,j,i}) + d_{l,j,i}.
    \end{equation} 
        To solve this optimization problem we first evaluate the activation function $\phi_{l,j,i}$ at $n_s$ sample points $x_{l,i}^{s}$ to obtain $\phi_{l,j,,i}^{s} = \phi_{l,j,,i}(x_{l,i}^{s})$. The sampling range from which the tuples $\{(x_{l,i}^{s},\phi_{l,j,i}^{s})\}_{s=1}^{n_s}$ are drawn, as well as the spacing between these tuples, is crucial for accurately identifying the symbolic function $f_{l,j,i}$ and the associated affine parameters. Details of the sampling strategy are discussed in  Appendix \ref{app:symbol_caveats}.
    After sampling, the affine parameters are determined for each $f_{l,j,i}\in\mathcal{B}$, by minimizing the loss function 
    \begin{equation}
        \mathcal{L}_\text{affine} = \frac{1}{n_s} \sum_{s=1}^{n_s} \left( \frac{\phi_{l,j,i}^{s} - y_{l,j,i}( x_{l,i}^s)}{\phi_{l,j,i}^{s}} \right)^2 + \frac{1}{n_s} \sum_{s=1}^{n_s} \left( \frac{{\phi'}_{l,j,i}^{s} - y'_{l,j,i}( x_{l,i}^s)}{{\phi'}_{l,j,i}^{s}} \right)^2,
    \end{equation}
    where the second term incorporates information on the derivative 
    \begin{equation}\label{eq:loss_affine}
        y'_{l,j,i}(x_{l,i}^s)= a_{l,j,i}\,c_{l,j,i}\,f'_{l,j,i}(a_{l,j,i}\, x_{l,i}^s + b_{l,j,i}),
    \end{equation}
    which can be computed analytically since $f_{l,j,i}'$ is available in symbolic form. The derivatives ${\phi'}_{l,j,i}^{s}$ can be computed either numerically or analytically using established formulas for spline derivatives. The goodness of the fit of all $f_{l,j,i}\in\mathcal{B}$ and the corresponding affine parameters is ranked according to a suitable performance metric, such as the coefficient of determination $R^2$, and the best is selected as the symbolic counterpart of $\phi_{l,j,i}^{s}$.
    
    In the original KAN \cite{Liu2024a} and the recent MultKAN \cite{Liu2024} paper, this fitting process is performed through an initial iterative grid search over $a_{l,j,i}$ and $b_{l,j,i}$, followed by linear regression to determine $c_{l,j,i}$ and $d_{l,j,i}$.
    In contrast, we simultaneously optimized these parameters using the BFGS-B optimizer. In our examples, this approach led to a more accurate fit of the symbolized activation functions and significantly improved computational speed, particularly when working with an extensive function library and a large number of extracted sample points $n_s$. Moreover, this method allows for applying individual box constraints to the affine parameters, enhancing flexibility and control in the optimization process.
    For instance, the monotonicity of the spline activation functions can be preserved in the symbolic counterpart by selecting candidate functions $f_{l,j,i}$ from a function library that contains only monotonic functions, and restricting $a_{l,j,i}\,,\,c_{l,j,i}\in\mathbb{R}_{\geq 0}$. As a result, the monotonicity of the strain energy function with respect to its inputs is also inherited by its symbolic counterpart, given that a KAN consists of a composition of univariate functions and positively weighted sums, cf. \eqref{eq:activation}.
    
    \section{Numerical examples} \label{sec:results}
    
    In this section, we assess the performance and capabilities of the proposed Constitutive Kolmogorov--Arnold Networks (CKANs) using four experimental datasets. First, we examine Treloar's \cite{Treloar1944} and Kawabata's \cite{Kawabata1981} experimental data for rubber materials, widely regarded as benchmark datasets in hyperelasticity. Next, we study human brain (cortex) tissue data \cite{Budday2017}, which exhibit complex nonlinear behavior and tension-compression asymmetry. Finally, we use data from Ecoflex silicone polymers with varying Shore hardness values \cite{Liao2020a} to demonstrate how CKANs cannot only describe but also predict the mechanical properties of materials. 
    
    To train a CKAN, we minimize the difference between the predicted and experimentally measured first Piola–Kirchhoff stresses. In Appendix~\ref{app:1PK_general}, we derive the general stress expressions for different functional bases. In all examples, we assume incompressibility, a common assumption for rubber and brain tissue. The hydrostatic pressure \(p\) is determined from the assumption of plane stress, which is justified by the loading conditions and the shape of the specimens in the experiments. The detailed stress equations for biaxial tensile and simple shear tests under homogeneous deformation, necessary to reproduce the numerical examples, can be found in Appendix~\ref{app:HomogenousModes}.
    
    For simplicity, we report in this section and the related appendices
    only \(\hat{\Psi}^\text{KAN}\), as the normalization terms of the strain energy $\Psi^\sigma$ and $\Psi^\varepsilon$ directly follow from \eqref{eq:stress_offset} and \eqref{eq:energy_free}, respectively. For an isotropic incompressible material, the stress-free reference configuration is guaranteed via the stress contribution of the hydrostatic pressure $p$. For brevity, the feature vector \(\vec{f}\) is dropped in the notation where not needed explicitly.
    
    \subsection{Multi-axial loading of vulcanized rubber: Treloar's and Kawabata's experiments}\label{sec:rubber}
    
    The performance of constitutive models for rubber-like materials is often evaluated using the classical experimental dataset on vulcanized rubber reported by Treloar \cite{Treloar1944}. Several studies have used this prototypical dataset of a highly nonlinear, elastic, and incompressible material to assess and rank the validity of material models \cite{Steinmann2012, Marckmann2006}. Treloar's dataset includes uniaxial tension, equibiaxial tension, and pure shear experiments. While many material models in the literature fit a single deformation mode well, most struggle to capture all three deformation modes simultaneously.
    
\begin{figure}[h]
            \centering
            \begin{subfigure}{0.46\linewidth}
                \hspace{0.25cm}
                \includegraphics[width=\linewidth]{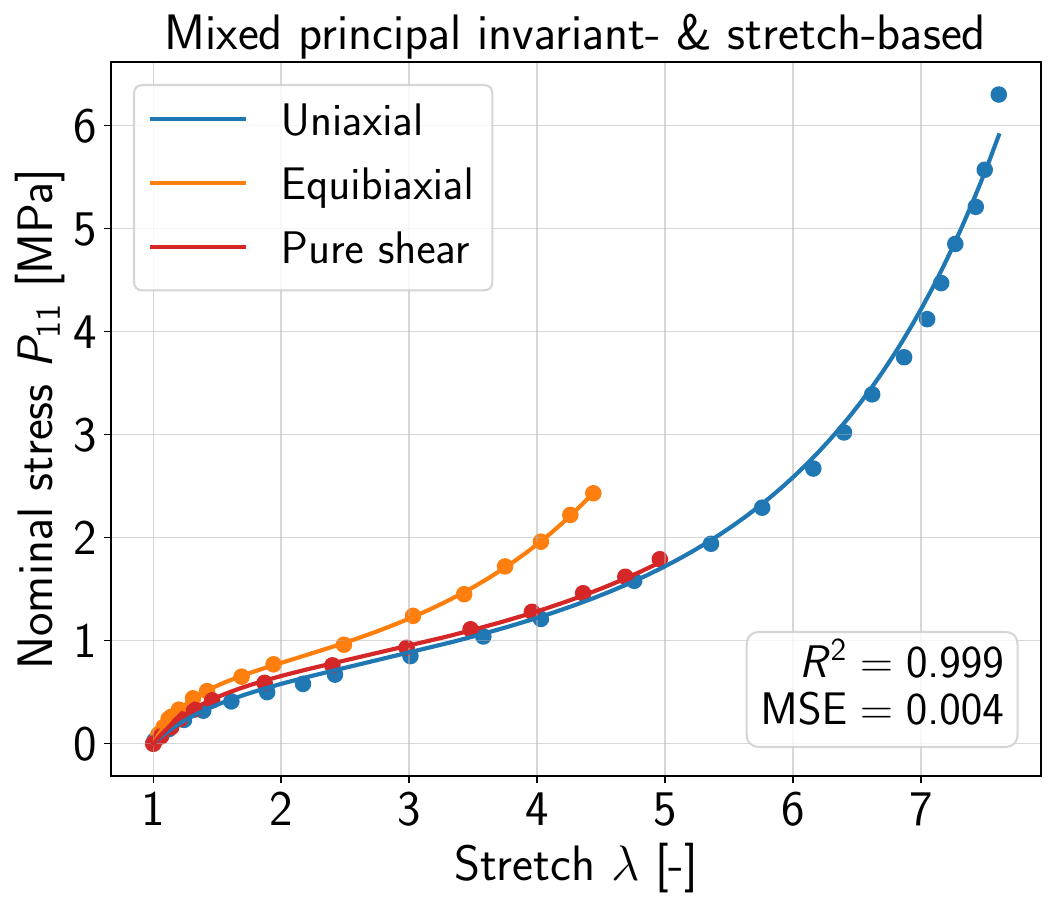}
                \caption{Fitting performance on Treloar's data.}
                \label{fig:result_treloar_invar_ps}
            \end{subfigure}
            \hfill
            \begin{subfigure}{0.46\linewidth}
                \centering
                \includegraphics[width=0.75\linewidth]{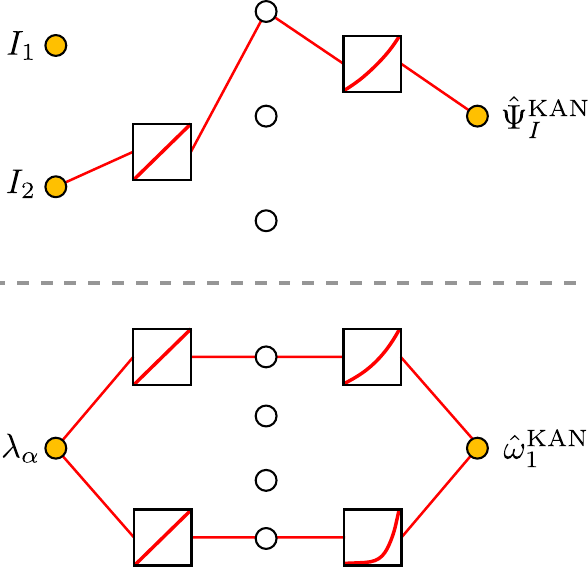}
                \vspace{0.8cm}
                \caption{CKAN architecture.}
                \label{fig:architecture_treloar_invar_ps}
            \end{subfigure}
            \\[2ex]
            \begin{subfigure}{0.48\linewidth}
                \includegraphics[width=\linewidth]{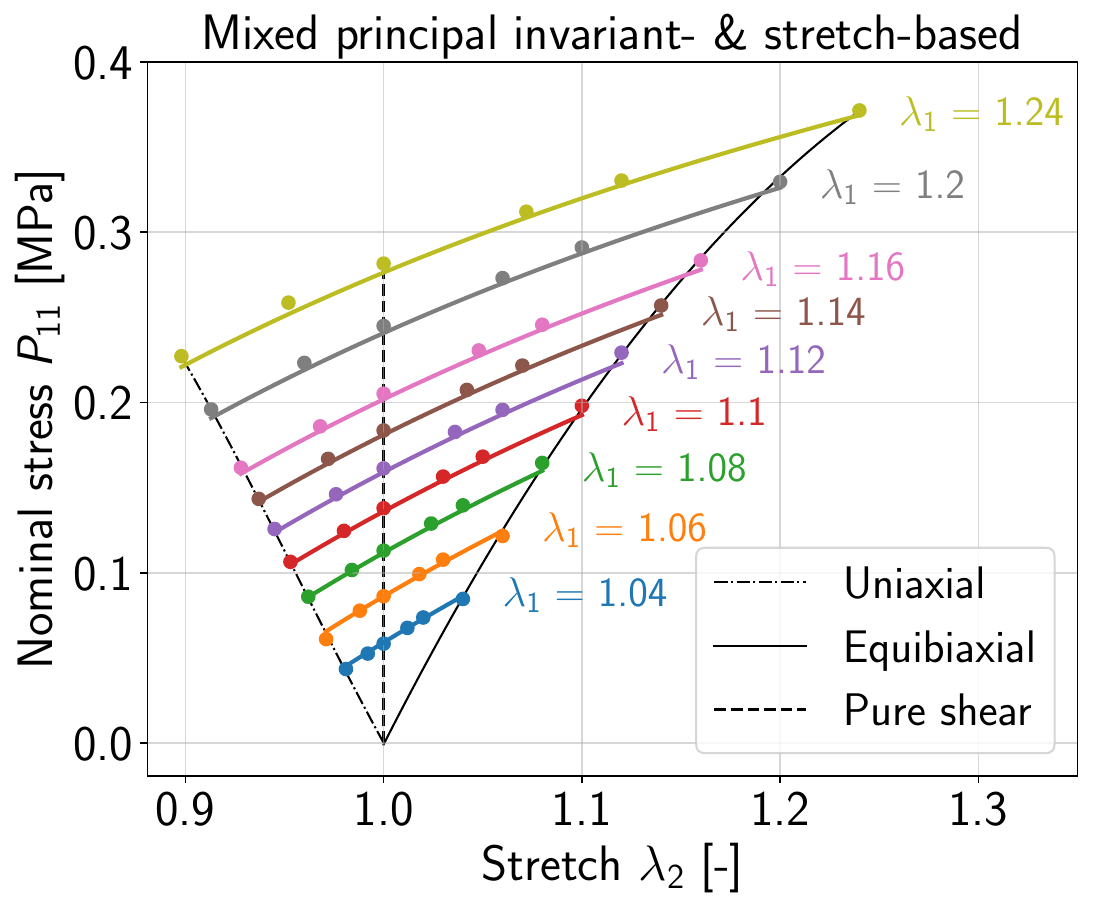}
                \caption{$P_{11}$ generalization: small stretch regime of Kawabata's data.}
                \label{fig:kawabata_P1_small_stretch_invar_ps}
            \end{subfigure}
            \hfill
            \begin{subfigure}{0.48\linewidth}
                \includegraphics[width=\linewidth]{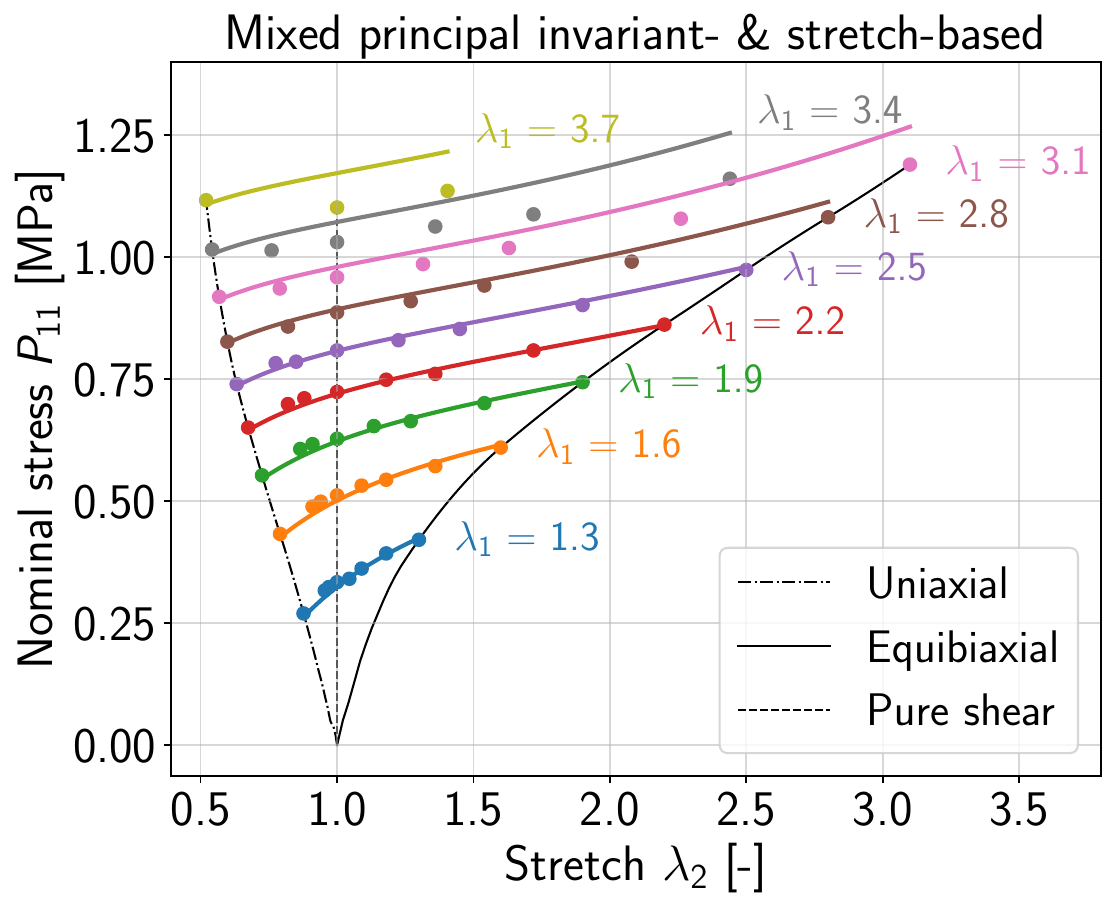}
                \caption{$P_{11}$ generalization: large stretch regime of Kawabata's data.}
                \label{fig:kawabata_P1_large_stretch_invar_ps}
            \end{subfigure}
            \\[2ex]
            \centering
            \begin{subfigure}{0.48\linewidth}
                \includegraphics[width=\linewidth]{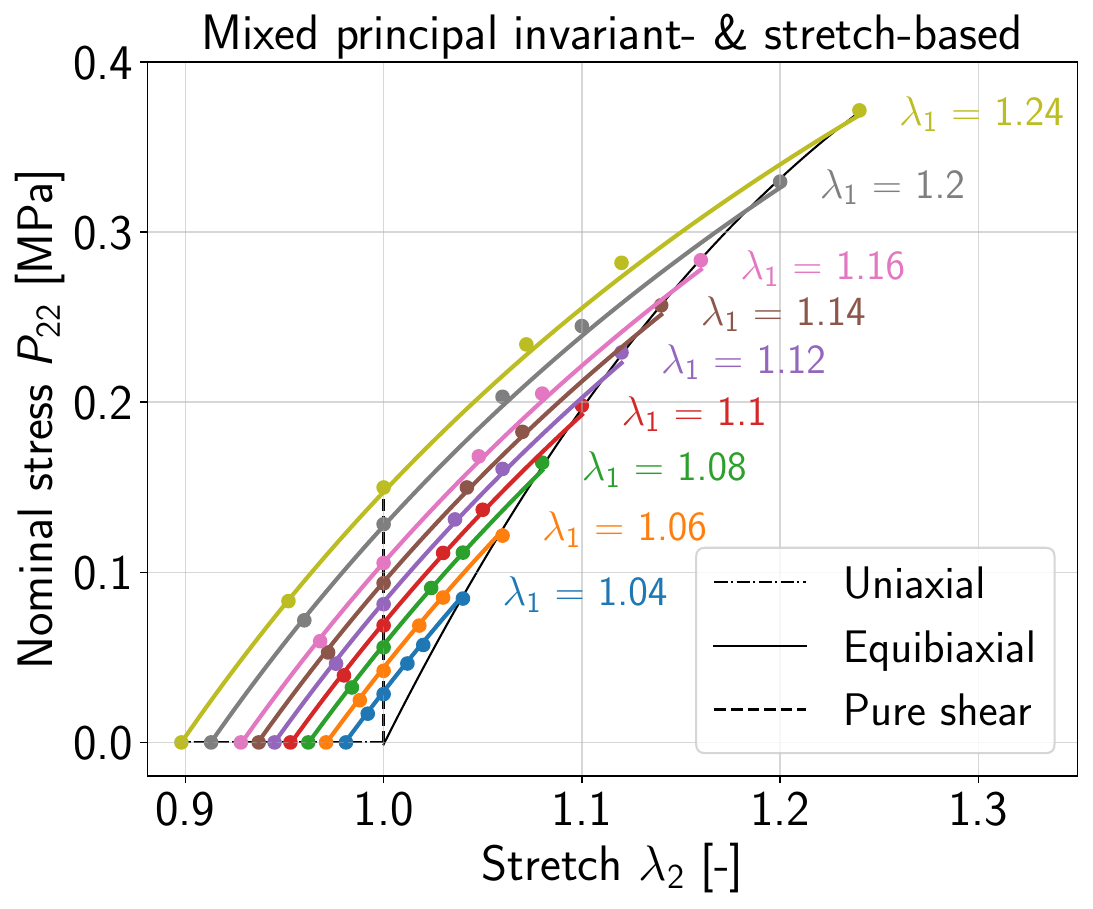}
                \caption{$P_{22}$ generalization: small stretch regime of Kawabata's data.}
                \label{fig:kawabata_P2_small_stretch_invar_ps}
            \end{subfigure}
            \hfill
            \begin{subfigure}{0.48\linewidth}
                \includegraphics[width=\linewidth]{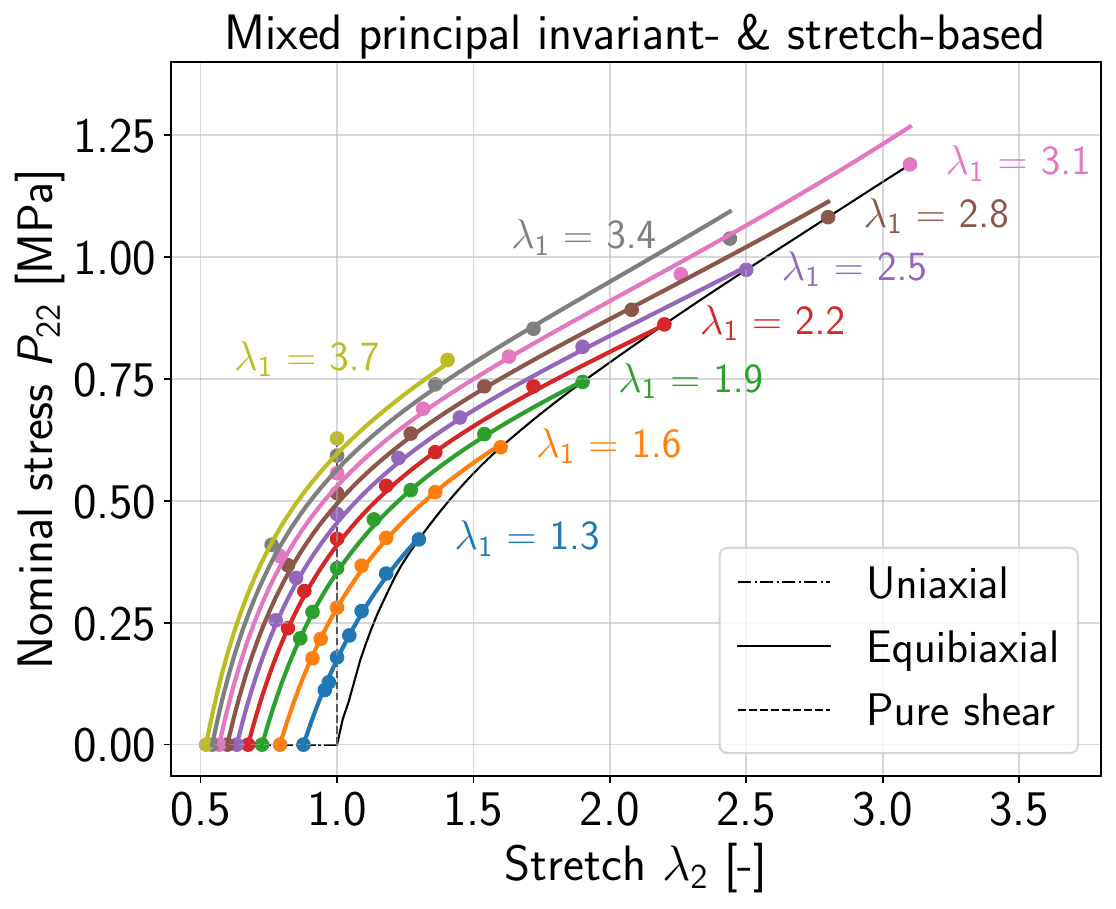}
                \caption{$P_{22}$ generalization: large stretch regime of Kawabata's data.}
                \label{fig:kawabata_P2_large_stretch_invar_ps}
            \end{subfigure}       
            \caption{\textbf{Training and validation results for vulcanized rubber}. (a) Fitting performance on Treloar's data \cite{Treloar1944}. (b) Architecture of the final CKAN after training; unconnected nodes indicate pruned activation functions of the network. (c)--(f) Generalization performance of the \emph{mixed principal invariant- and principal stretch-based} CKAN on the biaxial dataset by Kawabata et al. \cite{Kawabata1981}, after the CKAN was trained on Treloar's data only. The solid lines represent the model response. The dots represent the experimental data}
            \label{fig:Treloar_inv_stretch}
    \end{figure}

    Kawabata et al. \cite{Kawabata1981} conducted a series of general biaxial experiments. In these tests, the stretch $\lambda_1$ in one direction was fixed at various values while varying the stretch $\lambda_2$ in the orthogonal planar direction. During the experiments, the nominal stresses $P_{11}(\lambda_1, \lambda_2)$ and $P_{22}(\lambda_1,\lambda_2)$ in the first and second direction, respectively, were recorded. Due to the material's incompressibility, any kinematically admissible deformation is fully characterized by two independent variables, such as $\lambda_1$ and $\lambda_2$ or $I_1$ and $I_2$. Thus, Kawabata's dataset fully characterizes the material behavior under any multi-axial deformation mode within the prescribed stretch range. Indeed, the results of uniaxial, equibiaxial, and pure shear tests can in principle be extracted also from the general biaxial test data, as shown in Figure \ref{fig:Treloar_inv_stretch}.
    
    As reported in \cite{Marckmann2006}, Treloar and Kawabata et al. used vulcanized rubber of the same chemical composition, with only slight deviations observed in the experimental data \cite{Amores2020c}. Thus, following \cite{Marckmann2006, Amores2020c, Khiem2016}, we trained the CKAN architecture using Treloar's dataset alone and derived a symbolic strain energy function. This strain energy function was subsequently evaluated for generalization performance on Kawabata's dataset. We considered the four distinct functional bases for the strain energy function introduced above, that is, the principle stretches $\{\lambda_1,\lambda_2,\lambda_3\}$, the principal invariants $\{I_1, I_2, I_3\}$, the modified invariants $\{\iota_1,\iota_2,I_3\}$, and a mixed set of principal stretches and principal invariants $\{\lambda_1,\lambda_2,\lambda_3,I_1,I_2,I_3\}$. For each functional basis, the CKAN was trained, symbolic expressions for the strain energy function were derived, and the models were assessed for their training (fitting) performance on Treloar's data and their validation (prediction) performance on Kawabata's data.

    \begin{figure}[t]
        \centering
        \includegraphics[width=0.95\linewidth]{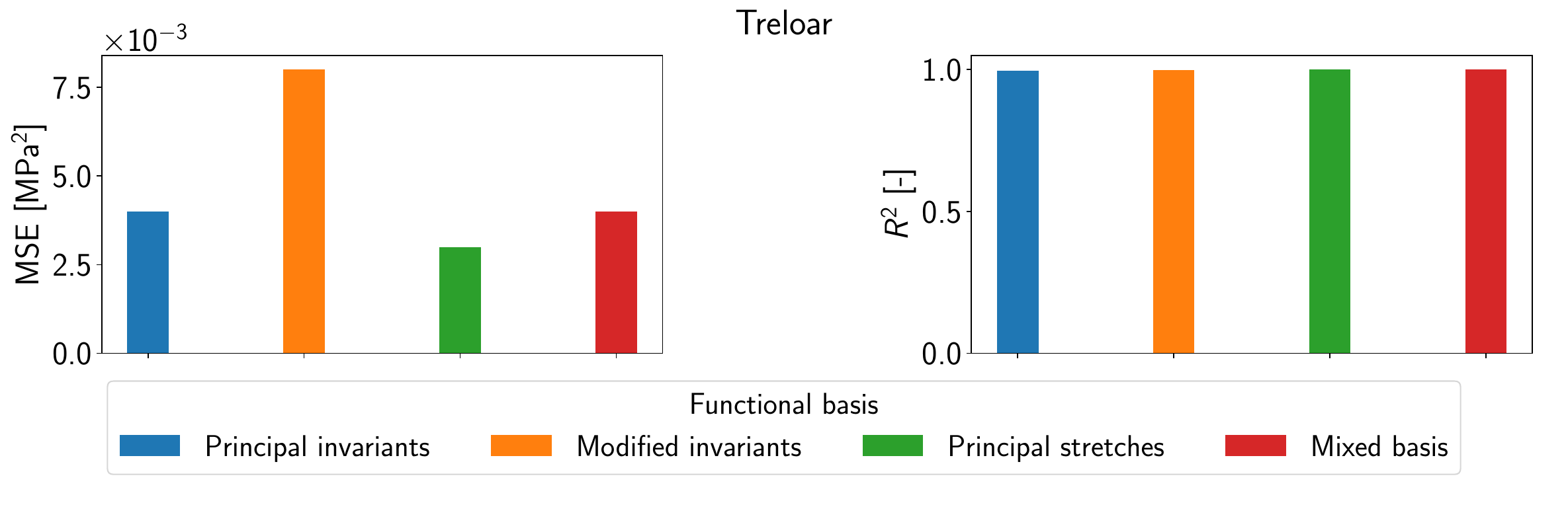}
        \caption{\textbf{Comparison of the descriptive performance on Treloar's data}: Mean squared error (MSE) and coefficient of determination ($R^2$) of each functional basis for predicting the stress component $P_{11}$.}
        \label{fig:error_bar_treloar}
    \end{figure}

    \begin{figure}[t]
        \centering
        \includegraphics[width=0.95\linewidth]{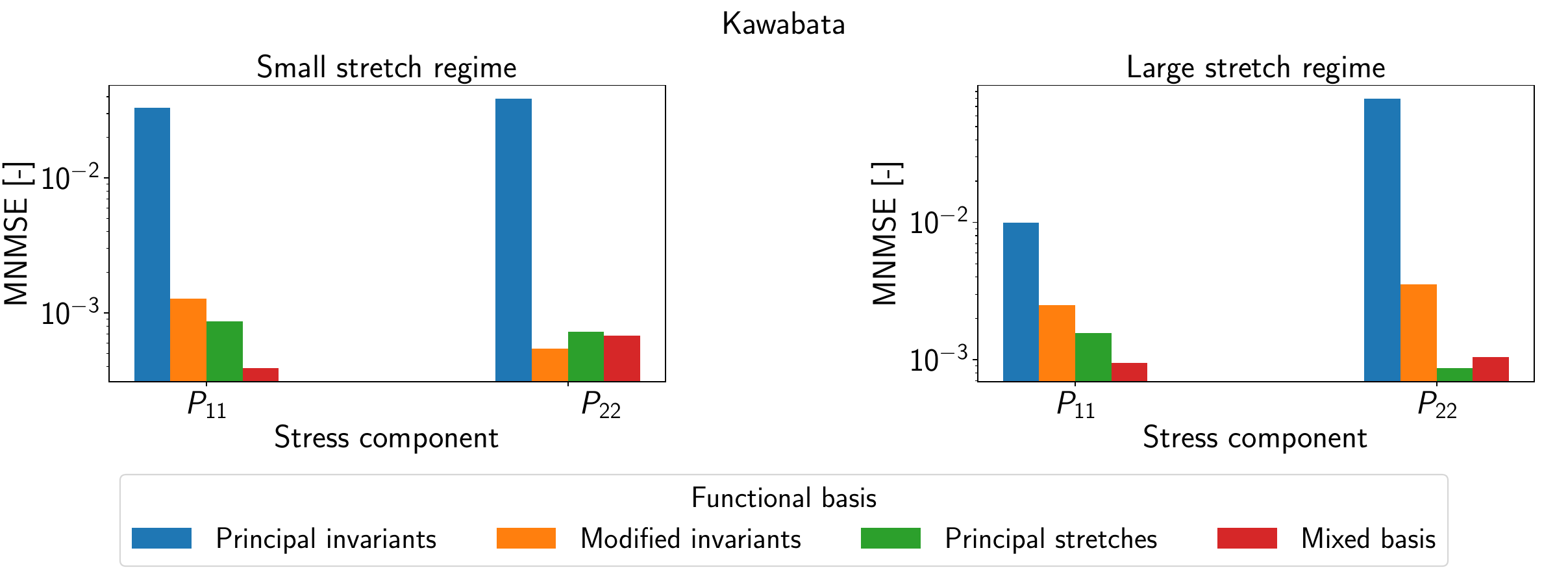}
        \caption{\textbf{Comparison of the generalization performance on Kawabata's data}: Mean normalized mean squared error (MNMSE) of each functional basis for predicting the stress components $P_{11}$ and $P_{22}$ in the small and large stretch regimes of Kawabata's data.}
        \label{fig:error_bar_kawabata}
    \end{figure}

    \paragraph{CKAN performance comparison across four functional bases}
    For each of the four functional bases, the trained CKAN exhibited high descriptive accuracy on Treloar’s dataset, all achieving \(R^2 \geq 0.996\), Figure \ref{fig:error_bar_treloar}. However, their ability to generalize beyond the training data differed substantially when tested on Kawabata’s biaxial data. In Figure~\ref{fig:error_bar_kawabata}, we report the mean normalized mean squared error\footnote{First, the normalized mean squared errors (NMSE) for each stretch level $\lambda_1$ are separately computed, then these NMSE are averaged for each functional basis and stress component in the small and large stretch regime individually.} (MNMSE) on Kawabata's dataset, demonstrating that the CKAN using the mixed functional basis outperformed all others. Figure \ref{fig:result_treloar_invar_ps} shows its fitting results on Treloar's data and Figures \ref{fig:kawabata_P1_small_stretch_invar_ps}--\ref{fig:kawabata_P2_large_stretch_invar_ps} the generalization performance on Kawabata's data. The CKAN using the principal-stretch basis was nearly as good as the one using the mixed basis. 
    The modified invariant basis ranked third, while the principal invariant basis exhibited the weakest generalization. These findings align with earlier studies noting that sole reliance on \(I_1\) and \(I_2\) may be insufficient for capturing multi-axial deformation in vulcanized rubber \cite{Dal2023, Marckmann2006}.

    \begin{table}[h]
        \renewcommand{\arraystretch}{1.45}    
        \centering       

                \begin{subtable}[]{\linewidth}
            \centering
            \begin{tabular}{@{}lC{13mm}C{20mm}C{20mm}C{13mm}C{13mm}C{13mm}@{}}
                \multicolumn{7}{c}{\textbf{Mixed principal invariant- and principal stretch-based}}                              \\ \toprule
                \multirow{2}{*}{Strain energy function}               & \multicolumn{6}{c}{$\hat{\Psi}^\text{KAN}= \hat{\Psi}^\text{KAN}_I(I_1,I_2) + \sum_{\alpha=1}^3 \hat{\omega}^\text{KAN}_1(\lambda_\alpha)$}  \\
                &\multicolumn{6}{c}{$\hat{\Psi}^\text{KAN}(I_1,I_2) = a(b\,I_2 + 1)^2, \quad \hat{\omega}^\text{KAN}_1(\lambda_\alpha) = c \, \left( d \, \lambda_{\alpha} + 1\right)^{32} + e \, \left(f \, \lambda_{\alpha} + 1\right)^{10}$} \\ \cmidrule{1-7}
                \multirow{2}{*}{Material parameters} & $a$ [MPa] & $b$ [-] &  $c$ [MPa] & $d$ [-] & $e$ [MPa] & $f$ [-] \\ \cmidrule{2-7}
                                                     &2.8333 & $7.6034 \cdot 10^{-4}$ & $4.9633 \cdot 10^{-3}$ & 0.0315 & 2.3312 & 0.0219 \\ \bottomrule
            \end{tabular}
            \caption{}
            \label{tab:params_SEF_Treloar_invar_ps}            
        \end{subtable}\\[2ex]

        \begin{subtable}[]{\linewidth}
            \centering
            \begin{tabular}{@{}lC{15mm}C{15mm}C{15mm}C{15mm}C{15mm}C{15mm}@{}}
                \multicolumn{7}{c}{\textbf{Principal stretch-based}}                              \\ \toprule
                \multirow{2}{*}{Strain energy function}               & \multicolumn{6}{c}{$\hat{\Psi}^\text{KAN}=\sum_{\alpha=1}^3 \hat{\omega}^\text{KAN}_1(\lambda_\alpha) + \sum_{\alpha=1}^3 \hat{\omega}^\text{KAN}_{-1}(\lambda_\alpha^{-1})$} \\
                & \multicolumn{6}{c}{$\hat{\omega}^\text{KAN}_1(\lambda_\alpha) = a \, ( b \, \lambda_{\alpha} + 1)^{30} + c \, (d \, \lambda_{\alpha} + 1)^{5},\quad \hat{\omega}^\text{KAN}_{-1}(\lambda_\alpha^{-1}) = e \, ( f \, \lambda_{\alpha}^{-1} + 1)^{15}$}  \\ \cmidrule{1-7}
                \multirow{2}{*}{Material parameters} & $a$ [MPa] & $b$ [-] &  $c$ [MPa] & $d$ [-] & $e$ [MPa] & $f$ [-] \\ \cmidrule{2-7}
                                                     & 0.0006 & 0.0441 & 1.1489 & 0.0736 & 0.3981 & 0.0067 \\ \bottomrule
            \end{tabular}
            \caption{}
            \label{tab:params_SEF_Treloar_ps}
        \end{subtable}\\[2ex]

        \begin{subtable}[]{\textwidth}
            \centering
            \begin{tabular}{@{}lC{15mm}C{15mm}C{15mm}C{15mm}C{15mm}C{15mm}@{}}
                \multicolumn{7}{c}{\textbf{Modified invariant-based}}                              \\ \toprule
                Strain energy function               & \multicolumn{6}{c}{$\hat{\Psi}^\text{KAN}(\iota_{1},\iota_{2}) = a \left(b\, \iota_{2} + 1\right)^{2} - c\, \atan\left(d (- e \, \iota_{1} - 1)^{3} + f \right) $}  \\\cmidrule{1-7}
                \multirow{2}{*}{Material parameters} & $a$ [MPa] & $b$ [MPa] &  $c$ [MPa] & $d$ [MPa] & $e$ [-] & $f$ [-]\\ \cmidrule{2-7}
                                                     & 2.1873 & 0.0868 & 1024.9 & 1.3653 & 0.4019 & 65.423 \\ \bottomrule
            \end{tabular}
            \caption{}
            \label{tab:params_SEF_Treloar_mod_invar}
        \end{subtable} \\[2ex]

        \begin{subtable}[]{\textwidth}
            \centering
            \begin{tabular}{@{}lC{18mm}C{20mm}C{20mm}C{18mm}C{18mm}@{}}
                \multicolumn{6}{c}{\textbf{Principal invariant-based}}                              \\ \toprule
                Strain energy function               & \multicolumn{5}{c}{$\hat{\Psi}^\text{KAN}(I_1, I_2) = a\left(b \, I_{1} + c \, I_{2} + 1 \right)^{3} + d \exp(e \, I_{1})$}  \\\cmidrule{1-6}
                \multirow{2}{*}{Material parameters} & $a$ [MPa] & $b$ [-] &  $c$ [-] & $d$ [MPa] & $e$ [-] \\ \cmidrule{2-6}
                                                     & 468.37 & $1.1034\cdot10^{-4}$ & $2.2632\cdot 10^{-6}$ & 0.0362 & 0.0774  \\ \bottomrule
            \end{tabular}
            \caption{}
            \label{tab:params_SEF_Treloar_invar}
        \end{subtable}

        \caption{Symbolic strain energy functions and material parameters for vulcanized rubber discovered by the (a) mixed principal invariant- and principal stretch-based (b) principal stretch-based, (c) modified invariant-based, and (d) principal invariant-based CKAN framework trained on Treloar's data \cite{Treloar1944}}
        \label{tab:my-table}
    \end{table}

    \paragraph{Details for each basis}
    
    \emph{Mixed basis.} 
    The mixed basis enabled top performance (average \(R^2=0.999\) on Treloar’s data). Its symbolic strain energy function (Table~\ref{tab:params_SEF_Treloar_invar_ps}) resembles an Ogden-type term alongside a quadratic dependence on \(I_2\). However, the larger network architecture prolongs training because separate sub-networks handle the contributions from \(I_1,I_2\) and from \(\lambda_\alpha,\lambda_\alpha^{-1}\). Despite this computational cost, the mixed approach excels in both descriptive and generalization performance (Figures~\ref{fig:result_treloar_invar_ps} and \ref{fig:kawabata_P1_small_stretch_invar_ps}--\ref{fig:kawabata_P2_large_stretch_invar_ps}).
    
    \emph{Principal stretch basis.}
    The purely principal stretch-based CKAN also demonstrated strong generalization, narrowly trailing the mixed basis in generalization performance on Kawabata’s data and maintaining speed similar to the mixed approach (due to multiple evaluations of \(\lambda_\alpha\) and \(\lambda_\alpha^{-1}\)). The recovered symbolic expression (Table~\ref{tab:params_SEF_Treloar_ps}) shows a close resemblance to the Ogden model~\cite{Ogden1972}, indicating the natural link between principal stretches and rubber material behavior (see Appendix, Figure~\ref{fig:Treloar_stretch}).
    
    \emph{Modified invariant basis.}
    Finally, the CKAN relying on the modified invariants achieved a descriptive performance comparable to the more complex bases on Treloar’s data (cf. Appendix, Figure~\ref{fig:Treloar_ModInv}) but fell short of the principal stretch-based and principal invariant-based formulations in generalization performance. The resulting symbolic function (Table~\ref{tab:params_SEF_Treloar_mod_invar}) better distinguishes between uniaxial tension and pure shear and thus outperforms the principal invariant basis when extrapolating to Kawabata’s biaxial scenario.
    
    \emph{Principal invariant basis.}
    Although the CKAN with the principal invariant-based formulation reached \(R^2=0.996\) on Treloar’s data (cf. Appendix, Figure~\ref{fig:Treloar_Inv}), it yielded the largest MNMSE on Kawabata’s dataset (Figure~\ref{fig:error_bar_kawabata}). This aligns with previous findings that principal invariants \(I_1\) and \(I_2\) alone may not adequately capture multi-axial deformation in vulcanized rubber. On the positive side, the simpler network architecture (only one sub-network) reduces training time, and the final symbolic expression (Table~\ref{tab:params_SEF_Treloar_invar}) incorporates polynomial terms in \(I_1, I_2\) and an exponential term in \(I_1\).

    Overall, while all formulations fit Treloar’s data well, their ability to generalize to Kawabata’s biaxial results varied considerably. The mixed basis and principal stretch-based CKANs exhibited the best generalization performance, while the modified invariant basis offered a balanced compromise between generalization performance and computational simplicity. The principal invariant basis, though fastest to train, proved least suitable for capturing complex deformations. As the examples underscore, the choice of functional basis should reflect both computational needs and the specific behaviors required for accurate material modeling.

    \FloatBarrier


    \subsection{Human brain tissue}\label{sec:brain}
    
    To demonstrate how the CKAN framework can handle complex biological tissues with limited experimental data and still produce interpretable strain energy functions, we consider human brain tissue experiments from \cite{Budday2017b}. Specifically, we examine homogeneous deformation modes of uniaxial tension (UT), uniaxial compression (UC), and simple shear (SS). These datasets were obtained from cubic samples of the cortex region, excised from ten different human brains. For UT and UC, the experimental measurements consist of the in-plane stretches and corresponding nominal stress, while simple shear data comprise shear displacement and nominal shear stress. All tests were performed within the elastic regime, but the dataset remains sparse, covering only a limited range of deformations. Unlike the more extensive rubber datasets of Sections~\ref{sec:rubber}, the primary focus here is on how effectively the identified strain energy functions describe each observed deformation mode rather than on wide-range extrapolation.

    \begin{figure}[h!]
        \centering
        \begin{subfigure}{\linewidth}
            \includegraphics[width=\linewidth]{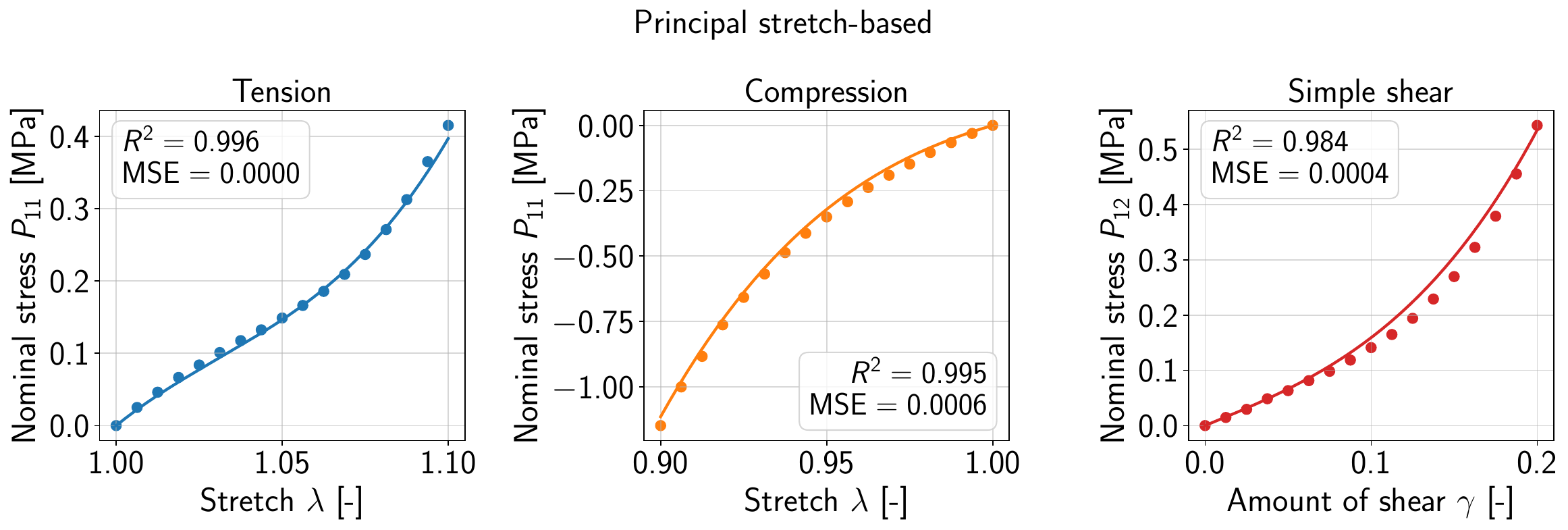}
            \caption{Fitting results: experimental data (dots) and performance of CKAN (solid lines)}
            \label{fig:results_brain_ps}
        \end{subfigure}     \\[4ex]
        \begin{subtable}[c]{0.59\linewidth}
            \renewcommand{\arraystretch}{1.4}
            \centering
            \begin{tabular}{@{}lcccc@{}}
                \toprule
                \multirow{3}{*}{Strain energy function} & \multicolumn{4}{c}{$\hat{\Psi}^\text{KAN} = \sum_{\alpha=1}^3 \hat{\omega}^\text{KAN}_{1}(\lambda_{\alpha}) + \sum_{\alpha=1}^3 \hat{\omega}^\text{KAN}_{-1}(\lambda_{\alpha}^{-1})$} \\
                & \multicolumn{4}{c}{$\hat{\omega}^\text{KAN}_1(\lambda_\alpha) = a\,(b\,\lambda_\alpha -1)^{4},$} \\
                                                        & \multicolumn{4}{c}{$\hat{\omega}^\text{KAN}_{-1}(\lambda_\alpha^{-1}) = c \, \exp(d \, \lambda_\alpha^{-1})$} \\ \cmidrule{1-5}
                \multirow{2}{*}{Material parameters} & $a$ [MPa] & $b$ [-] &  $c$ [MPa] & $d$ [-] \\ \cmidrule{2-5}
                                                        & 104.0147 & 0.9639 & 14.4781 & 0.1410 \\ \bottomrule
            \end{tabular}
            \caption{Discovered strain energy function and material parameters.}
            \label{tab:params_SEF_Brain_ps}
        \end{subtable} 
        \hfill
        \begin{subfigure}[c]{0.31\linewidth}
            \centering
            \includegraphics[width=\linewidth]{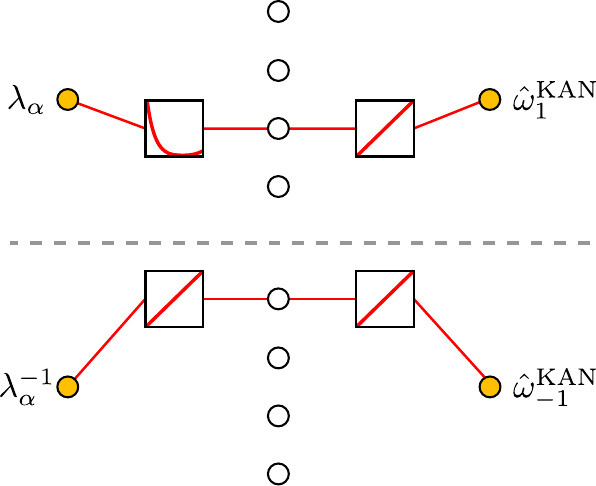}
            \caption{CKAN architecture.}
            \label{fig:architecture_brain_ps}
        \end{subfigure}
        \caption{\textbf{Descriptive performance of CKAN based on principal stretches for human brain (cortex) tissue} \cite{Budday2017b}: (a) Performance of CKAN simultaneously fitted to uniaxial tension, compression, and simple shear data. (b) Resulting strain energy functions and material parameters. (c) Architecture of the final CKAN after training; unconnected nodes indicate pruned activation functions of the network.}
        \label{fig:Brain_ps}
    \end{figure}   
    \begin{figure}[h]
        \centering
        \includegraphics[width=\linewidth]{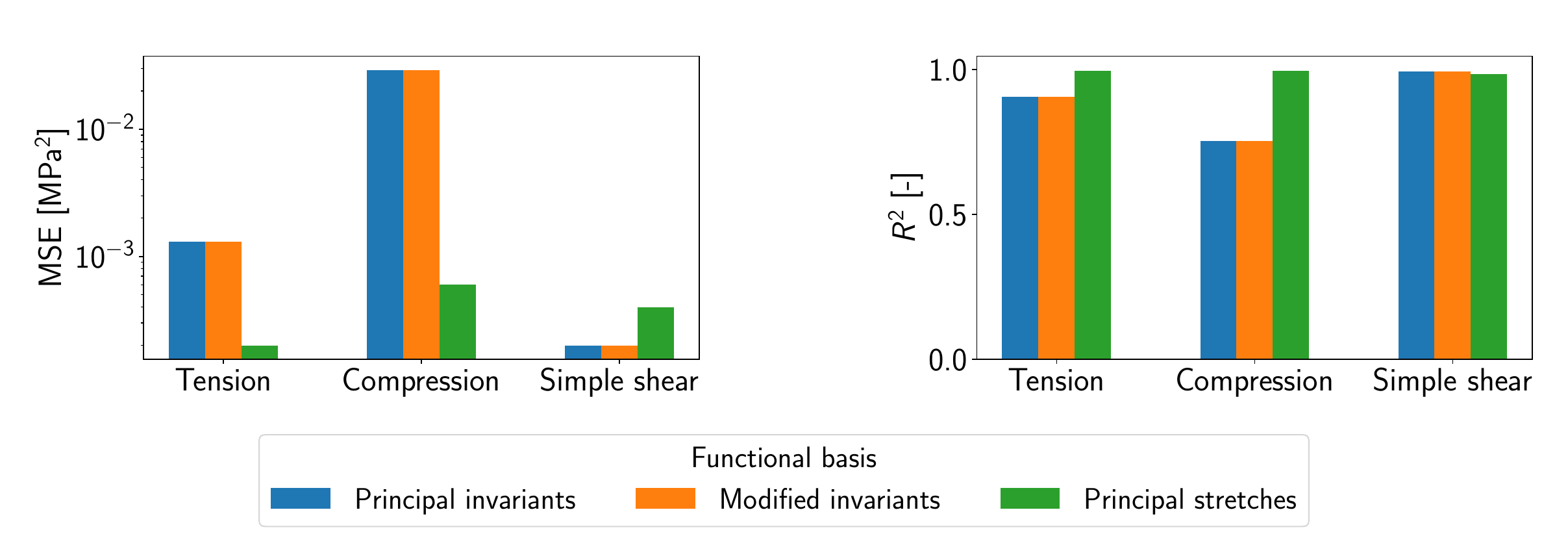}
        \caption{\textbf{Comparison of the descriptive performance on the brain tissue data.} Mean squared error (MSE) and coefficient of determination ($R^2$) of the symbolic strain energy functions for the human brain (cortex) each using a different functional basis.}
        \label{fig:error_brain}
    \end{figure}

\paragraph{Principal stretches as the most effective basis for brain tissue}
Among the three functional bases (principal stretches, principal invariants, and modified invariants) tested for this example, the principal stretch-based formulation provided the best descriptive performance for UT, UC, and SS. Figure \ref{fig:Brain_ps} illustrates this performance. Figure~\ref{fig:error_brain} compares the mean squared error (MSE) and \(R^2\) metrics for each basis, illustrating that principal stretches capture both tension-compression asymmetry and shear response more accurately. By default, the network included terms involving both \(\lambda_\alpha\) and their reciprocals \(\lambda_\alpha^{-1}\), mirroring the behavior noted in Ogden-type models \cite{Budday2017,Mihai2017,Anssari-Benam2022}. In fact, increasing sparsity (Section~\ref{sec:sparse_symbol}) often compressed the strain energy to a single function of the reciprocal stretches,
\[
    \hat{\Psi}^\text{KAN} 
    = \sum_{\alpha=1}^3 
      \hat{\omega}^\text{KAN}_{-1}\bigl(\lambda_{\alpha}^{-1}\bigr),
    \quad\text{with}\quad 
    \hat{\omega}^\text{KAN}_{-1}(\lambda_{\alpha}^{-1}) 
    = a\,(\lambda_{\alpha}^{-1}+b)^{18},
\]
highlighting the pivotal role of \(\lambda_{\alpha}^{-1}\) in capturing UC and SS. However, this single-term model struggles with the S-shaped stress response of UT, underlining the need for additional terms to accurately depict tension-compression asymmetry. The fitting results of the one-term strain energy function, along with the corresponding material parameters, are presented in the Appendix in Figure \ref{fig:Brain_PS_one_term}.

\paragraph{Alternative bases: principal and modified invariants}
The principal invariant-based CKAN converged to a strain energy function solely dependent on the second invariant \(I_2\). This choice agrees with earlier observations that the brain’s distinctive shape-change-dominated behavior can be effectively modeled by functions of \(I_2\)~\cite{kuhl2024too}. Although this approach performed well for uniaxial tension and simple shear, it struggled with uniaxial compression, as shown in the Appendix in Figure~\ref{fig:Brain_inv}. Similarly, the modified invariant-based formulation produced nearly identical results, relying exclusively on \(\iota_2\) (cf. Appendix, Figure~\ref{fig:Brain_Modinv}). Consequently, both invariant-based approaches captured certain deformation modes but fell short in describing the full tension-compression asymmetry observed in brain tissue.

Altogether, these findings demonstrate that the choice of functional basis critically influences model performance, particularly for materials exhibiting pronounced tension-compression asymmetry. While invariant-based and modified invariant-based CKANs suffice for tension and shear, the principal stretch-based formulation offers a more comprehensive description of brain tissue mechanics. Unlike the rubber datasets of Sections~\ref{sec:rubber}, this dataset does not permit a rigorous evaluation of generalizability due to the lack of multi-axial or biaxial tension data in brain tissue. Nevertheless, the significant differences in descriptive performance across the three functional bases considered underscore the importance of selecting an appropriate representation for the targeted material and deformation modes.

    \FloatBarrier


    \subsection{Ecoflex silicone polymer}\label{sec:exoflex}
    Silicone elastomers are highly versatile materials extensively used in biomedical engineering \cite{Bernardi2017}, soft robotics \cite{Case2015}, and wearable devices \cite{Jiang2018} due to their biocompatibility, thermal stability, and stretchability. Ecoflex, a commercial silicone elastomer, is available in various Shore hardness levels, which directly correlate with its stiffness. In \cite{Liao2020a}, extensive uniaxial tension tests were conducted, including uniaxial cyclic loading-unloading tests on Ecoflex with varying Shore hardness levels. 

    To demonstrate the utility of auxiliary features in constitutive modeling, we analyzed virgin loading curves from these cyclic tests, incorporating Shore hardness as an additional input to the CKAN. Specifically, Shore hardness values $s \in \{$OO-10, OO-20, OO-30, OO-50$\}$ and maximal uniaxial stretch $\lambda_{\max}=6$ were considered. Due to the small dataset, we employed leave-one-out cross-validation, withholding one Shore hardness level for validation while training on the others. This setup allowed us to evaluate the model's predictive capability in interpolating and extrapolating material behavior across different Shore hardness levels.
    
    \paragraph{Feature augmentation for predictive constitutive modeling} 
    In many elastomeric and collagenous tissue models, the influence of additional parameters, such as filler volume fraction \cite{Mullins1965} or cross-link density \cite{Linka2018}, is often captured through \emph{strain amplification factors} $\chi$. These factors modify the principal invariants as follows:
    \begin{equation}\label{eq:amplified_invars}
        \hat{I}_i = \chi(s) \bigl(I_i - 3\bigr) + 3, \quad i = 1,2,
    \end{equation}
    allowing for direct parametrization of the constitutive model by an external feature. However, determining an appropriate functional form for $\chi$ can be challenging and is often highly material-specific.
    In this study, we adopt a simpler approach to incorporate the Shore hardness $s$. First, $s$ is normalized using
    \begin{equation}
    \bar{s} = \frac{s - s_{\min}}{s_{\max} - s_{\min}},
    \end{equation}
    where $s_{\max} = 50$ and $s_{\min} = 10$. The resulting dimensionless variable $\bar{s}$ is then included via the feature vector $\vec{f}$, as input to the CKAN. A basic 2-layer CKAN architecture with the topology $\bm{\mathcal{N}}=[3,1,1]$ is utilized to learn a strain energy function that explicitly incorporates the normalized Shore hardness $\bar{s}$. Given that the dataset consists exclusively of uniaxial tension data and the primary objective is to evaluate generalizability across different Shore hardness levels, we adopt the computationally efficient principal invariant-based CKAN formulation. Due to the small dataset and the well-documented relationship that higher Shore hardness generally corresponds to stiffer material behavior \cite{Liao2020a}, we additionally impose a monotonicity constraint on the relation between $\bar{s}$ and the strain energy, see Appendix \ref{app:mono_kan} for details.

\begin{figure}[H]
        \centering
        \includegraphics[width=\linewidth]{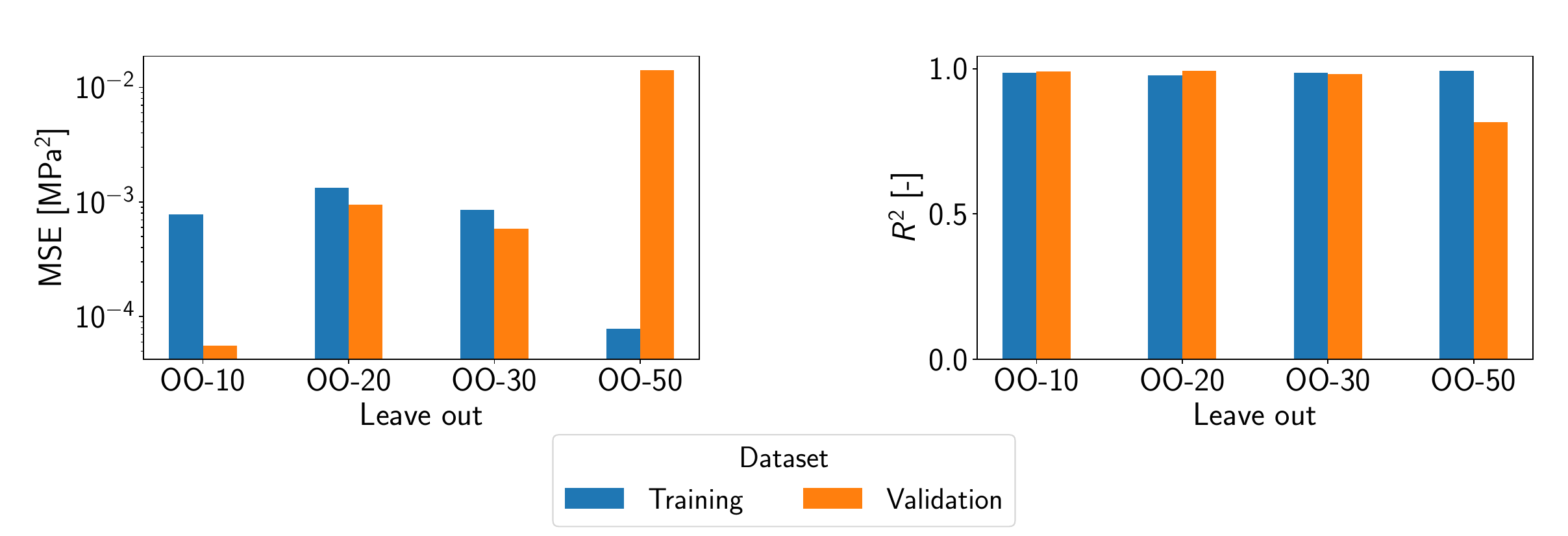}
        \caption{\textbf{Comparison of the predictive performance on the Ecoflex silicone polymer data.} Mean squared error (MSE) and coefficient of determination ($R^2$) of the stress curves generated from the discovered symbolic principal invariant-based strain energy functions for Ecoflex silicone on the training and validation set using leave-one-out cross-validation.}
        \label{fig:enter-label}
    \end{figure}

\begin{figure}
        \centering
        \includegraphics[width=\linewidth]{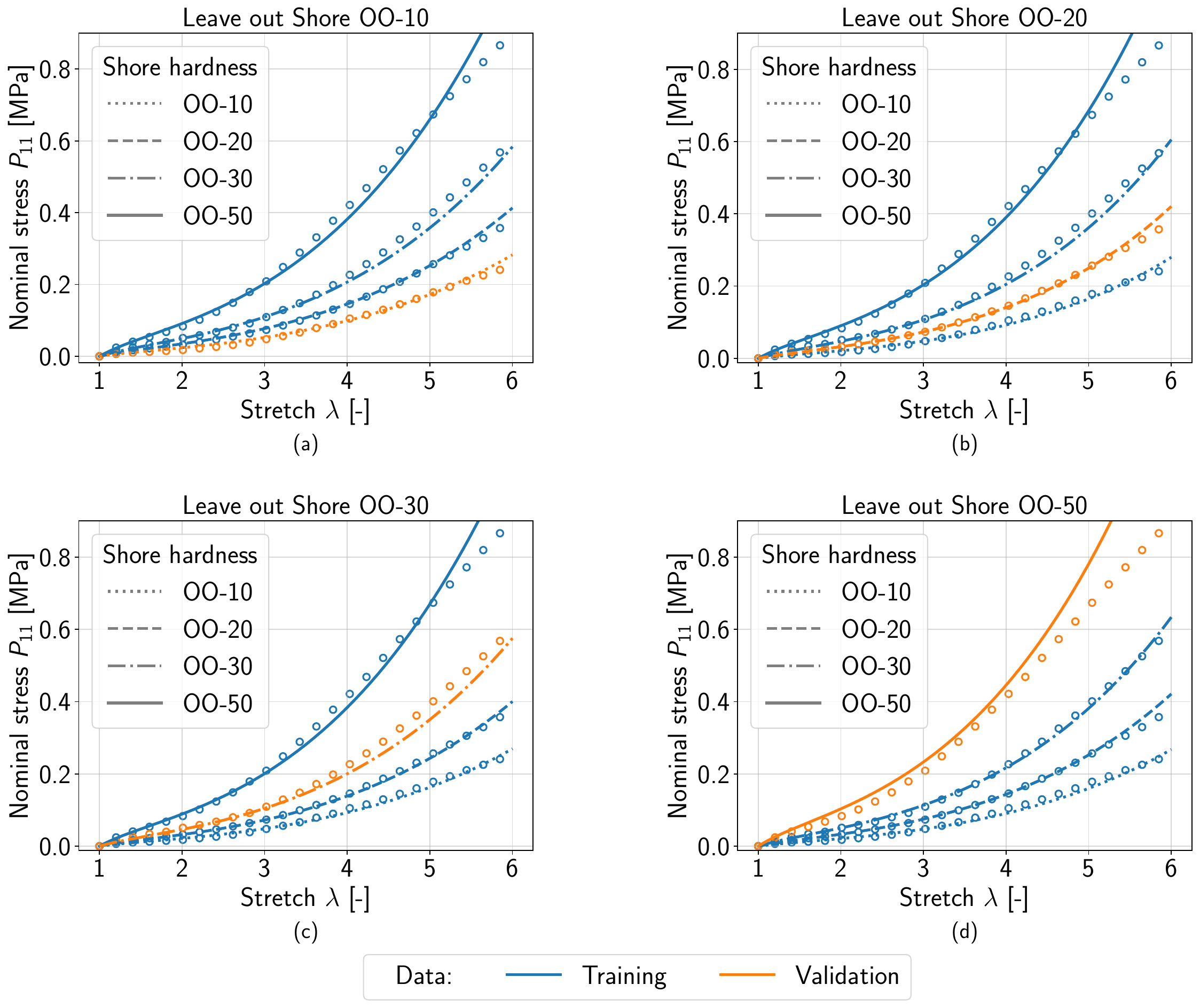}
        \caption{\textbf{Training results (descriptive performance) and validation results (predictive performance) for Ecoflex silicone polymer data from \cite{Liao2020a}} for principal invariant-based symbolic CKANs applied in leave-one-out cross-validation scheme. The lines represent the model response, and the dots the experimental data. Blue indicates data provided during the training (and fitting performance of the model), and orange indicates data withheld during the respective training process. The orange lines thus indicate the ability of the trained CKAN to predict the behavior of a material type (with a different Shore hardness level) for which no testing data was provided within the training data.}
        \label{fig:loocv_ecoflex_results_symbolic}
    \end{figure}

    \paragraph{Training results (descriptive performance) and validation results (predictive performance)} Figure \ref{fig:enter-label} and \ref{fig:loocv_ecoflex_results_symbolic} visualize the results of the identified symbolic strain energy function on the training and validation dataset. The symbolic strain energies demonstrate excellent agreement with the experimental data, achieving accurate fits to the training dataset and reliable predictions for the Shore hardness levels withheld in the respective training process. Especially, in extrapolating the stress-stretch curve for the withheld Shore hardness levels OO-10 and OO-20 in Figure \ref{fig:loocv_ecoflex_results_symbolic}(a) and \ref{fig:loocv_ecoflex_results_symbolic}(b), respectively, precise predictions are achieved.

    \begin{figure}[]
        \centering
        \begin{subfigure}{0.2\linewidth}
            \includegraphics[width=\linewidth]{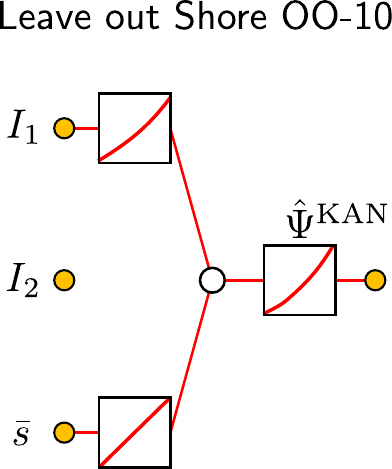}
            \caption{}
            \label{fig:loo_10}
        \end{subfigure}
        \hfill
        \begin{subfigure}{0.2\linewidth}
            \includegraphics[width=\linewidth]{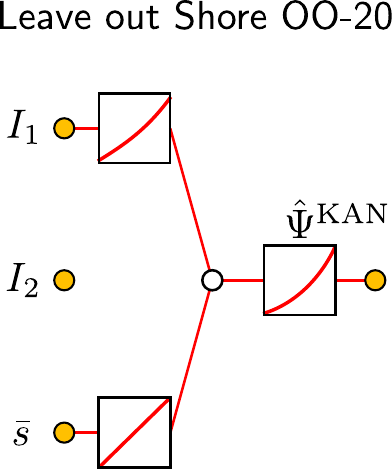}
            \caption{}
            \label{fig:loo_20}
        \end{subfigure}
        \hfill
        \begin{subfigure}{0.2\linewidth}
            \includegraphics[width=\linewidth]{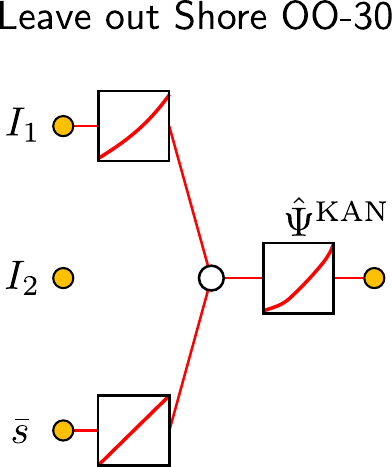}
            \caption{}
            \label{fig:loo_30}
        \end{subfigure}
        \hfill
        \begin{subfigure}{0.2\linewidth}
            \includegraphics[width=\linewidth]{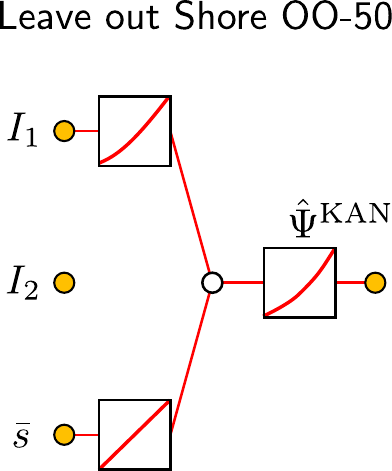}
            \caption{}
            \label{fig:loo_50}
        \end{subfigure} 
        \caption{\textbf{CKAN architectures obtained during the leave-one-out cross-validation test for Ecoflex silicone polymer data.} The red graphs illustrate the symbolic activation functions $f_{i,j,k}$, and the red lines are the only remaining connections after the training. Unconnected nodes indicate activation functions identical to zero. The principal invariants $I_1$ and $I_2$, along with the normalized Shore hardness $\bar{s}$ as an additional feature formed the input to the CKANs. However, the second principal invariant $I_2$ was discarded during training across all datasets. That is, it was identified as irrelevant.}
        \label{fig:loocv_ecoflex_architecture}
    \end{figure}
    
    \paragraph{Symbolic strain energy functions} The symbolic strain energy functions independently identified by the CKANs across the four training datasets are listed in Table \ref{tab:symbolic_psi_ecoflex}. Notably, the identified strain energy functions share the same structure, with a hyperbolic cosine of $I_1$ and a quadratic function of $\bar{s}$. This similarity is also visible in the learned network topologies and activation functions illustrated in Figure \ref{fig:loocv_ecoflex_architecture}. The following characteristics of the identified strain energy functions stand out: First, through the training process, the second principal invariant $I_2$ is effectively excluded from the input space, probably because of its strong correlation with $I_1$ in uniaxial tensile tests. In fact, in such tests, $I_2$ can be uniquely expressed as a function of $I_1$. Second, the identified material parameters $d$ and $e$, which operate directly on $I_1$ within the hyperbolic cosine function, are very similar across all training data sets.
    
    The discovered symbolic expression underscores the capability of the CKAN framework to identify interpretable and predictive material models. Additionally, our findings demonstrate the efficacy of feature augmentation in enabling predictive modeling of materials with variable properties, such as Ecoflex silicone elastomers.

    \begin{table}[]
        \centering
        {\def\arraystretch{1.2}%
        \begin{tabular}{cccccccc}
        \toprule
        \multirow{4}{*}{Left out} & \multicolumn{6}{c}{Strain energy function}                                                                                                                \\
                               & \multicolumn{6}{c}{$\hat{\Psi}^\text{KAN} = a \left[\left(b \, \bar{s} + 1\right)^{2} + c \cosh{\left(d \, I_{1} + e \right)} + f \right]^{3}$}                                                                                                                                \\ \cmidrule(l){2-7} 
                               & \multicolumn{6}{c}{Material parameters}                                                                                                                   \\
                               & \multicolumn{1}{c}{$a$ [MPa]} & \multicolumn{1}{c}{$b$ [-]} & \multicolumn{1}{c}{$c$ [-]} & \multicolumn{1}{c}{$d$ [-]} & \multicolumn{1}{c}{$e$ [-]} & \multicolumn{1}{c}{$f$ [-]} \\ \midrule
                               Shore OO-10   & 9.6046 & 0.3722 & 0.2392 & 0.0092 & 0.0805 & -0.3287 \\
                               Shore OO-20   & 6.4998 & 0.3427 & 0.4190 & 0.0100 & 0.0800 & -0.6707 \\
                               Shore OO-30   & 6.4998 & 0.3427 & 0.4190 & 0.0100 & 0.0800 & -0.6707 \\
                               Shore OO-50   & 5.4116 & 0.4997 & 0.1585 & 0.0131 & 0.1045 & -0.1422
 \\ \bottomrule
        \end{tabular}}
        \caption{Strain energy function and material parameters discovered by the principal invariant-based CKAN trained on Ecoflex silicone polymer data from \cite{Liao2020a} using leave-one-out cross-validation.}
        \label{tab:symbolic_psi_ecoflex}
    \end{table}
        

    \section{Conclusions} \label{sec:discussion}

    Existing approaches to mechanical constitutive modeling often require a trade-off between accuracy (fitting performance), interpretability, and extrapolability, while also struggling to predict the properties of unknown materials.    
    In this paper, we introduced Constitutive Kolmogorov--Arnold Networks (CKANs), a new framework for discovering symbolic mechanical constitutive laws for materials. CKANs build on Kolmogorov--Arnold Networks (KANs) and incorporate symbolic regression as a post-processing step. This combination achieves high accuracy, interpretability, and extrapolability simultaneously. Furthermore, thanks to their flexible neural network architecture, CKANs demonstrate strong generalization capabilities, making them well-suited for predicting the behavior of previously unseen materials.
    
    \paragraph{Interpretability}
    CKANs provide symbolic expressions for the mechanical constitutive behavior of materials. These symbolic expressions are as easy to interpret as traditional, manually derived symbolic constitutive equations. In particular, CKANs can automatically recover from experimental data insights previously obtained through theoretical derivations. For instance, in our example on human brain tissue, CKANs using principal and modified invariants as input identified the second invariant as the key variable for constitutive modeling, aligning with \cite{kuhl2024too, Linka2023}. Similarly, CKANs based on principal stretches recognized an Ogden-type formulation, consistent with prior findings \cite{Budday2017b,Mihai2017}. In contrast, previous data-driven \emph{model discovery} approaches rely on a predefined library of building block expressions \cite{Flaschel2021a, Linka2022a, brunton2016discovering, brunton2022data}, which limits the variety of possible constitutive models.
    
    \paragraph{Computational efficiency}
    Using sparsification techniques during training, the CKAN architecture effectively eliminates non-essential connections, resulting in sparse symbolic expressions. This sparsity enhances computational efficiency, making CKANs well-suited for finite element software implementation. 
    
    \paragraph{Extrapolability}
    CKANs build on the concept of constitutive artificial neural networks (CANNs) originally introduced in \cite{Linka2021, Abdolazizi2023} inheriting their strong extrapolation capabilities. For instance, when trained solely on Treloar’s data (uniaxial tension, equibiaxial tension, pure shear), CKANs accurately predict stress-strain curves in Kawabata’s multi-axial dataset. 
    
    \paragraph{Predictive modeling}
    Like CANNs, CKANs are very flexible regarding their input. By incorporating additional information beyond stress-strain data, they can predict the mechanical behavior of unknown materials. We demonstrated this by using Shore hardness data for Ecoflex polymers. However, as shown for CANNs, CKANs could also leverage microstructural imaging data or processing information to enhance material property predictions \cite{Linka2022}. 
    
    \paragraph{Limitations and future directions}
    Despite the advantages shown here, certain limitations remain. Some manual intervention is required for hyperparameter tuning and symbolic regression, highlighting the need for automated solutions. AI-driven meta-optimization \cite{hospedales2021meta} could streamline these steps, accelerating the discovery of near-optimal models. Another avenue for future research is the extension of CKANs to more complex materials, e.g., anisotropic \cite{Linka2021, Kalina2025} or inelastic materials \cite{Abdolazizi2023, holthusen2024theory}, and multi-physics problems \cite{Franke2023}, potentially aided by a next-generation neural architecture (KAN~2.0) \cite{Liu2024}. Finally, extensive validation under real-world conditions and incorporation of uncertainty quantification will be vital to firmly establish CKANs as a robust tool for constitutive modeling.
    
    \paragraph{Summary}
    Reliable constitutive equations are crucial in computational mechanics, especially for high-stakes applications such as biomedical or aerospace engineering. While black-box models can provide high predictive accuracy, their lack of transparency limits their use in settings where interpretability and reliability are essential. The CKAN framework presented here offers a compelling alternative, combining robust performance and predictive capabilities with symbolically interpretable strain energy functions. This makes CKANs a tailor-made tool for future virtual material design and the development of digital twins of complex systems in high-stakes applications.

	\section*{Acknowledgments}
    We gratefully acknowledge ZiSheng Liao (South China University of Technology, Guangzhou) for providing the experimental data of Ecoflex silicone polymer. Funded by the Deutsche Forschungsgemeinschaft (DFG, German Research Foundation) – 533187597, 517243167.

	\appendix
    \renewcommand{\thedefinition}{\Alph{section}.\arabic{definition}}

    \section{First Piola--Kirchhoff stresses}\label{app:1PK_stress}
        
    Here, we summarize the derivation of the first Piola-Kirchhoff stresses $\tns{P}$ for strain energy functions expressed in terms of the different functional bases considered in this paper. Subsequently, the general expressions of $\tns{P}$ are specified for the homogeneous deformation modes of biaxial tension and simple shear, which are essential for training our CKANs.

    \subsection{General} \label{app:1PK_general}

    \subsubsection{Principal invariant-based strain energy functions} \label{app:general_invar}
    Inserting the principal invariant-based strain energy functions \eqref{eq:psi}$_1$ and \eqref{eq:psi_iso}$_1$ into \eqref{eq:stress_hyper}, and applying the chain rule along with the differentiation relations
    \begin{equation}\label{eq:diff_rel_inv}
        \partialder{I_1}{\tns{C}} = \tns{I}, \qquad \partialder{I_2}{\tns{C}} = I_1\tns{I} - \tns{C}, \qquad \partialder{I_3}{\tns{C}} = I_3\tns{C}^{-1},
    \end{equation}
    yields the following first Piola-Kirchhoff stresses for compressible and incompressible materials, respectively:
    \begin{align}\label{eq:1PK_invar_comp}
        \tns{P} &= 2 \left[ \left(\partialder{\Psi}{I_1} + I_1 \partialder{\Psi}{I_2} \right) \tns{F} - \partialder{\Psi}{I_2}\tns{F}\tns{C} + I_3 \partialder{\Psi}{I_3} \tns{F}^{-\mathrm{T}} \right],  \\ \label{eq:1PK_invar_incomp}
        \tns{P} &= - p \tns{F}^{-\mathrm{T}} + 2 \left[ \left(\partialder{\Psi}{I_1} + I_1 \partialder{\Psi}{I_2} \right) \tns{F} - \partialder{\Psi}{I_2}\tns{F}\tns{C}  \right].
    \end{align}
    \subsubsection{Modified invariant-based strain energy functions} \label{app:general_invar_mod}
    For a strain energy function expressed in terms of the modified invariants $\iota_1$ and $\iota_2$, inserting the differentiation relations
    \begin{equation}\label{eq:dIotadI}
        \partialder{\hat{\Psi}}{I_1} = \partialder{\hat{\Psi}}{\iota_1} \partialder{\iota_1}{I_1} = \frac{1}{6\iota_1} \partialder{\hat{\Psi}}{\iota_1} , \qquad
        \partialder{\hat{\Psi}}{I_2} = \partialder{\hat{\Psi}}{\iota_2} \partialder{\iota_2}{I_2} =  \frac{1}{9\iota_2^2} \partialder{\hat{\Psi}}{\iota_2}, 
    \end{equation}
    into \eqref{eq:1PK_invar_comp} and \eqref{eq:1PK_invar_incomp} yields the following first Piola-Kirchhoff stresses for compressible and incompressible materials, respectively:
    \begin{align}\label{eq:1PK_mod}
        \tns{P} &= 2 \left[ \left(\frac{1}{6\iota_1} \partialder{\Psi}{\iota_1} + I_1 \frac{1}{9\iota_2^2} \partialder{\Psi}{\iota_2} \right) \tns{F} - \frac{1}{9\iota_2^2} \partialder{\hat{\Psi}}{\iota_2} \tns{F}\tns{C} + I_3 \partialder{\Psi}{I_3} \tns{F}^{-\mathrm{T}} \right],  \\     
        \tns{P} &= - p \tns{F}^{-\mathrm{T}} + 2 \left[ \left( \frac{1}{6\iota_1} \partialder{\hat{\Psi}}{\iota_1} + I_1 \frac{1}{9\iota_2^2} \partialder{\hat{\Psi}}{\iota_2} \right) \tns{F} - \frac{1}{9\iota_2^2} \partialder{\hat{\Psi}}{\iota_2} \tns{F}\tns{C}  \right].
    \end{align}

    \subsubsection{Principal stretch-based strain energy functions} \label{app:general_ps}
    We substitute the principal stretch-based strain energy functions \eqref{eq:psi}$_2$ and \eqref{eq:psi_iso}$_2$ into \eqref{eq:stress_hyper}. Applying the chain rule along with the relations 
    \begin{equation}\label{eq:rel_ps}
        \partialder{\lambda_\alpha}{\tns{C}} = \frac{1}{2\lambda_\alpha} \vec{N}_\alpha \otimes \vec{N}_\alpha, \qquad \partialder{J}{\lambda_a} =  \frac{J}{\lambda_\alpha}, \qquad \tns{F}\vec{N}_\alpha = \lambda_\alpha \vec{n}_\alpha,
    \end{equation}
    yields the spectral representation of the first Piola-Kirchhoff stress for compressible and incompressible materials, respectively,
    \begin{align}\label{eq:1PK_spec_comp}
        \tns{P} &= \sum_{\alpha=1}^3 \underbrace{\partialder{\Psi}{\lambda_\alpha}}_{={P_\alpha}} \, \vec{n}_\alpha \otimes \vec{N}_\alpha, \\ \label{eq:1PK_spec_incomp}
        \tns{P} &= \sum_{\alpha=1}^3 \underbrace{\left( -\frac{p}{\lambda_\alpha} + \partialder{\Psi}{\lambda_\alpha} \right)}_{\hat{P}_\alpha} \, \vec{n}_\alpha \otimes \vec{N}_\alpha,
    \end{align}
    where $P_\alpha$ and $\hat{P}_\alpha$ denote the principal stresses.
    
    Finally, we specialize the general expressions \eqref{eq:1PK_spec_comp} and \eqref{eq:1PK_spec_incomp} for the compressible and incompressible separable strain energy functions \eqref{eq:separable_comp} and \eqref{eq:separable_incomp}. In the compressible case,
    \begin{equation}
        \Psi(\lambda_1,\lambda_2,\lambda_3, \vec{f}) = \sum_{\alpha=1}^{3} \omega_1(\lambda_\alpha, \vec{f}) + \sum_{\alpha=1}^{3} \omega_{-1}(\nu_\alpha, \vec{f}) + \Omega(\lambda_1\lambda_2\lambda_3,\vec{f})
    \end{equation}
    and the corresponding first Piola--Kirchhoff stress is
    \begin{equation}
        \tns{P} = 2\tns{F}\partialder{\Psi}{\tns{C}} = 2 \tns{F} \left[ \left( \sum_{\alpha=1}^{3} \partialder{\omega_1}{\lambda_\alpha} \partialder{\lambda_\alpha}{\tns{C}} + \sum_{\alpha=1}^{3} \partialder{\omega_{-1}}{\nu_\alpha}\partialder{\nu_\alpha}{\tns{C}} \right) +  \sum_{\alpha=1}^{3} \partialder{\Omega}{J} \partialder{J}{\lambda_\alpha}\partialder{\lambda_\alpha}{\tns{C}} \right].
    \end{equation}
    Using \eqref{eq:rel_ps} and 
    \begin{equation}
        \partialder{\nu_\alpha}{\tns{C}} = \frac{\nu_\alpha}{2}\left( \tns{C}^{-1} - \frac{1}{\lambda_\alpha^2} \vec{N}_\alpha\otimes\vec{N_\alpha} \right),
        \quad 
        \tns{C}^{-1} = \sum_{\alpha=1}^{3} \frac{1}{\lambda_\alpha^2} \vec{N}_\alpha \otimes \vec{N}_\alpha,
    \end{equation}
    we obtain
    \begin{align}
        \begin{split}
        \tns{P} =&\tns{F} \Bigg[\sum_{\alpha=1}^{3} \frac{1}{\lambda_\alpha} \partialder{\omega_1}{\lambda_\alpha}  \vec{N}_\alpha \otimes \vec{N}_\alpha + \sum_{\beta=1}^{3} \nu_\beta\partialder{\omega_{-1}}{\nu_\beta} \Bigg( \sum_{\alpha=1}^{3} \frac{1}{\lambda_\alpha^2} \vec{N}_\alpha \otimes \vec{N}_\alpha \Bigg)   \\ 
        & - \sum_{\alpha=1}^{3} \frac{\nu_\alpha}{\lambda_\alpha^2} \partialder{\omega_{-1}}{\nu_\alpha}   \vec{N}_\alpha\otimes\vec{N_\alpha}  +  \sum_{\alpha=1}^{3} \partialder{\Omega}{J}  \frac{J}{\lambda_\alpha^2} \vec{N}_\alpha \otimes \vec{N}_\alpha \Bigg] \end{split}\\
        = &\sum_{\alpha=1}^{3} \Bigg[ \partialder{\omega_1}{\lambda_\alpha} + \frac{1}{\lambda_\alpha} \sum_{\beta=1}^{3} \nu_\beta \partialder{\omega_{-1}}{\nu_\beta}  - \frac{\nu_\alpha}{\lambda_\alpha} \partialder{\omega_{-1}}{\nu_\alpha} +  \nu_\alpha \partialder{\Omega}{J} \Bigg] \vec{n}_\alpha \otimes \vec{N}_\alpha . \label{eq:1PK_explicit}
    \end{align}
    Comparing \eqref{eq:1PK_explicit} and \eqref{eq:1PK_spec_comp} finally yields
    \begin{equation}
        P_\alpha = \partialder{\Psi}{\lambda_\alpha} = \partialder{\omega_1}{\lambda_\alpha} + \frac{1}{\lambda_\alpha} \sum_{\beta=1}^{3} \nu_\beta\partialder{\omega_{-1}}{\nu_\beta}  - \frac{\nu_\alpha}{\lambda_\alpha} \partialder{\omega_{-1}}{\nu_\alpha} + \nu_\alpha \partialder{\Omega}{J}.
    \end{equation}
    For the incompressible case, the strain energy function takes on the form  
    \begin{equation}
        \Psi = \hat{\Psi}(\lambda_1,\lambda_2,\lambda_3, \vec{f}) -p(J-1)= \sum_{\alpha=1}^{3} \hat{\omega}_1(\lambda_\alpha, \vec{f}) + \sum_{\alpha=1}^{3} \hat{\omega}_{-1}(\lambda_\alpha^{-1}, \vec{f})  - p(J-1).
    \end{equation}
    Using
    \begin{equation}
        \partialder{\lambda_\alpha^{-1}}{\tns{C}} = - \frac{1}{2\lambda_\alpha^3} \vec{N}_\alpha\otimes\vec{N_\alpha}
    \end{equation}
    and following the same steps as in the compressible case, we obtain  
    \begin{equation}
        P_\alpha = -\frac{p}{\lambda_\alpha} + \partialder{\hat{\Psi}}{\lambda_\alpha} = -\frac{p}{\lambda_\alpha} + \partialder{\hat{\omega}_1}{\lambda_\alpha}  - \frac{1}{\lambda_\alpha^2} \partialder{\hat{\omega}_{-1}}{\lambda_\alpha^{-1}}.
    \end{equation}

    \subsection{Special deformation modes} \label{app:HomogenousModes}
    In the numerical examples presented in Section \ref{sec:results}, we use experimental stress data from biaxial tensile and simple shear tests on rubber and human brain tissue, conducted under conditions of homogeneous deformation. During training, the error between the experimental stress data and the stress predicted by the CKAN is minimized, thereby implicitly learning the strain energy function. Consequently, we specify the stress tensor associated with the strain energy function for these special load cases.
    
\begin{figure}[ht!]
        \centering
        \begin{subfigure}[b]{0.22\linewidth}
            \centering
            \includegraphics[width=\linewidth]{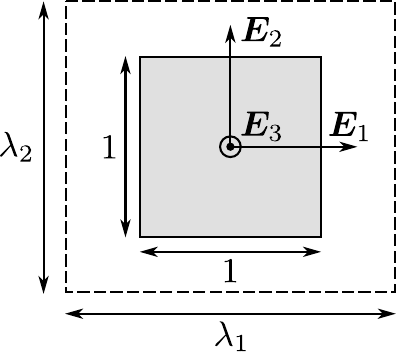}
            \caption{General biaxial tensile test.}
            \label{fig:biaxial_test}
        \end{subfigure}
        \hfill
        \begin{subtable}[b]{.5\linewidth}
            \centering
            \begin{tabular}{@{}llll@{}}
                \toprule
                Deformation mode  &  $\lambda_{1}$ & $\lambda_{2}$ & $\lambda_{3}$ \\
                \midrule
                Uniaxial tension & $\lambda$ & $\lambda^{-1/2}$ & $\lambda^{-1/2}$ \\
                Equibiaxial tension &  $\lambda$ & $\lambda$ & $\lambda^{-2}$ \\
                Pure shear & $\lambda$ & $1$ & $\lambda^{-1}$ \\
                \bottomrule
            \end{tabular}
            \vspace*{0.7cm}
            \caption{Special biaxial tensile test.}
            \label{tab:protocols}
        \end{subtable}
        \hfill
        \begin{subfigure}[b]{0.22\linewidth}
            \centering
            \includegraphics[width=\linewidth]{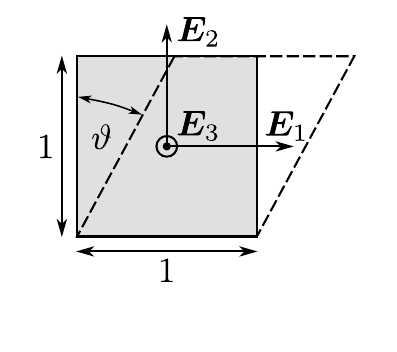}
            \caption{Simple shear test.}
            \label{fig:simple_shear}
        \end{subfigure}
        \caption{Homogeneous deformations. (a) In a general biaxial tensile test, a thin incompressible material specimen is stretched in the $\vec{E}_1$- and $\vec{E}_2$-directions by two independent stretches $\lambda_1$ and $\lambda_2$, respectively. The stretch $\lambda_3$ in the thickness direction $\vec{E}_3$ is determined by the incompressibility condition. (b) Common special cases of the general biaxial tensile test are characterized by fixed ratios between the stretches $\lambda_1$ and $\lambda_2$. (c) In a simple shear test, the top and bottom surfaces of an incompressible material specimen are sheared relative to each other by an angle $\vartheta$, while maintaining their initial distance in the $\vec{E}_2$-direction.}
        \label{fig:experiments}
    \end{figure}

    In the following, the component indices of tensors refer to the fixed Cartesian  coordinate system with basis vectors $\{ \vec{E}_1,\vec{E}_2,\vec{E}_3 \}$ illustrated in Figure \ref{fig:experiments}. For the ease of notation, we simply equate in the following tensors with the matrices representing their components. This notation is unambiguous and thus justified because we always refer to the above fixed and well-defined basis.

    \subsubsection{Biaxial tension}
    In biaxial tensile tests, we stretch a thin material specimen in two orthogonal directions $\vec{E}_1$ and $\vec{E}_2$ by stretches $\lambda_1$ and $\lambda_2$, Figure \ref{fig:biaxial_test}. The incompressibility condition determines the stretch in the thickness direction $\vec{E}_3$: $\lambda_3 = (\lambda_1 \lambda_2 )^{-1}$. Although $\lambda_1$ and $\lambda_2$ can be independently controlled in a general biaxial tension test \cite{Kawabata1981}, specific test protocols in which $\lambda_1$ and $\lambda_2$ follow predefined functional relationships are often employed, Figure \ref{tab:protocols}. For a biaxial tensile test, the deformation gradient in matrix notation and the principal invariants are
    \begin{equation}\label{eq:kin_biax}
        \tns{F} =
        \begin{bmatrix}
            \lambda_1 & 0 & 0 \\ 0 & \lambda_2 & 0 \\ 0 & 0 & (\lambda_1\lambda_2)^{-2} \\  
        \end{bmatrix}, \qquad I_1 = \lambda_1^2+\lambda_2^2 + (\lambda_1\lambda_2)^{-2}, \quad I_2 = \lambda_1^2\lambda_2^2 + \lambda_1^{-2} + \lambda_2^{-2}, \quad I_3=1.
    \end{equation}
    The only non-zero stress components are $\mathrm{P}_{11}$ and $\mathrm{P}_{22}$ and depending on the employed functional basis for the strain energy function, we arrive at the following expressions:
    
    \paragraph{Principal invariant-based strain energy function} Inserting the deformation gradient and first principal invariant of a biaxial tensile deformation from \eqref{eq:kin_biax}, along with the incompressibility condition, $\lambda_3 = (\lambda_1 \lambda_2 )^{-1}$, into \eqref{eq:1PK_invar_incomp} allows us to determine the hydrostatic pressure $p$ from the plane stress assumption, $\mathrm{P}_{33}=0$:
    \begin{equation}
        p = \frac{2}{\lambda_{1}^{2} \lambda_{2}^{2}} \bigg[\partialder{\hat{\Psi}}{I_1} + \partialder{\hat{\Psi}}{I_2} \left( \lambda_{1}^{2} + \lambda_{2}^{2} \right) \bigg],
    \end{equation}
    Substituting $p$ back into \eqref{eq:1PK_invar_incomp} results in
    \begin{equation}\label{eq:1PK_biax_incomp}
        \tns{P} = 
        \begin{bmatrix}2 \left( \lambda_{1} - \frac{1}{\lambda_{1}^{3} \lambda_{2}^{2}} \right) \left( \partialder{\hat{\Psi}}{I_1} +  \lambda_{2}^{2} \partialder{\hat{\Psi}}{I_2} \right) & 0 & 0 \\
        0 & 2 \left( \lambda_{2} - \frac{1}{\lambda_{1}^{2} \lambda_{2}^{3}} \right) \left( \partialder{\hat{\Psi}}{I_1} +  \lambda_{1}^{2} \partialder{\hat{\Psi}}{I_2} \right) & 0
        \\0 & 0 & 0\end{bmatrix}.
    \end{equation}
    
    \paragraph{Modified invariant-based strain energy function} Inserting the differentiation relations \eqref{eq:dIotadI} into \eqref{eq:1PK_biax_incomp} yields the expression for the modified invariant formulation:
    \begin{equation}
        \tns{P} = 
        \begin{bmatrix}2 \left( \lambda_{1} - \frac{1}{\lambda_{1}^{3} \lambda_{2}^{2}} \right) \left( \frac{1}{6\iota_1} \partialder{\hat{\Psi}}{\iota_1} +  \lambda_{2}^{2} \frac{1}{9\iota_2^2} \partialder{\hat{\Psi}}{\iota_2} \right) & 0 & 0 \\
        0 & 2 \left( \lambda_{2} - \frac{1}{\lambda_{1}^{2} \lambda_{2}^{3}} \right) \left( \frac{1}{6\iota_1} \partialder{\hat{\Psi}}{\iota_1} +  \lambda_{1}^{2} \frac{1}{9\iota_2^2} \partialder{\hat{\Psi}}{\iota_2} \right) & 0
        \\0 & 0 & 0\end{bmatrix}.
    \end{equation}

    \paragraph{Principal stretch-based strain energy function} In a biaxial tensile test of a homogeneous isotropic material, only normal stresses arise, and consequently, the principal directions $\vec{N}_\alpha$ and $\vec{n}_\alpha$, and the basis vectors $\vec{E}_\alpha$ coincide. In this case, the Cartesian stress components directly equate with the principal stresses. Utilizing the plane stress assumption, $\mathrm{P}_{33}=0$, to determine the hydrostatic pressure $p$ from \eqref{eq:1PK_spec_incomp} gives:
    \begin{equation}
        p = \frac{1}{\lambda_1\lambda_2} \partialder{\hat{\Psi}}{\lambda_3},
    \end{equation}
    therefore, substituting $p$ back into \eqref{eq:1PK_spec_incomp}, gives:
    \begin{equation}
        \tns{P} = 
        \begin{bmatrix}
            \partialder{\hat{\Psi}}{\lambda_1} - \frac{1}{\lambda_1^2\lambda_2} \partialder{\hat{\Psi}}{\lambda_3} & 0 & 0 \\
            0 & \partialder{\hat{\Psi}}{\lambda_2} - \frac{1}{\lambda_1\lambda_2^2} \partialder{\hat{\Psi}}{\lambda_3} & 0 \\
            0 & 0 & 0
        \end{bmatrix}.
    \end{equation}
        
    \subsubsection{Simple shear}
    In a simple shear test, we shear the top and bottom surfaces of the specimen relative to each other by the angle $\vartheta=\tan(\gamma)$, while maintaining their initial distance in the $\vec{E}_2$-direction, Figure \ref{fig:simple_shear}. Here, $\gamma$ denotes the shear angle. We obtain
    \begin{equation}\label{eq:kin_ss}
        \tns{F}  = 
        \begin{bmatrix}
            1 & \gamma & 0 \\ 0 & 1 & 0 \\ 0 & 0 & 1 \\  
        \end{bmatrix},
        \qquad \lambda_{1,2} = \frac{\sqrt{4+\gamma^2} \pm \gamma}{2}, \quad \lambda_3 = 1, \qquad I_1 = I_2 = 3 + \gamma^2, \quad I_3=1.
    \end{equation}   
    Depending on the employed functional basis for the strain energy function, we arrive at the following expressions for the first Piola--Kirchhoff stress tensor.

    \paragraph{Principal invariant-based strain energy function} Inserting the deformation gradient and first principal invariant of a simple shear deformation from \eqref{eq:kin_ss}, along with the incompressibility condition, $\lambda_3 = (\lambda_1 \lambda_2 )^{-1}=1$, into \eqref{eq:1PK_invar_incomp} allows us to determine the hydrostatic pressure $p$ from the plane stress assumption, $\mathrm{P}_{33}=0$:
    \begin{equation}
        p = 2 \bigg[ \partialder{\hat{\Psi}}{I_1} + \partialder{\hat{\Psi}}{I_2} (\gamma^2+2) \bigg].
    \end{equation}
    Substituting $p$ back into \eqref{eq:1PK_invar_incomp} results in
    \begin{equation}\label{eq:P_ss_ps}
        \tns{P} = \left[\begin{matrix}- 2 \partialder{\hat{\Psi}}{I_2} \gamma^{2} & 2 \gamma \left(\partialder{\hat{\Psi}}{I_1} + \partialder{\hat{\Psi}}{I_2}\right) & 0 \\
        2 \gamma \left( \partialder{\hat{\Psi}}{I_1} + \partialder{\hat{\Psi}}{I_2} (\gamma^{2} + 1) \right) & - 2 \partialder{\hat{\Psi}}{I_2} \gamma^{2} & 0 
        \\0 & 0 & 0\end{matrix}\right].
    \end{equation}

    \paragraph{Modified invariant-based strain energy function} Substituting the differentiation relations \eqref{eq:dIotadI} into \eqref{eq:P_ss_ps} gives:
    \begin{equation}
        \tns{P} = \left[\begin{matrix}- 2 \frac{1}{9\iota_2^2} \partialder{\hat{\Psi}}{\iota_2} \gamma^{2} & 2 \gamma \left(\frac{1}{6\iota_1} \partialder{\hat{\Psi}}{\iota_1} + \frac{1}{9\iota_2^2} \partialder{\hat{\Psi}}{\iota_2} \right) & 0 \\
        2 \gamma \left( \frac{1}{6\iota_1} \partialder{\hat{\Psi}}{\iota_1} + \frac{1}{9\iota_2^2} \partialder{\hat{\Psi}}{\iota_2} (\gamma^{2} + 1) \right) & - 2 \frac{1}{9\iota_2^2} \partialder{\hat{\Psi}}{\iota_2} \gamma^{2} & 0 
        \\0 & 0 & 0\end{matrix}\right].
    \end{equation}
    
    \paragraph{Principal stretch-based strain energy function} In a simple shear test, the principal directions $\vec{N}_\alpha$ and $\vec{n}_\alpha$ do not align with the basis vectors $\vec{E}_\alpha$, except in the out-of-plane direction. The principal directions in the reference configuration $\vec{N}_\alpha$ are obtained by solving the eigenvector problem of $\tns{C}$:
    \begin{align}\label{eq:eigenvec_ref}
        \vec{N}_1 &= 
        \begin{bmatrix}
            \frac{1}{\sqrt{\lambda_{1}^{2} + 1}}, & \frac{\lambda_{1}}{\sqrt{\lambda_{1}^{2} + 1}}, & 0 
        \end{bmatrix}^\mathrm{T},
        &\vec{N}_2 &= 
        \begin{bmatrix}
            \frac{1}{\sqrt{\lambda_{2}^{2} + 1}}, & - \frac{\lambda_{2}}{\sqrt{\lambda_{2}^{2} + 1}}, & 0
        \end{bmatrix}^\mathrm{T},
        &\vec{N}_3 &= 
        \begin{bmatrix}
            0, & 0, & 1
        \end{bmatrix}^\mathrm{T}.
    \end{align}
    Utilizing \eqref{eq:rel_ps}$_2$, we can calculate the principal directions in the current configuration $\vec{n}_\alpha$:
    \begin{align}\label{eq:eigenvec_cur}
        \vec{n}_1 &= 
        \begin{bmatrix}
            \frac{\gamma \lambda_{1} + 1}{\lambda_{1} \sqrt{\lambda_{1}^{2} + 1}}, & \frac{1}{\sqrt{\lambda_{1}^{2} + 1}}, & 0 
        \end{bmatrix}^\mathrm{T},
        &\vec{n}_2 &= 
        \begin{bmatrix}
            \frac{- \gamma \lambda_{2} + 1}{\lambda_{2} \sqrt{\lambda_{2}^{2} + 1}}, & - \frac{1}{\sqrt{\lambda_{2}^{2} + 1}}, & 0
        \end{bmatrix}^\mathrm{T},
        &\vec{n}_3 &= 
        \begin{bmatrix}
            0, & 0, & 1
        \end{bmatrix}^\mathrm{T}.
    \end{align}
    Inserting the principal directions $\vec{N}_\alpha$ and $\vec{n}_\alpha$ into \eqref{eq:1PK_spec_incomp}, and applying the plane stress assumption, $\mathrm{P}_{33}=0$, enables us to determine the hydrostatic pressure:
    \begin{equation}
        p = \partialder{\hat{\Psi}}{\lambda_3}.
    \end{equation}
    Substituting $p$ along with $\vec{N}_\alpha$ and $\vec{n}_\alpha$ back into \eqref{eq:1PK_spec_comp}, and observing that $1 + \gamma \lambda_{1}  = \lambda_1^2$ and $1 -\gamma \lambda_{2} = \lambda_2^2$, leads to:
    \begin{equation}
        \tns{P} = 
        \begin{bmatrix}
            \frac{ 1}{\lambda_{1}^{2} + 1} \partialder{\hat{\Psi}}{\lambda_1} + \frac{1}{\lambda_{2}^{2} + 1} \partialder{\hat{\Psi}}{\lambda_2} - \partialder{\hat{\Psi}}{\lambda_3} & \frac{\lambda_{1}^2}{\lambda_{1}^{2} + 1} \partialder{\hat{\Psi}}{\lambda_1} - \frac{ \lambda_{2}^2}{\lambda_{2}^{2} + 1} \partialder{\hat{\Psi}}{\lambda_2} & 0 \\
            \frac{1}{\lambda_{1}^{2} + 1} \partialder{\hat{\Psi}}{\lambda_1} - \frac{1}{\lambda_{2}^{2} + 1} \partialder{\hat{\Psi}}{\lambda_2} + \gamma \partialder{\hat{\Psi}}{\lambda_3} & \frac{\lambda_{1}}{\lambda_{1}^{2} + 1} \partialder{\hat{\Psi}}{\lambda_1} + \frac{\lambda_{2}}{\lambda_{2}^{2} + 1} \partialder{\hat{\Psi}}{\lambda_2} - \partialder{\hat{\Psi}}{\lambda_3} & 0\\ 
            0 & 0 & 0
        \end{bmatrix}.
    \end{equation}

    \section{Internal architecture of a principal stretch-based CKAN} \label{app:architecture_prin_stretch}
        \setcounter{figure}{0}
    Figure \ref{fig:internal_architecture} schematically shows the internal architecture of a principal stretch-based CKAN. To preserve the symmetry of the strain energy function $\Psi^\text{KAN}$, parameter sharing is employed.

    \begin{figure}[h]
        \centering
        \includegraphics[width=0.78\linewidth]{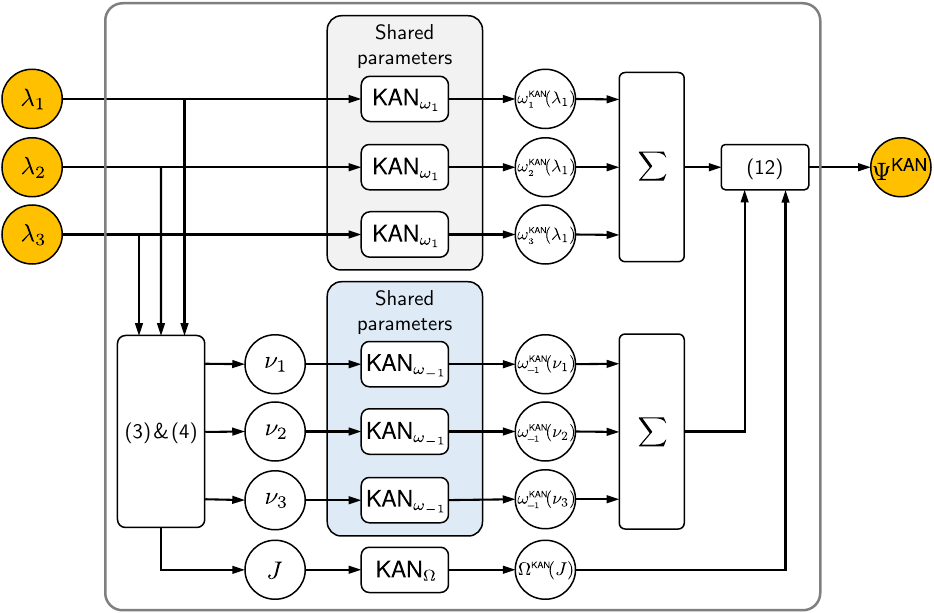}
        \caption{\textbf{Internal architecture of a principal stretch-based Constitutive Kolmogorov--Arnold Network (CKAN)}: Following the ansatz of separable functions from \eqref{eq:separable_comp} and \eqref{eq:separable_incomp}, the $\lambda_\alpha$ and the $\nu_\alpha$ are processed by distinct KANs, denoted as $\textsf{KAN}_{\omega_1}$ and $\textsf{KAN}_{\omega_{-1}}$. However, to ensure material symmetry (isotropy), all the three $\lambda_\alpha$ have to pass through the same mapping $\textsf{KAN}_{\omega_1}$ and all the three $\nu_\alpha$ through the same mapping $\textsf{KAN}_{\omega_1}$. This is implemented through KANs with shared parameters. For clarity, the feature vector $\vec{f}$ is not shown, but it would be processed by each KAN, namely $\textsf{KAN}_{\omega_{1}}$, $\textsf{KAN}_{\omega_{-1}}$, and $\textsf{KAN}_{\Omega}$.}
        \label{fig:internal_architecture}
    \end{figure}
    
    \section{Normalization terms for a stress-free reference configuration} \label{app:normalization}
            \setcounter{figure}{0}

     As outlined in Section \ref{sec:CANN-KAN}, the stress $\tns{S}^\text{KAN}$ associated with $\Psi^\text{KAN}$ does not in general vanish in the undeformed reference configuration. Here, we provide the specific stress normalization terms $\Psi^\sigma$ and the corresponding stress tensors $\tns{S}^\sigma$ to offset $\tns{S}^\text{KAN}$ in the reference configuration for each CKAN formulation considered in this work. By construction, $\Psi^\sigma$ vanishes in the undeformed reference configuration.
    
    \subsection{Principal invariant-based strain energy function}

    In view of \eqref{eq:1PK_invar_comp}, the second Piola--Kirchhoff stress is
    \begin{equation}
        \tns{S}^\textbf{KAN} = 2 \partialder{\Psi^\text{KAN}}{\tns{C}} = 2 \left[ \left(\partialder{\Psi^\text{KAN}}{I_1} + I_1 \partialder{\Psi^\text{KAN}}{I_2} \right) \tns{I} - \partialder{\Psi^\text{KAN}}{I_2}\tns{C} + I_3 \partialder{\Psi^\text{KAN}}{I_3} \tns{C}^{-1} \right],
    \end{equation}
    and becomes in the undeformed reference configuration $\tns{C}=\tns{I}$, $I_1=I_2=3$, $I_3=1$ 
    \begin{equation}\label{eq:2PK_inv_ref}
        \tns{S}^\text{KAN}|_{\tns{C}=\tns{I}} = 2 \left[ \partialder{\Psi^\text{KAN}}{I_1} + 2 \partialder{\Psi^\text{KAN}}{I_2} + \partialder{\Psi^\text{KAN}}{I_3}  \right]\bigg|_{\tns{C}=\tns{I}}\tns{I}.
    \end{equation}
    To offset this generally non-zero stress, we introduce the stress normalization term
    \begin{equation}
        \Psi^\sigma(J,\vec{f}) \coloneqq -\Xi(\vec{f})(J-1)
    \end{equation}
    where
    \begin{equation}
        \Xi(\vec{f})\coloneqq 2\left[ \partialder{\Psi^\text{KAN}}{I_1} + 2 \partialder{\Psi^\text{KAN}}{I_2} + \partialder{\Psi^\text{KAN}}{I_3}\right]\bigg|_{\tns{C}=\tns{I}},
    \end{equation}
    is determined by the derivatives of $\Psi^\text{KAN}$ with respect to the principal invariants, evaluated in the undeformed reference configuration. However, during training, $\Xi$ continuously adapts as $\Psi^\text{KAN}$ evolves with each update of the trainable spline parameters and is \emph{not} computed in a post-processing step. Evaluating the corresponding stress contribution
    \begin{equation} \label{eq:stress_offest_inv}
        \tns{S}^\sigma = - \Xi\, J \tns{C}^{-1}.
    \end{equation}
    for $\tns{C}=\tns{I}$ confirms that $\tns{S}^\sigma$ indeed corrects \eqref{eq:2PK_inv_ref}, ensuring a stress-free reference configuration.
    For an isotropic incompressible material, the 
        
    \subsection{Modified invariant-based strain energy function}
    Following the same reasoning, the stress normalization term for the modified invariant-based formulation is defined as
    \begin{equation}
        \Psi^{\sigma}(J,\vec{f}) \coloneqq - \Xi(\vec{f}) (J-1),
    \end{equation}
    where
    \begin{equation}
        \Xi(\vec{f}) \coloneqq 2 \left[ \frac{1}{6} \partialder{\Psi^\text{KAN}}{\iota_1} + \frac{2}{9} \partialder{\Psi^\text{KAN}}{\iota_2} + \partialder{\Psi^\text{KAN}}{I_3} \right]\bigg|_{\tns{C}=\tns{I}}
    \end{equation}
    The corresponding stress contribution, ensuring a stress-free reference configuration, is given by
    \begin{equation} 
        \tns{S}^\sigma = - \Xi J \tns{C}^{-1}.
    \end{equation}

    \subsection{Principal stretch-based strain energy function}
    In view of \eqref{eq:1PK_explicit}, the second Piola--Kirchhoff stress tensor is
    \begin{equation}
        \tns{S}^\text{KAN} = \sum_{\alpha=1}^{3} \Bigg[ \frac{1}{\lambda_\alpha} \partialder{\omega^\text{KAN}_1}{\lambda_\alpha} + \frac{1}{\lambda_\alpha^2} \sum_{\beta=1}^{3} \nu_\beta \partialder{\omega^\text{KAN}_{-1}}{\nu_\beta}  - \frac{\nu_\alpha}{\lambda_\alpha^2} \partialder{\omega^\text{KAN}_{-1}}{\nu_\alpha} + \frac{\nu_\alpha}{\lambda_\alpha} \partialder{\Omega^\text{KAN}}{J} \Bigg] \vec{N}_\alpha \otimes \vec{N}_\alpha
    \end{equation}   
    and becomes in the undeformed reference configuration $\tns{C}=\tns{I}$, $\lambda_\alpha = \nu_\alpha =J=1$
    \begin{align}
        \tns{S}^\text{KAN}|_{\tns{C}=\tns{I}} 
        & = \Bigg[ \partialder{\omega^\text{KAN}_1}{\lambda_\alpha} + 2  \partialder{\omega^\text{KAN}_{-1}}{\nu_\alpha} + \partialder{\Omega^\text{KAN}}{J} \Bigg] \Bigg|_{\tns{C}=\tns{I}} \tns{I},
    \end{align} 
    where we exploited that $\sum_{\alpha=1}^{3} \vec{N}_\alpha \otimes \vec{N}_\alpha = \tns{I}$. Thus, we define the normalization term
    \begin{equation}
        \Psi^\sigma(J, \vec{f}) = -\Xi(\vec{f}) (J-1), 
    \end{equation}
    where
    \begin{equation}
        \Xi(\vec{f}) \coloneqq \Bigg[ \partialder{\omega^\text{KAN}_1}{\lambda_\alpha} + 2  \partialder{\omega^\text{KAN}_{-1}}{\nu_\alpha} + \partialder{\Omega^\text{KAN}}{J} \Bigg] \Bigg|_{\tns{C}=\tns{I}}.
    \end{equation}
    The corresponding stress contribution ensuring a stress-free reference configuration is
    \begin{equation}
        \tns{S}^\sigma = -\Xi J \tns{C}^{-1}.
    \end{equation}

    \FloatBarrier
    \section{Implementation} \label{sec:implement}

    The computational framework presented here builds on KAN 1.0 \cite{Liu2024a}, implemented in PyTorch, and extends it with the MonoKAN architecture, which enables partial monotonicity constraints \cite{polo2024monokan}. Adapting KANs for constitutive modeling of hyperelastic materials required several modifications. Key implementation aspects include designing the loss function, Appendix \ref{app:loss}, enforcing monotonicity constraints, Appendix~\ref{app:mono_kan}, and---as discussed in Sections~\ref{sec:sparse} and \ref{sec:prune}---applying sparsification and pruning.
    While we described the general symbolification process already in Section \ref{sec:symbolification}, we discuss in Appendix \ref{app:symbol_caveats}  caveats of the symbolification specific to the functional bases considered herein. 
    
    \subsection{Loss function} \label{app:loss}
    The objective of CKANs is, in a first step, to learn the spline parameters that minimize the error between the stress prediction of the CKAN $\tns{P}$ and the experimental stress data $\tilde{\tns{P}}$. In all numerical examples in Section \ref{sec:results}, we consider the special biaxial load cases uniaxial tension (UT) and compression (UC), equibiaxial tension (ET), and pure shear (PS), Figure \ref{tab:protocols}. Each load case comprises $n_a$ data tuples: $\{(\lambda_{1,i}^a, \lambda_{2,i}^a, \tilde{\mathrm{P}}_{11,i}^a)\}_{i=1}^{n_a}$, where $a \in \{\mathrm{UT,UC,ET,PS}\}$. Additionally, $n_\mathrm{SS}$ data pairs are obtained from simple shear (SS) tests: $\{(\gamma_i, \tilde{P}_{12,i}^\mathrm{SS})\}_{i=1}^{n_{\mathrm{SS}}}$.
    
    Then, the objective is to minimize the following general (data) loss function:
    \begin{equation}\label{eq:loss_pred}
        \mathcal{L}_\text{data} = \sum_{a \in \{\mathrm{UT,UC,ET,PS}\}} \, \frac{1}{n_a} \sum_{i=1}^{n_a} \left( \frac{\mathrm{P}_{11,i}^a(\lambda_{1,i}^a,\lambda_{2,i}^a) - \tilde{\mathrm{P}}_{11,i}^a}{ \tilde{\mathrm{P}}_{11,i}^a} \right)^2 + \frac{1}{n_{\mathrm{SS}}} \sum_{i=1}^{n_{\mathrm{SS}}} \left( \frac{\mathrm{P}_{12,i}^\mathrm{SS}(\gamma_{i}) - \tilde{\mathrm{P}}_{12,i}^\mathrm{SS}}{ \tilde{\mathrm{P}}_{12,i}^\mathrm{SS}} \right)^2.
    \end{equation}
    Often, not all load cases are considered simultaneously during training, but only a subset is used, while the withheld load cases validate the model's generalization capabilities. In addition, the loss function could be augmented with other stress components, if available.  However, typically only the $\mathrm{P}_{11}$ and $\mathrm{P}_{12}$ components are reported for the special biaxial load cases and simple shear, respectively, and used in this work.

    \subsection{Incorporating (partially) monotonic splines into KANs} \label{app:mono_kan}
    To maintain a monotonic strain energy function in hyperelastic material modeling, we build on the MonoKAN framework introduced by Polo-Molina et al.~\cite{polo2024monokan}. This approach extends the Kolmogorov--Arnold Network (KAN) by imposing (partial) monotonicity constraints on the spline activation function, thereby enforcing that the strain energy increases with strain in physically meaningful ways. Such monotonicity mitigates numerical pathologies like stress oscillations or material instabilities, thus enhancing both robustness and extrapolation capabilities.

    While the original KAN formulation \cite{Liu2024a} uses B-splines as activation, MonoKAN \cite{polo2024monokan} uses cubic Hermite splines and imposes constraints on their control points to ensure monotonicity. Let \(\mathbf{X} = \{x_k\}_{k=1}^K\) and \(\mathbf{Y} = \{y_k\}_{k=1}^K\) be the knots and their associated control points, respectively. For an increasing spline, we require \(y_k \leq y_{k+1}\) so that the spline value does not decrease between adjacent knots. Similarly, the derivatives at each knot, \(m_k\), must satisfy \(m_k \geq 0\). Where consecutive control points are equal (\(y_k = y_{k+1}\)), the corresponding derivative is set to zero to preserve monotonicity. The slope of the secant line between two knots is denoted \(d_k = \frac{y_{k+1} - y_k}{x_{k+1} - x_k}\), and the ratios 
    \begin{equation}
        \alpha_k = \frac{m_k}{d_k}, \quad \beta_k = \frac{m_{k+1}}{d_k}
    \end{equation}    
    must satisfy \(\alpha_k^2 + \beta_k^2 \leq 9\) to ensure monotonic segments \cite{fritsch1980monotone}. The cubic Hermite spline for a segment is then defined as
    \begin{equation}
        p_k(t) = h_{00}(t)\,y_k + h_{10}(t)\,(x_{k+1} - x_k)\,m_k + h_{01}(t)\,y_{k+1} + h_{11}(t)\,(x_{k+1} - x_k)\,m_{k+1},
    \end{equation}
    where \(t = \frac{x - x_k}{x_{k+1} - x_k}\) and the Hermite basis functions \(h_{ij}(t)\) are:
    \begin{align}
        h_{00}(t) &= 2t^3 - 3t^2 + 1, &h_{10}(t) &= t^3 - 2t^2 + t,\\
        h_{01}(t) &= -2t^3 + 3t^2, &h_{11}(t) &= t^3 - t^2.
    \end{align}    
    
    During training, the parameters \(\mathbf{Y}\) and \(\mathbf{M} = \{m_k\}_{k=1}^K\) are updated at each optimization step. First, we enforce the ordering of the control points (\(y_k \leq y_{k+1}\) or \(y_k \geq y_{k+1}\) for decreasing monotonicity) by projecting invalid values. Second, we adjust \(m_k\) to satisfy the non-negativity or non-positivity constraint. Finally, if \(\alpha_k^2 + \beta_k^2 > 9\), we scale both \(\alpha_k\) and \(\beta_k\) by $\tau_k = 3/(\alpha_k^2 + \beta_k^2)^{1/2}$. By imposing monotonic behavior, the CKAN framework respects essential physical requirements of hyperelastic modeling while retaining the predictive and interpretive advantages of the KAN architecture.

    \subsection{Functional basis-specific symbolification} \label{app:symbol_caveats}
    When an input vector $\vec{x}_0^s$ is passed to a CKAN, information propagates from the input layer to the output layer by computing the node outputs $x_{l,i}$ which serve as inputs to the activation functions $\phi_{l,j,i}$, cf. \eqref{eq:activation}. The values of $x_{l,i}^s$ and $\phi_{l,j,i}(x_{l,i}^s)$ based on the input vector $\vec{x}_0^s$ are stored in the corresponding layers during the final forward pass of $\vec{x}_0^s$ and subsequently serve as sample points for the symbolification process outlined in Section  \ref{sec:symbolification}. Thus, selecting $\vec{x}_0^s$ wisely in the final forward pass before symbolizing the activation functions is crucial. To this end, dedicated sampling strategies are proposed for principal and modified invariant-based CKANs, as well as for principal stretch-based CKANs. For CKANs relying on a mixed functional basis, these sampling strategies are applied individually to the respective sub-networks.
    
    \paragraph{Principal and modified invariant-based CKANs} 
    If the CKAN is trained using mini-batches, where each mini-batch contains data from a single load case or a subset thereof, the symbolification process relies on only a small portion of the training data. As a result, the symbolic form and affine parameters derived from this limited sample set may be suboptimal.
    For CKANs that use principal invariants or modified invariants as the functional basis, a straightforward solution to this issue is to pass the entire training dataset as a full batch during the final forward pass through the CKAN, immediately before the symbolification process. This ensures that the symbolification captures a more comprehensive representation of the training data, leading to improved accuracy and robustness of the symbolic model.

    \paragraph{Principal stretch-based CKANs} For principal stretch-based CKANs, passing the entire training set as a full batch does not solve the problem. Considering the separable strain energy function \eqref{eq:separable_incomp} and following the standard operator precedence in \texttt{Python}, where the two sums are evaluated from left to right, the strain energy functions $\omega_1(\lambda_\alpha)$ and $\omega_{-1}(\nu_\alpha)$ are always evaluated last with $\lambda_3$ and $\nu_3$ as inputs, respectively. For example, in a uniaxial tension test of an incompressible material, as shown in Figure \ref{fig:biaxial_test}, where the specimen is stretched in the $\vec{E}_1$-direction by $\lambda_1\geq1$, the lateral contractions are $0<\lambda_3, \lambda_2\leq1$. Performing symbolic regression of $\hat{\omega}_1$ using $x_{l,i}$ and $\phi_{l,j,i}(x_{l,i})$ obtained by evaluating the spline activation with $\lambda_3\leq1$ is unlikely to yield accurate results if the symbolic forms are later evaluated with $\lambda_1\geq1$. To ensure that the symbolification accounts for all $\lambda_\alpha$ and $\nu_\alpha$ present in the training data, we stack all these values and pass them as a full batch to $\omega_1$ and $\omega_{-1}$, respectively. Alternatively, one could identify the minimum and maximum values, $\lambda_{\min} = \min\{ \lambda_{\alpha,i}^a\}$ and $\lambda_{\max} = \max\{\lambda_{\alpha,i}^a\}$, from the training dataset and use \texttt{torch.linspace} to generate a sequence from $\lambda_{\min}$ to $\lambda_{\max}$ with $n_s$ points. This sequence is then passed to $\omega_1$ before symbolification. This approach allows flexibility in choosing the number of sampling points $n_s$, independent of the actual training data size, which can be advantageous if the derivatives ${\phi'}^s_{l,j,i}$ in \eqref{eq:loss_affine} are numerically computed, for instance, using finite differences. The same procedure applies to the area stretches $\nu_\alpha$ and the strain energy function $\omega_{-1}$.\\

    After symbolifying an activation function, only its affine parameters remain trainable during subsequent training. However, this can lead to violations of constraints---such as monotonicity---that were imposed on the spline activation function and initially inherited by its symbolic counterpart through constraints on the affine parameters.
    To address this issue, we adopted a pragmatic approach by fixing the affine parameters after symbolification.\footnote{This can be implemented, for example, by defining the \texttt{affine} attribute of the \texttt{SymbolicCKAN} layer as a \texttt{torch.Tensor} instead of a trainable \texttt{torch.nn.Parameter}.} Symbolifying one activation function at a time and continuing training can often be beneficial, particularly if the affine parameters are fixed post-symbolification. This strategy allows the remaining splines to adjust to the previously symbolized activation function. Once all activation functions in the network are symbolified,  all affine parameters can be optimized simultaneously in a final step.
    For this final step, optimizers such as BFGS-B are particularly useful, as they can enforce constraints directly on the affine parameters. This ensures that any constraints imposed on the strain energy function---such as monotonicity---are satisfied. Incorporating this final optimization step substantially improved the model's overall performance in all considered numerical examples in Section \ref{sec:results}.

    \section{Treloar’s and Kawabata’s experiments}
    \setcounter{figure}{0}
    
    \begin{figure}[ht!]
        \centering
            \begin{subfigure}{0.46\linewidth}
            \hspace{0.25cm}
            \includegraphics[width=\linewidth]{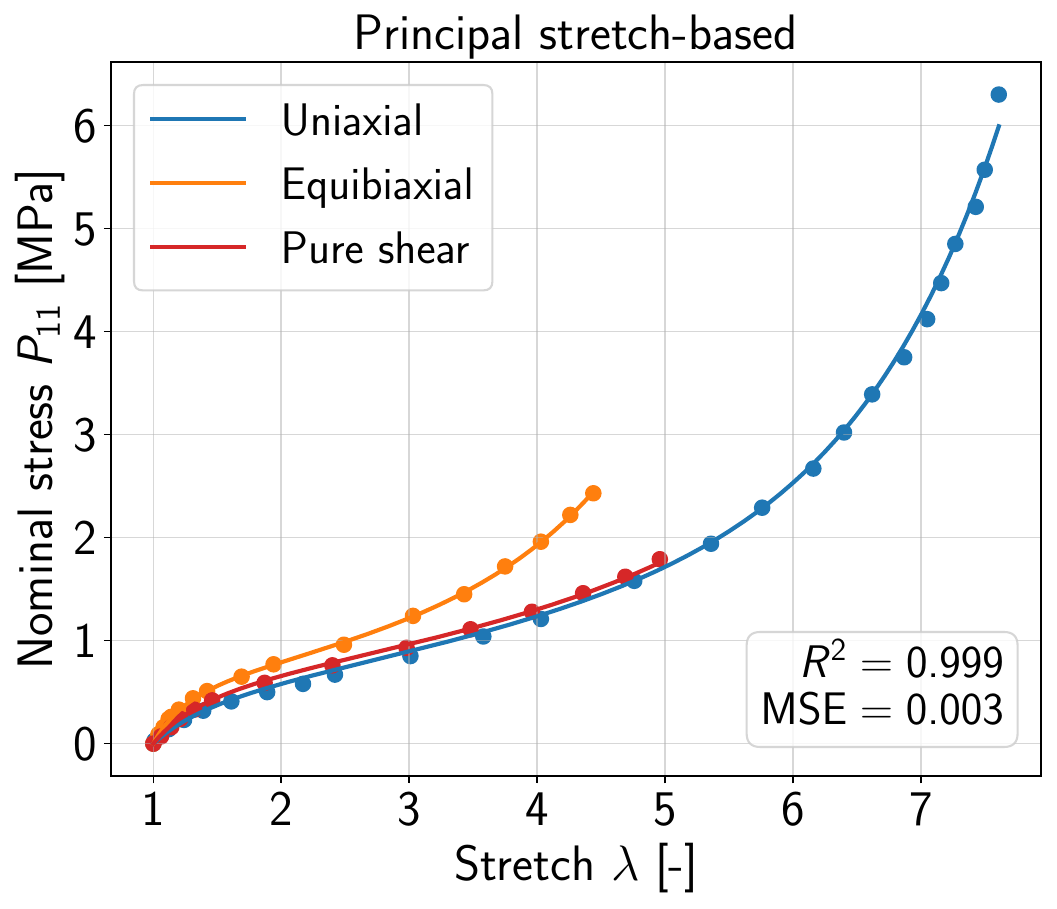}
            \caption{Fitting performance on Treloar's data.}
            \label{fig:result_treloar_ps}
        \end{subfigure}
        \hfill
        \begin{subfigure}{0.46\linewidth}
            \centering
            \includegraphics[width=0.75\linewidth]{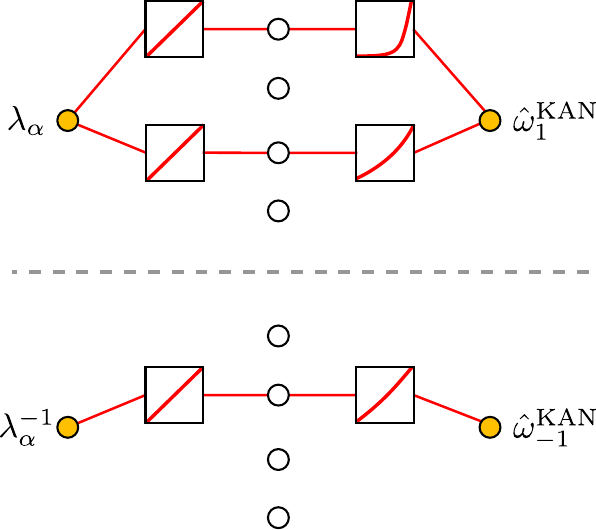}
            \vspace{0.8cm}
            \caption{\emph{Principal stretch-based} CKAN architecture.}
            \label{fig:architecture_treloar_ps}
        \end{subfigure}
        \\[2ex]
        \begin{subfigure}{0.48\linewidth}
            \includegraphics[width=\linewidth]{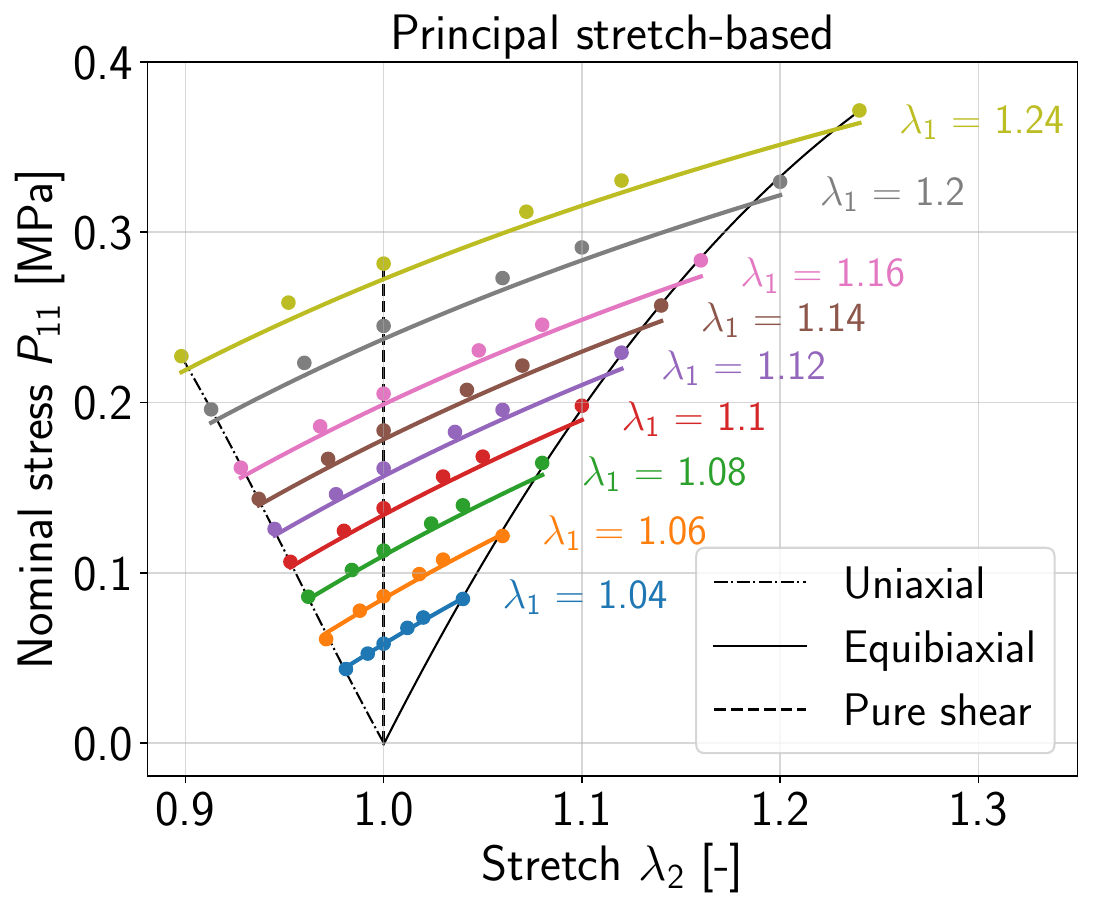}
            \caption{$P_{11}$ generalization: small stretch regime of Kawabata's data.}
            \label{fig:kawabata_P1_small_stretch_ps}
        \end{subfigure}
        \hfill
        \begin{subfigure}{0.48\linewidth}
            \includegraphics[width=\linewidth]{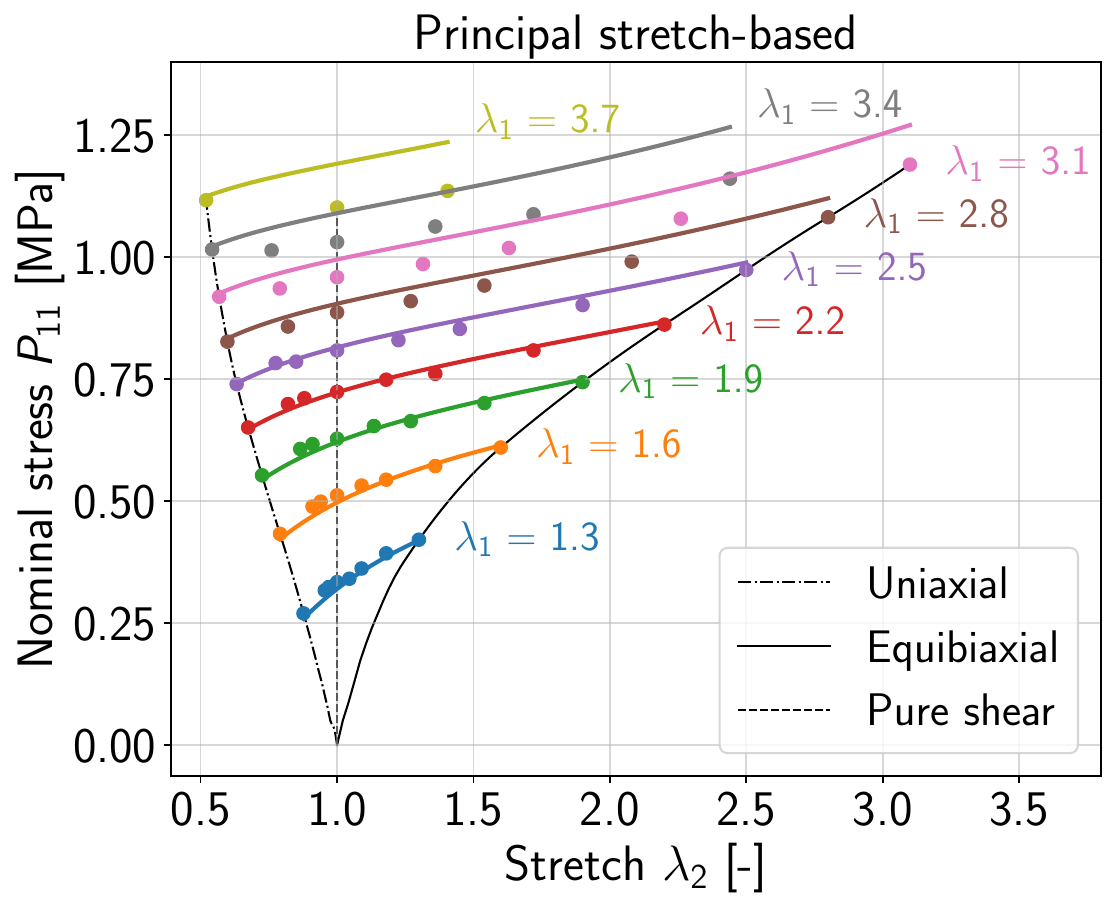}
            \caption{$P_{11}$ generalization: large stretch regime of Kawabata's data.}
            \label{fig:kawabata_P1_large_stretch_ps}
        \end{subfigure}
        \\[2ex]
        \centering
        \begin{subfigure}{0.48\linewidth}
            \includegraphics[width=\linewidth]{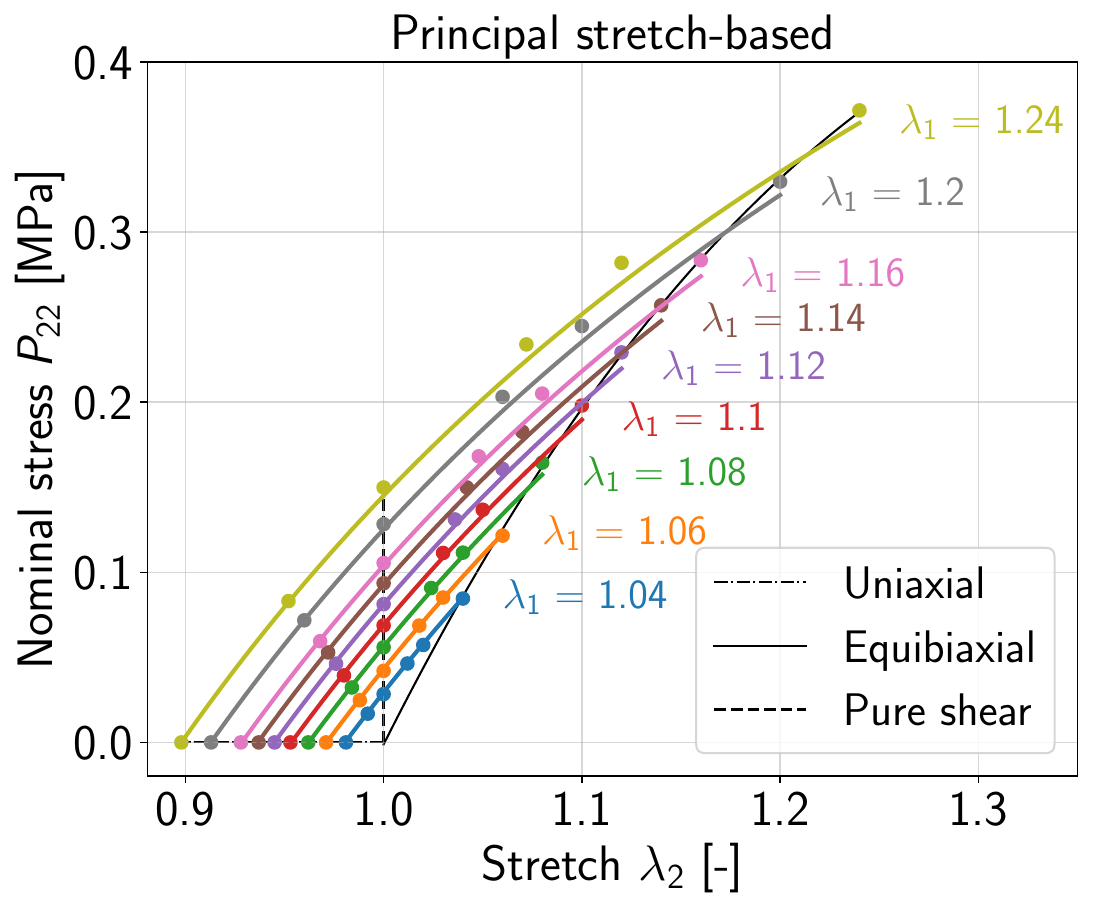}
            \caption{$P_{22}$ generalization: small stretch regime of Kawabata's data.}
            \label{fig:kawabata_P2_small_stretch_ps}
        \end{subfigure}
        \hfill
        \begin{subfigure}{0.48\linewidth}
            \includegraphics[width=\linewidth]{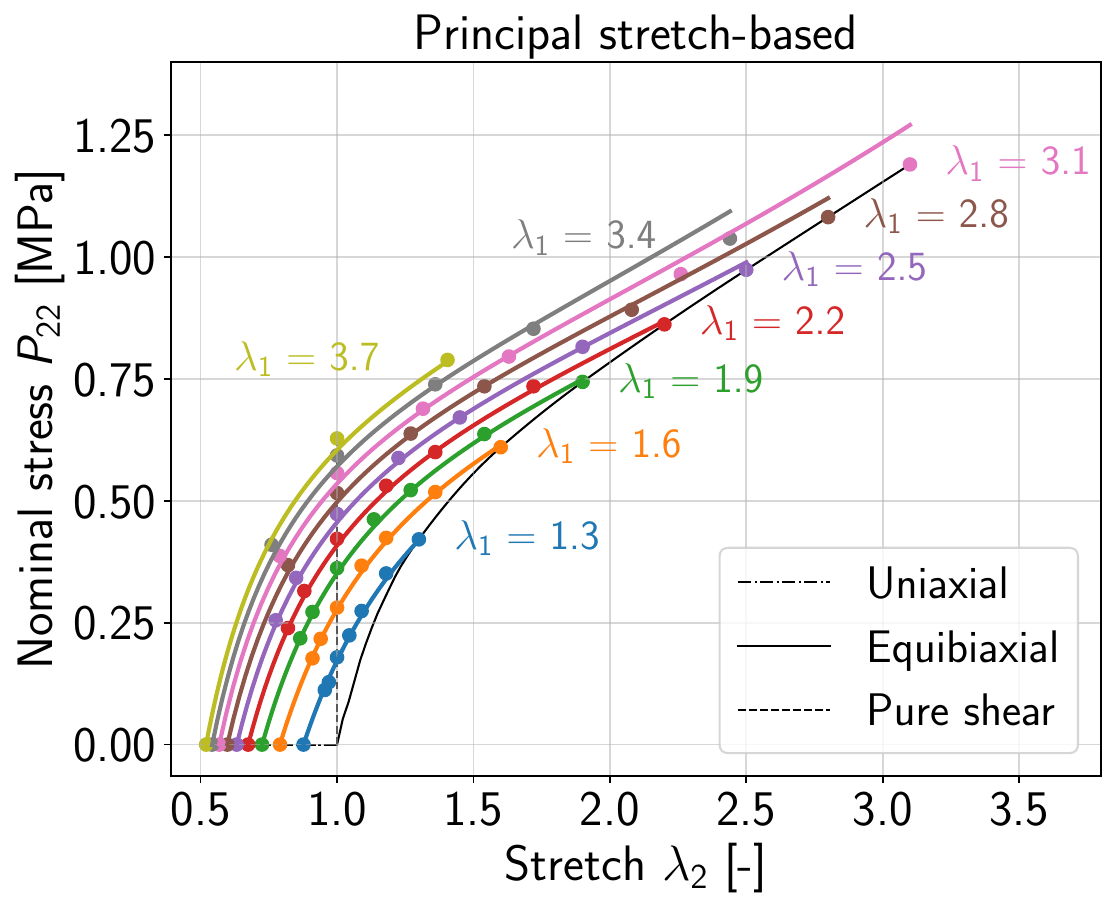}
            \caption{$P_{22}$ generalization: large stretch regime of Kawabata's data.}
            \label{fig:kawabata_P2_large_stretch_ps}
        \end{subfigure}
        \caption{\textbf{Principal stretch-based CKAN for rubber}. (a) Fitting performance on Treloar's data \cite{Treloar1944}. (b) Architecture of the final CKAN after training; unconnected nodes indicate pruned activation functions of the network. (c)--(f) Predictions on the biaxial dataset by Kawabata et al. \cite{Kawabata1981} using the CKAN trained on Treloar's data only. The solid lines represent the model response. The dots represent the experimental data.}
        \label{fig:Treloar_stretch}
    \end{figure}

    \begin{figure}[ht!]
        \centering
        \begin{subfigure}{0.46\linewidth}
            \hspace{0.25cm}
            \includegraphics[width=\linewidth]{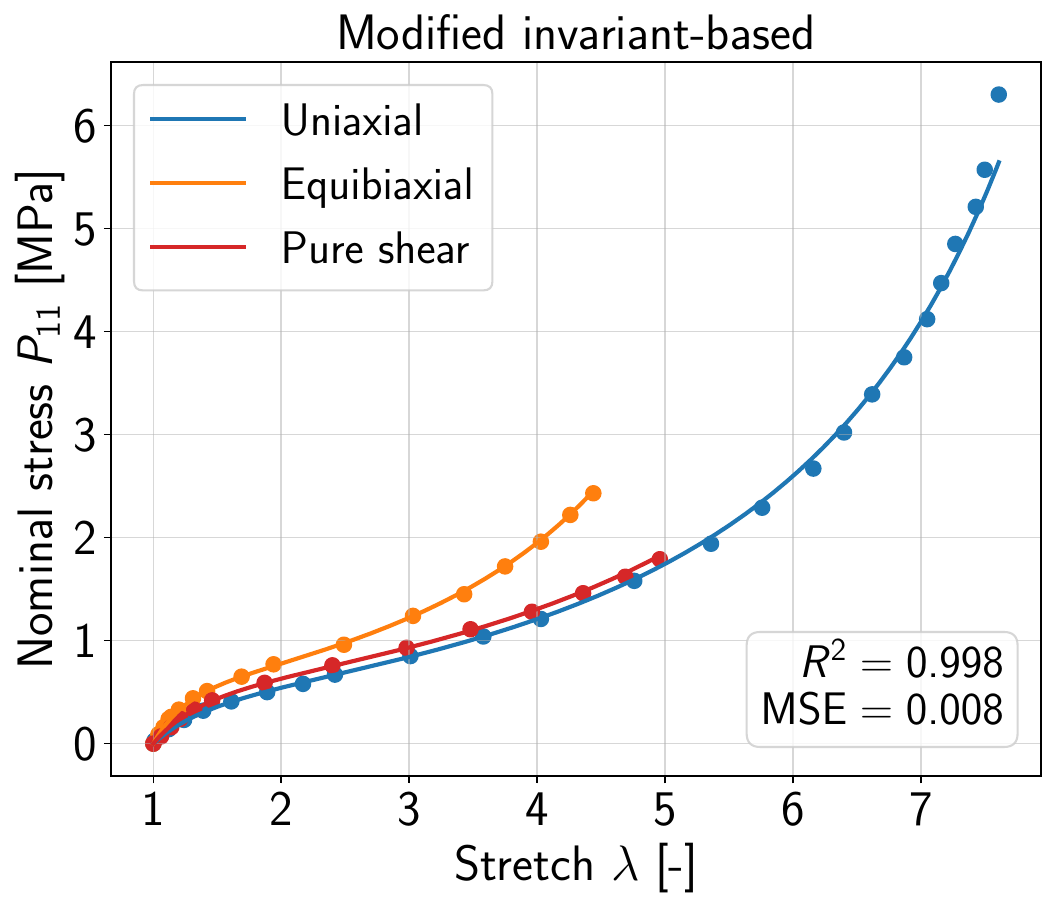}
            \caption{Fitting performance on Treloar's data.}
            \label{fig:result_treloar_modInvar}
        \end{subfigure}
        \hfill
        \begin{subfigure}{0.42\linewidth}
        \centering
            \includegraphics[width=0.85\linewidth]{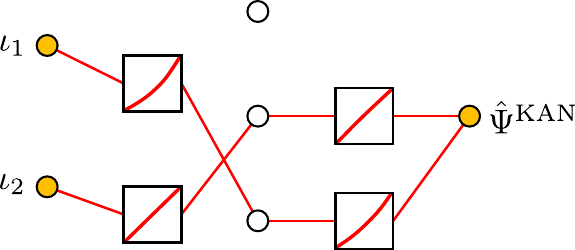}
            \vspace{2cm}
            \caption{\emph{Modified invariant-based} CKAN architecture.}
            \label{fig:architecture_treloar_modInvar}
        \end{subfigure}
        \\[2ex]
        \begin{subfigure}{0.48\linewidth}
            \includegraphics[width=\linewidth]{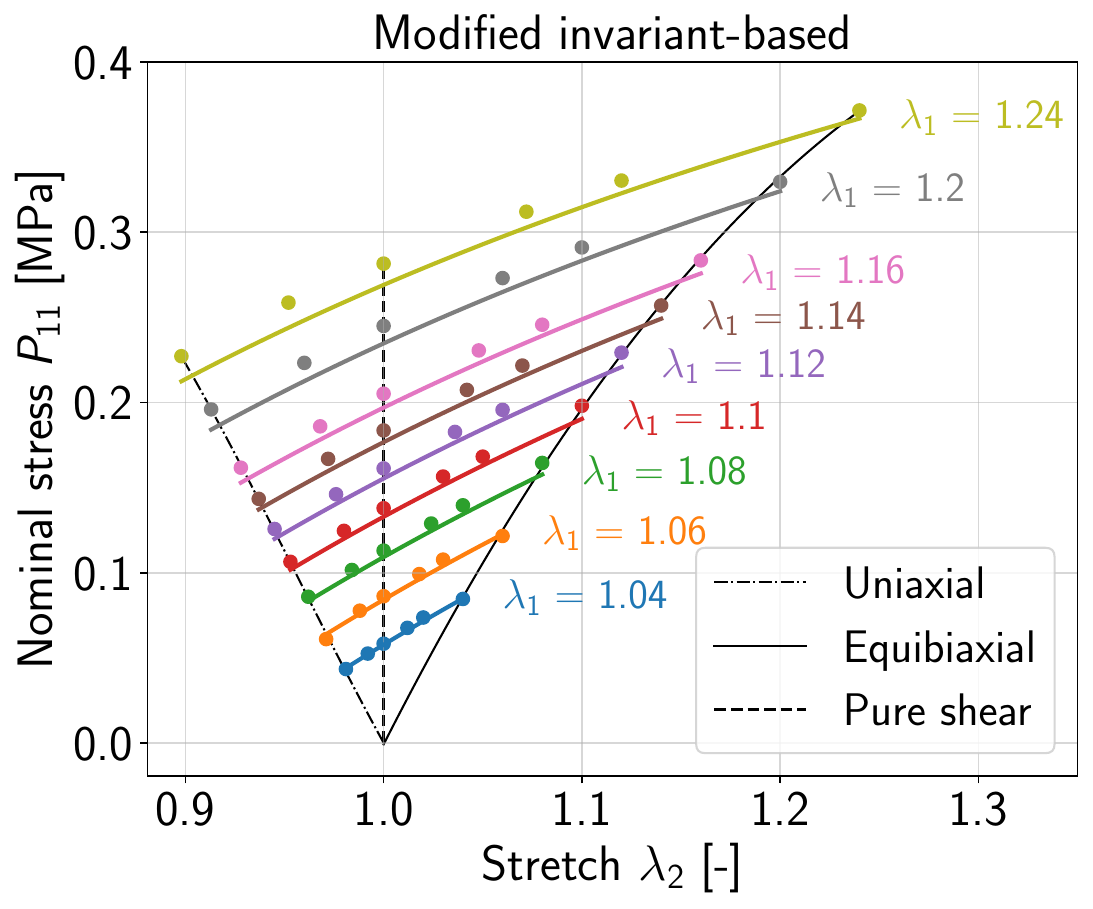}
            \caption{$P_{11}$ generalization: small stretch regime of Kawabata's data.}
            \label{fig:kawabata_P1_small_stretch_modInvar}
        \end{subfigure}
        \hfill
        \begin{subfigure}{0.48\linewidth}
            \includegraphics[width=\linewidth]{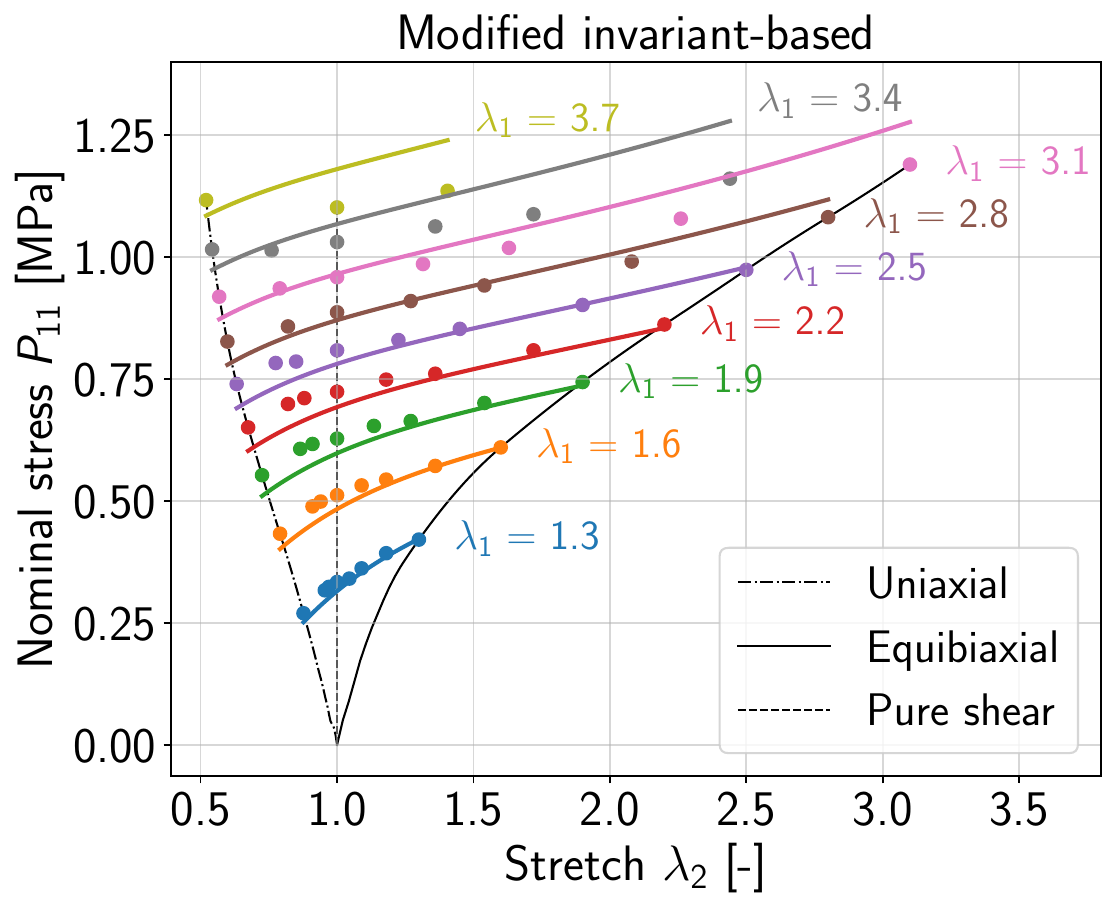}
            \caption{$P_{11}$ generalization: large stretch regime of Kawabata's data.}
            \label{fig:kawabata_P1_large_stretch_modInvar}
        \end{subfigure}
        \\[2ex]
        \centering
        \begin{subfigure}{0.48\linewidth}
            \includegraphics[width=\linewidth]{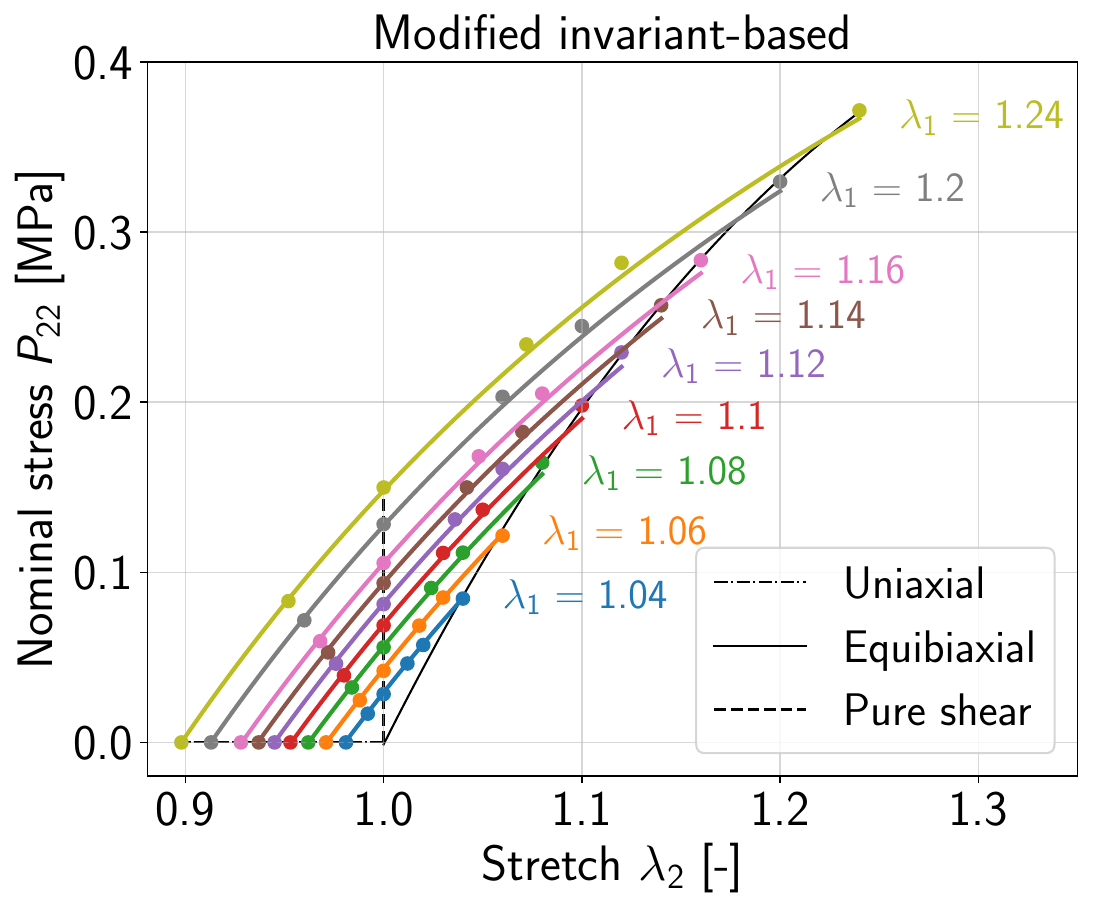}
            \caption{$P_{22}$ generalization: small stretch regime of Kawabata's data.}
            \label{fig:kawabata_P2_small_stretch_modInvar}
        \end{subfigure}
        \hfill
        \begin{subfigure}{0.48\linewidth}
            \includegraphics[width=\linewidth]{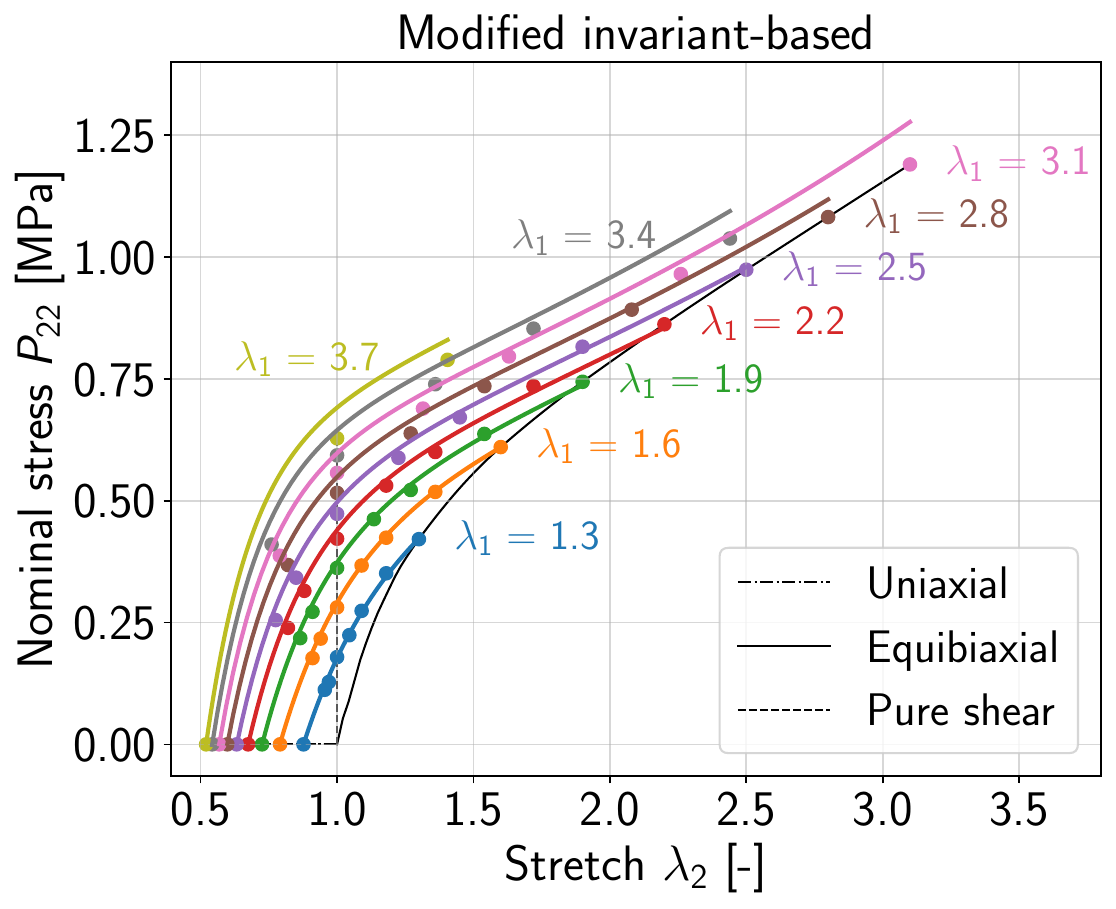}
            \caption{$P_{22}$ generalization: large stretch regime of Kawabata's data.}
            \label{fig:kawabata_P2_large_stretch_modInvar}
        \end{subfigure}
        \caption{\textbf{Modified invariant-based CKAN for rubber}. (a) Fitting performance on Treloar's data \cite{Treloar1944}. (b) Architecture of the CKAN after training; unconnected nodes indicate pruned activation functions of the network. (c)--(f) Predictions on the biaxial dataset by Kawabata et al. \cite{Kawabata1981} using the CKAN trained on Treloar's data only. The solid lines represent the model response. The dots represent the experimental data.}
        \label{fig:Treloar_ModInv}
    \end{figure}

    \begin{figure}[ht!]
        \centering
            \begin{subfigure}{0.46\linewidth}
            \hspace{0.25cm}
            \includegraphics[width=\linewidth]{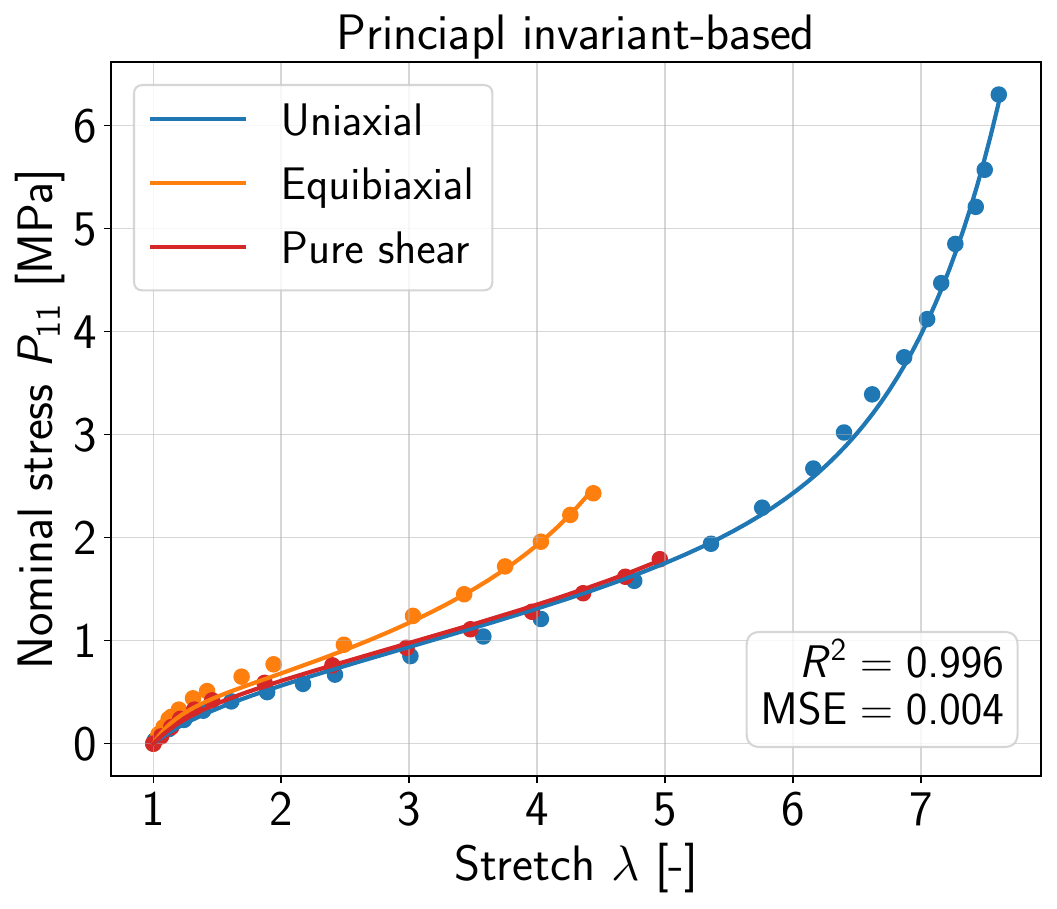}
            \caption{Fitting performance on Treloar's data.}
            \label{fig:result_treloar_invar}
        \end{subfigure}
        \hfill
        \begin{subfigure}{0.42\linewidth}
            \centering
            \includegraphics[width=0.85\linewidth]{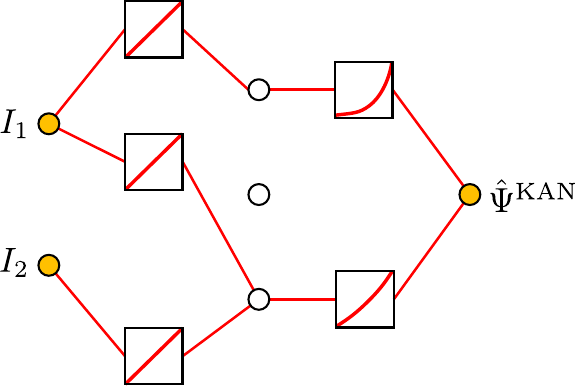}
            \vspace{1.5cm}
            \caption{\emph{Principal invariant-based} CKAN architecture.}
            \label{fig:architecture_treloar_invar}
        \end{subfigure}
        \\[2ex]
        \begin{subfigure}{0.48\linewidth}
            \includegraphics[width=\linewidth]{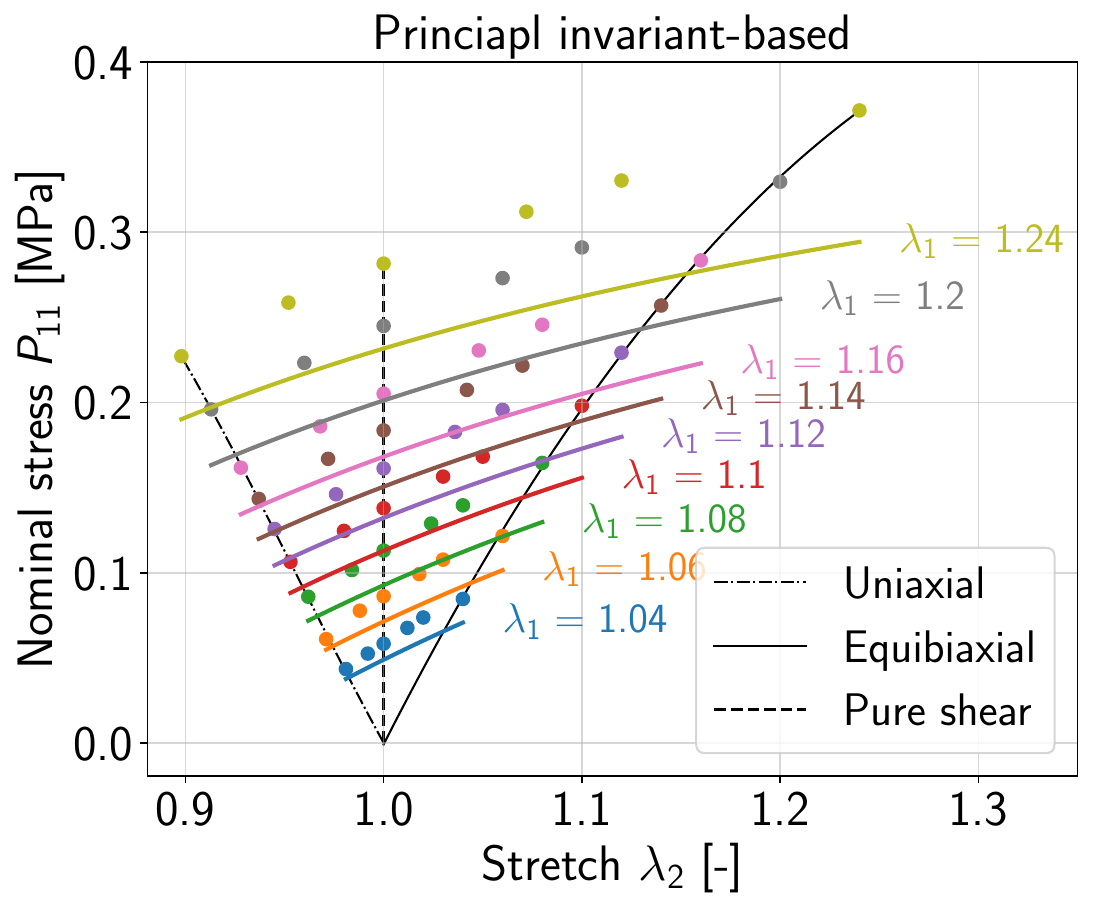}
            \caption{$P_{11}$ generalization: small stretch regime of Kawabata's data.}
            \label{fig:kawabata_P1_small_stretch_invar}
        \end{subfigure}
        \hfill
        \begin{subfigure}{0.48\linewidth}
            \includegraphics[width=\linewidth]{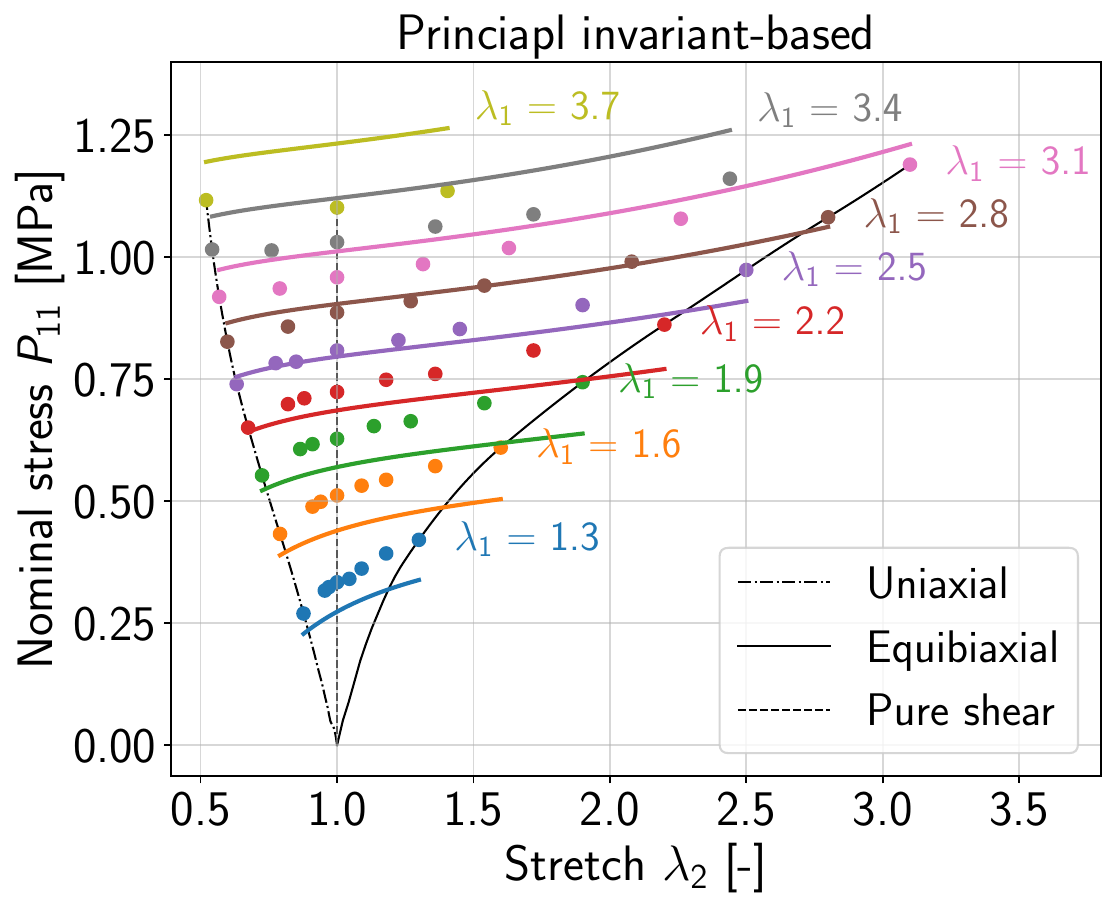}
            \caption{$P_{11}$ generalization: large stretch regime of Kawabata's data.}
            \label{fig:kawabata_P1_large_stretch_invar}
        \end{subfigure}
        \\[2ex]
        \centering
        \begin{subfigure}{0.48\linewidth}
            \includegraphics[width=\linewidth]{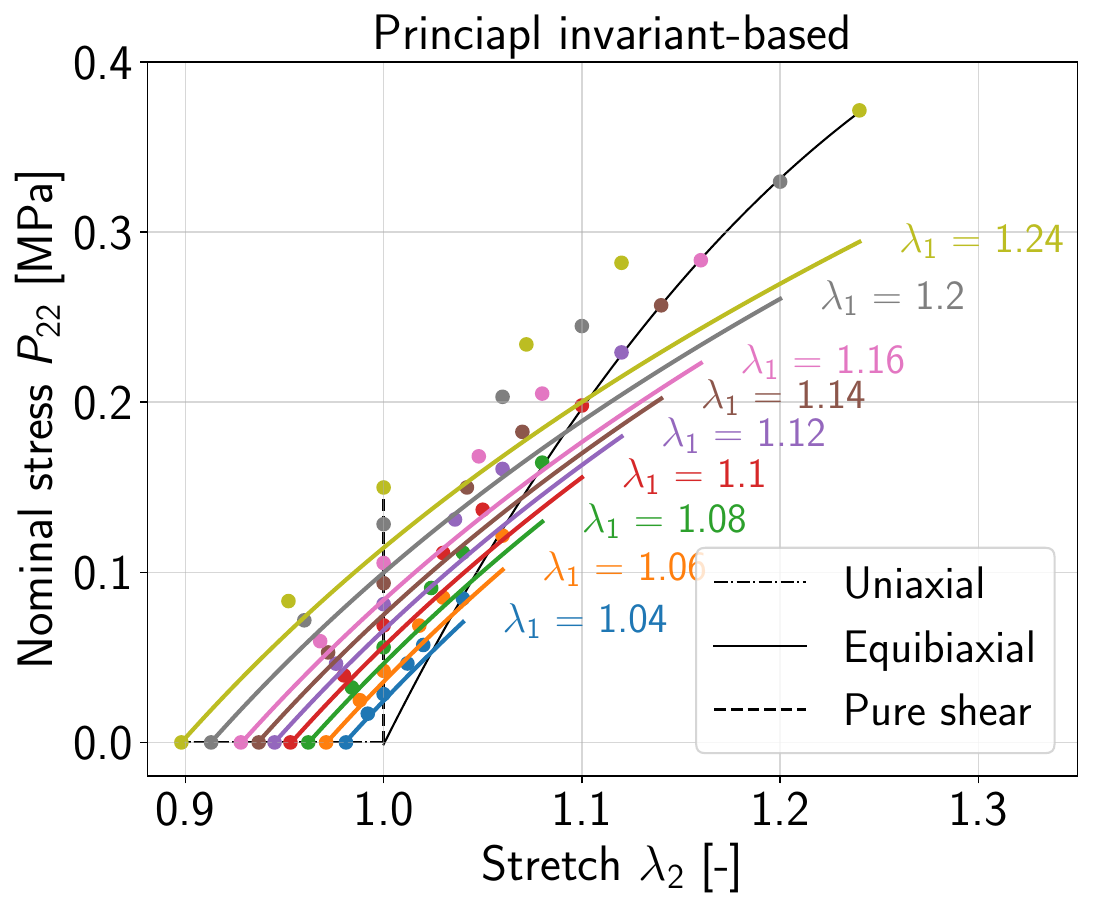}
            \caption{$P_{22}$ generalization: small stretch regime of Kawabata's data.}
            \label{fig:kawabata_P2_small_stretch_invar}
        \end{subfigure}
        \hfill
        \begin{subfigure}{0.48\linewidth}
            \includegraphics[width=\linewidth]{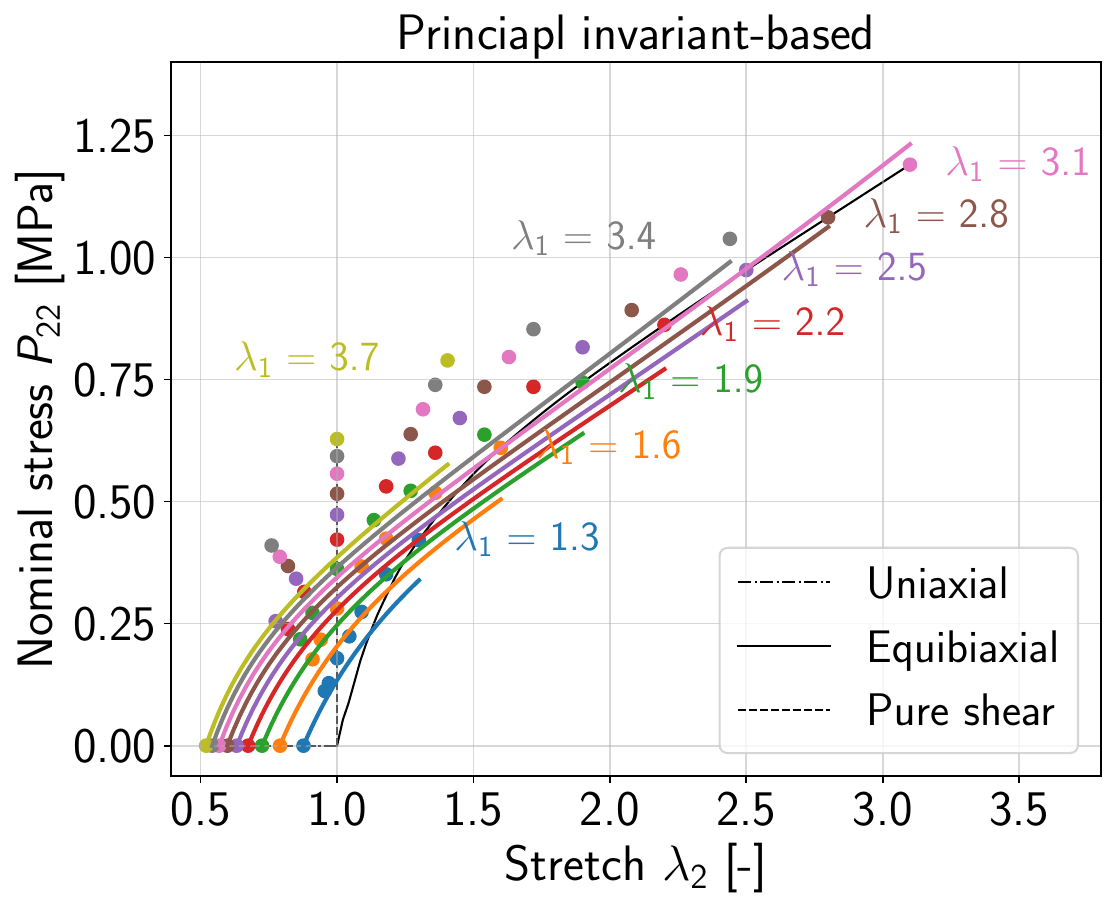}
            \caption{$P_{22}$ generalization: large stretch regime of Kawabata's data.}
            \label{fig:kawabata_P2_large_stretch_invar}
        \end{subfigure}
        \caption{\textbf{Principal invariant-based CKAN for rubber}. (a) Fitting performance on Treloar's data \cite{Treloar1944}. (b) Architecture of the final CKAN after training; unconnected nodes indicate pruned activation functions of the network. (c)--(f) Predictions on the biaxial dataset by Kawabata et al. \cite{Kawabata1981} using the CKAN trained on Treloar's data only. The solid lines represent the model response. The dots represent the experimental data.}
         \label{fig:Treloar_Inv}
    \end{figure}
  
    \FloatBarrier
    \section{Human brain tissue}\label{app:brain_ps}
    \setcounter{figure}{0}

    \subsection{One-term principal stretch-based CKAN }
    
    \begin{figure}[h]
        \centering
        \begin{subfigure}{\linewidth}
            \centering
            \includegraphics[width=\linewidth]{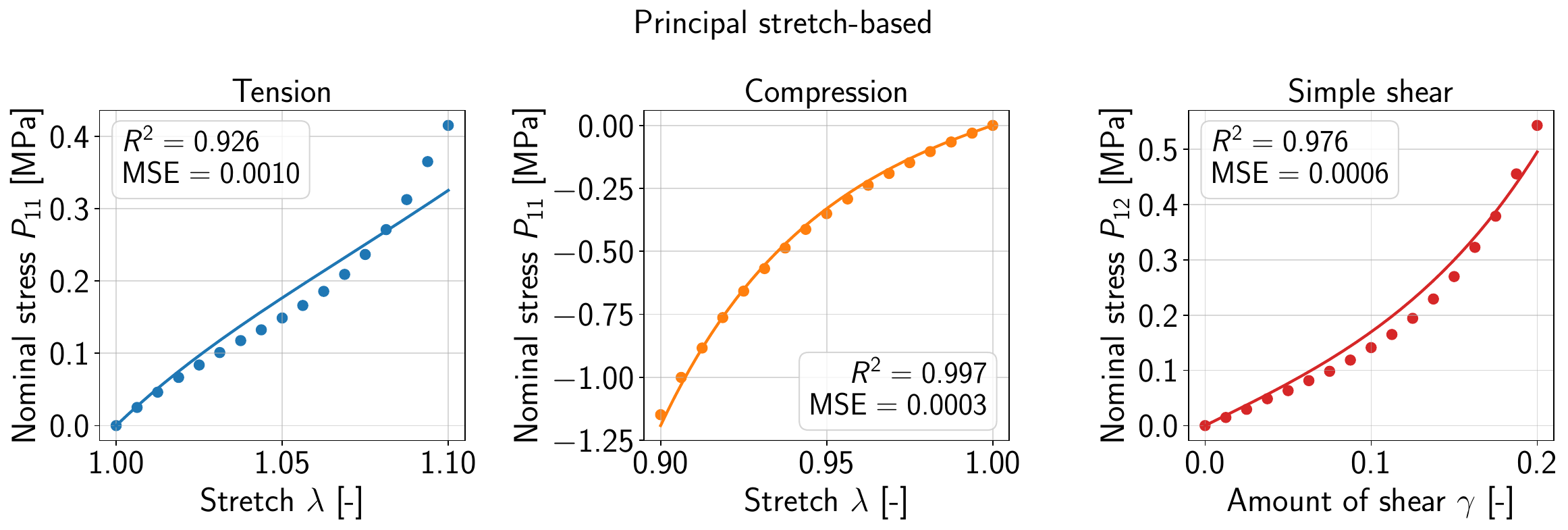}
            \caption{Fitting results; the solid lines represent the model response, the dots represent the experimental data.}
            \label{fig:results_brain_ps_poly}
        \end{subfigure}    \\[4ex]
        \begin{subtable}[c]{0.6\textwidth}
            \renewcommand{\arraystretch}{1.4}
            \centering
            \begin{tabular}{ccc}
                \toprule
                \multirow{2}{*}{Strain energy function} & \multicolumn{2}{c}{$\hat{\Psi}^\text{KAN} = \sum_{\alpha=1}^3 \hat{\omega}^\text{KAN}_{1}(\lambda_{\alpha}) + \sum_{\alpha=1}^3 \hat{\omega}^\text{KAN}_{-1}(\lambda_{\alpha}^{-1})$}  \\ 
                & \multicolumn{2}{c}{$\hat{\omega}^\text{KAN}_1(\lambda_\alpha) = 0, \quad \hat{\omega}^\text{KAN}_{-1}(\lambda_\alpha^{-1}) = a\, (\lambda_\alpha^{-1} + b )^{18}$} \\\cmidrule{1-3}
                \multirow[c]{2}{*}{ Material parameters} & $a$ [MPa] & $b$ [-] \\ \cmidrule{2-3}
                                                         & 0.01872 & -0.04393 \\ \bottomrule
            \end{tabular}
            \caption{Discovered strain energy function and material parameters.}
            \label{tab:params_SEF_Brain_ps_poly}
        \end{subtable}
        \hfill
        \begin{subfigure}[c]{0.32\linewidth}
            \centering
            \includegraphics[width=1\linewidth]{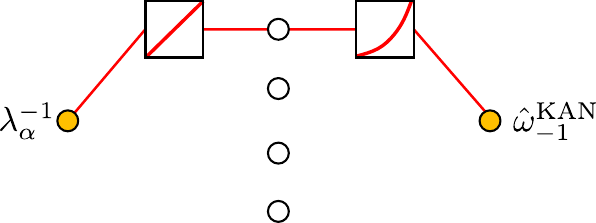}
            \caption{CKAN architecture.}
            \label{fig:architecture_brain_ps_poly}
        \end{subfigure}
        \caption{ \textbf{Results of the {one-term principal stretch-based} CKAN}. (a) simultaneously fitted to uniaxial tension, compression, and simple shear data of human brain (cortex) tissue data reported in \cite{Budday2017b}. (b) Resulting strain energy functions and material parameters. (c) Architecture of the final CKAN after training; unconnected nodes indicate pruned activation functions of the network.}
        \label{fig:Brain_PS_one_term}
    \end{figure}
    
    \newpage
    
    \subsection{Principal invariant-based CKAN}
    \begin{figure}[h]
        \centering
        \begin{subfigure}{\linewidth}
            \includegraphics[width=\linewidth]{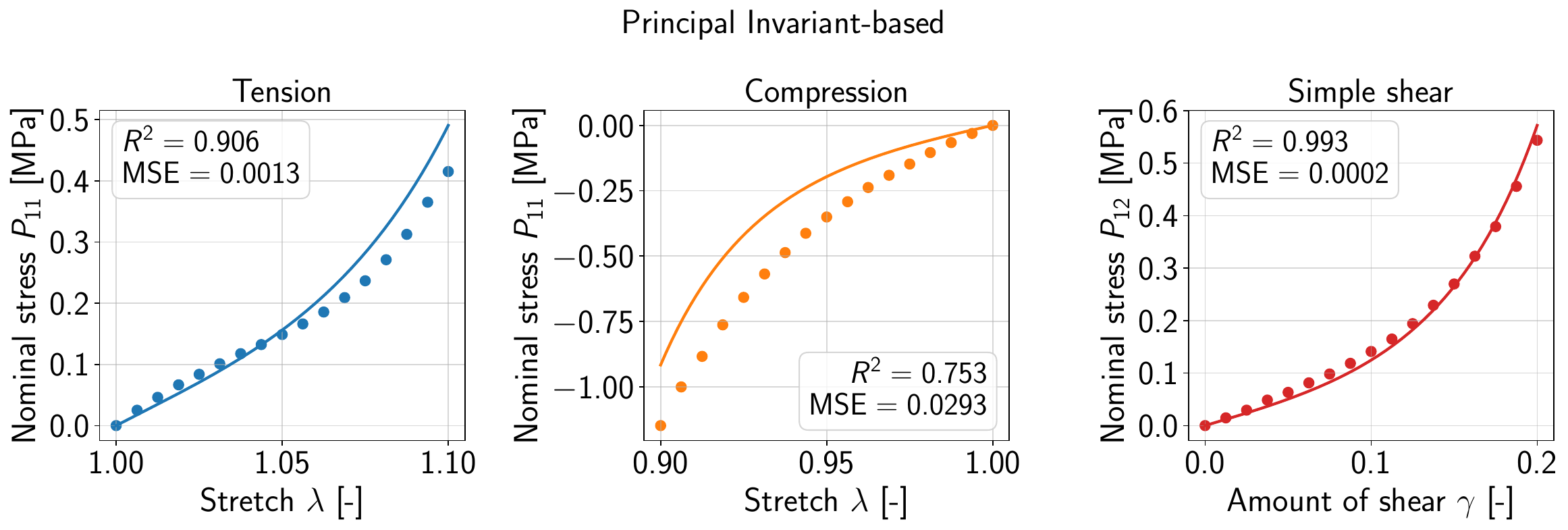}
            \caption{Fitting results; the solid lines represent the model response, the dots represent the experimental data.}
            \label{fig:results_brain_invar}
        \end{subfigure}     \\[4ex]
        \begin{subtable}[c]{0.6\textwidth}
            \renewcommand{\arraystretch}{1.4}
            \centering
            \begin{tabular}{@{}lcc@{}}
                \toprule
                Strain energy function & \multicolumn{2}{c}{$\hat{\Psi}^\text{KAN}(I_2) = a\,( b\, I_2 -1)^{24}$}  \\ \cmidrule{1-3}
                \multirow[c]{2}{*}{Material parameters} & $a$ [MPa] & $b$ [-] \\ \cmidrule{2-3}
                                                        & $4.26077\cdot10^8$    & 0.45586  \\ \bottomrule
            \end{tabular}
            \caption{Discovered strain energy function and material parameters.}
            \label{tab:params_SEF_Brain_inv}
        \end{subtable}
        \hfill
        \begin{subfigure}[c]{0.35\linewidth}
            \centering
            \includegraphics[width=\linewidth]{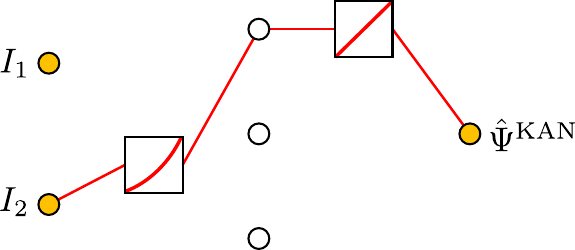}
            \caption{CKAN architecture.}
            \label{fig:architecture_brain_invar}
        \end{subfigure}
        \caption{ \textbf{Results of the {principal invariant-based} CKAN}. (a) simultaneously fitted to uniaxial tension, compression, and simple shear data of human brain (cortex) tissue data reported in \cite{Budday2017b}. (b) Resulting strain energy functions and material parameters. (c) Architecture of the final CKAN after training; unconnected nodes indicate pruned activation functions of the network.}
        \label{fig:Brain_inv}
    \end{figure}

    \newpage
    
    \subsection{Modified invariant-based CKAN}
    \begin{figure}[h]
        \centering
        \begin{subfigure}{\linewidth}
            \includegraphics[width=\linewidth]{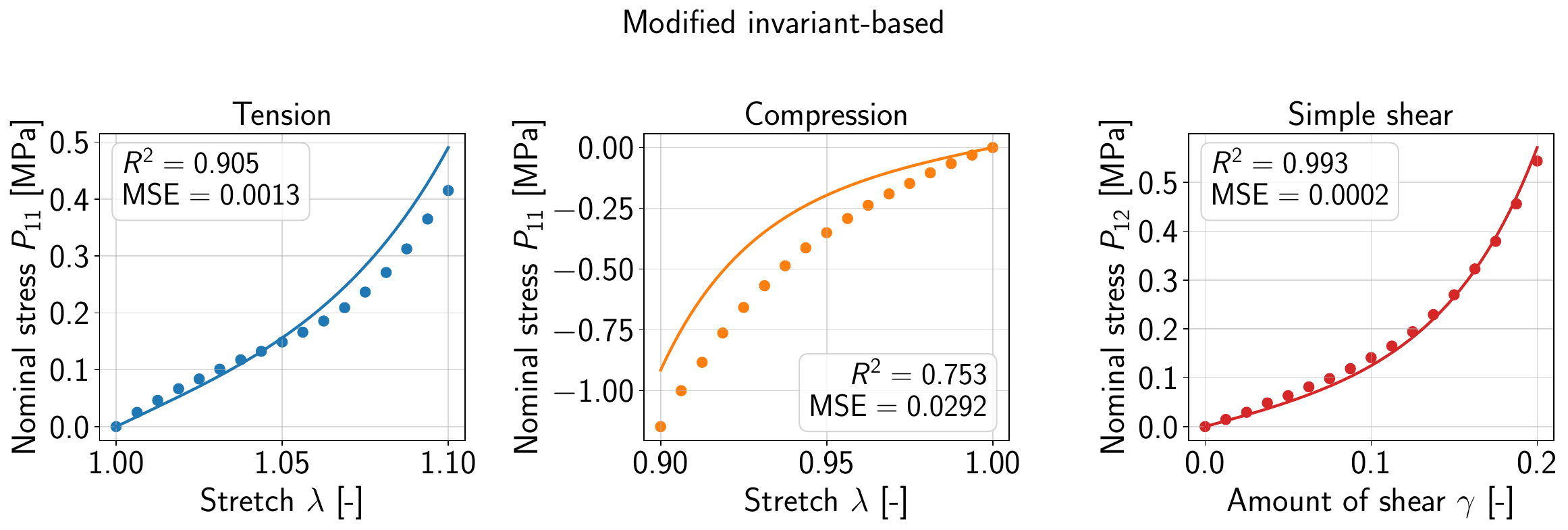}
            \caption{Fitting results; the solid lines represent the model response, the dots represent the experimental data.}
            \label{fig:results_brain_mod_invar}
        \end{subfigure}     \\[4ex]
        \begin{subtable}[c]{0.6\textwidth}
            \renewcommand{\arraystretch}{1.3}
            \centering
            \begin{tabular}{@{}lcc@{}}
                \toprule
                Strain energy function & \multicolumn{2}{c}{$\hat{\Psi}^\text{KAN}(\iota_2) =  a\,(\iota_2 + b)^{23}$}  \\ \cmidrule{1-3}
                \multirow[c]{2}{*}{
                    Material parameters} & $a$ [MPa] & $b$ [-] \\ \cmidrule{2-3}
                                         & $7.27523\cdot10^{22}$  & -0.91540128 \\ \bottomrule
            \end{tabular}
            \caption{Discovered strain energy function and material parameters.}
            \label{tab:params_SEF_Brain_mod_inv}
        \end{subtable}
        \hfill
        \begin{subfigure}[c]{0.35\linewidth}
            \centering
            \includegraphics[width=\linewidth]{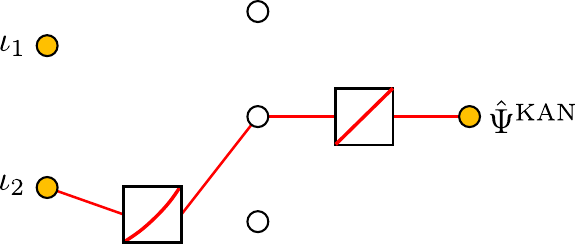}
            \caption{CKAN architecture.}
            \label{fig:architecture_brain_mod_invar}
        \end{subfigure}
        \caption{\textbf{Results of the {modified invariant-based} CKAN.} (a) simultaneously fitted to uniaxial tension, compression, and simple shear data of human brain (cortex) tissue data reported in \cite{Budday2017b}. (b) Resulting strain energy functions and material parameters. (c) Architecture of the final CKAN after training; unconnected nodes indicate pruned activation functions of the network.}
         \label{fig:Brain_Modinv}
    \end{figure}

    \FloatBarrier


	\bibliography{library.bib}

\begin{thebibliography}{10}

\bibitem{Treloar1943}
Leslie R.~G. Treloar.
\newblock {The Elasticity of a Network of Long-Chain Molecules. II}.
\newblock {\em Rubber Chemistry and Technology}, 17(2):296--302, 1944.

\bibitem{Rivlin1951}
Ronald~S. Rivlin and D.~W. Saunders.
\newblock {Large elastic deformations of isotropic materials VII. Experiments on the deformation of rubber}.
\newblock {\em Philosophical Transactions of the Royal Society of London. Series A, Mathematical and Physical Sciences}, 243(865):251--288, 1951.

\bibitem{Ogden1972}
Ray~W. Ogden.
\newblock {Large deformation isotropic elasticity – on the correlation of theory and experiment for incompressible rubberlike solids}.
\newblock {\em Proceedings of the Royal Society of London. A. Mathematical and Physical Sciences}, 326(1567):565--584, 1972.

\bibitem{Fung1981}
Yuan-Cheng Fung.
\newblock {\em {Biomechanics - Mechanical Properties of Living Tissues}}.
\newblock Springer New York, New York, NY, 1981.

\bibitem{Holzapfel2000}
Gerhard~A. Holzapfel, T.~Christian Gasser, and Ray~W. Ogden.
\newblock {A new constitutive framework for arterial wall mechanics and a comparative study of material models}.
\newblock {\em Journal of Elasticity}, 61(1-3):1--48, 2000.

\bibitem{Ehret2007}
Alexander~E. Ehret and Mikhail Itskov.
\newblock {A polyconvex hyperelastic model for fiber-reinforced materials in application to soft tissues}.
\newblock {\em Journal of Materials Science}, 42(21):8853--8863, 2007.

\bibitem{Kirchdoerfer2016}
Trenton Kirchdoerfer and Michael Ortiz.
\newblock {Data-driven computational mechanics}.
\newblock {\em Computer Methods in Applied Mechanics and Engineering}, 304:81--101, 2016.

\bibitem{Carrara2020}
Pietro Carrara, Laura {De Lorenzis}, Laurent Stainier, and Michael Ortiz.
\newblock {Data-driven fracture mechanics}.
\newblock {\em Computer Methods in Applied Mechanics and Engineering}, 372, 2020.

\bibitem{Ghaboussi1991}
Jamshid Ghaboussi, James~H. Garrett, and Xiping Wu.
\newblock {Knowledge‐Based Modeling of Material Behavior with Neural Networks}.
\newblock {\em Journal of Engineering Mechanics}, 117(1):132--153, 1991.

\bibitem{Hashash2004}
Youssef.~M.A. Hashash, Sungmoon Jung, and Jamshid Ghaboussi.
\newblock {Numerical implementation of a neural network based material model in finite element analysis}.
\newblock {\em International Journal for Numerical Methods in Engineering}, 59(7):989--1005, 2004.

\bibitem{Sussman2009}
Theodore Sussman and Klaus~J. Bathe.
\newblock {A model of incompressible isotropic hyperelastic material behavior using spline interpolations of tension-compression test data}.
\newblock {\em Communications in Numerical Methods in Engineering}, 25(1):53--63, 2009.

\bibitem{Latorre2013}
Marcos Latorre and Francisco~Javier Mont{\'{a}}ns.
\newblock {Extension of the Sussman-Bathe spline-based hyperelastic model to incompressible transversely isotropic materials}.
\newblock {\em Computers and Structures}, 122:13--26, 2013.

\bibitem{Latorre2014}
Marcos Latorre and Francisco~Javier Mont{\'{a}}ns.
\newblock {What-You-Prescribe-Is-What-You-Get orthotropic hyperelasticity}.
\newblock {\em Computational Mechanics}, 53(6):1279--1298, 2014.

\bibitem{Crespo2017}
Jos{\'{e}} Crespo, Marcos Latorre, and Francisco~Javier Mont{\'{a}}ns.
\newblock {WYPIWYG hyperelasticity for isotropic, compressible materials}.
\newblock {\em Computational Mechanics}, 59(1):73--92, 2017.

\bibitem{Dal2023}
H{\"{u}}sn{\"{u}} Dal, Funda~Aksu Denli, Alp~Kağan A{\c{c}}an, and Michael Kaliske.
\newblock {Data-driven hyperelasticity, Part I: A canonical isotropic formulation for rubberlike materials}.
\newblock {\em Journal of the Mechanics and Physics of Solids}, 179:105381, 2023.

\bibitem{Wiesheier2024}
Simon Wiesheier, Miguel~Angel Moreno-Mateos, and Paul Steinmann.
\newblock {Versatile data-adaptive hyperelastic energy functions for soft materials}.
\newblock {\em Computer Methods in Applied Mechanics and Engineering}, 430:117208, 2024.

\bibitem{goldberg2024no}
Carey~B. Goldberg et~al.
\newblock {To do no harm---and the most good---with AI in health care}.
\newblock {\em Nature Medicine}, 30(3):623--627, 2024.

\bibitem{fuhg2024review}
Jan~N. Fuhg, Govinda Anantha~Padmanabha, Nikolaos Bouklas, Bahador Bahmani, WaiChing Sun, Nikolaos~N. Vlassis, Moritz Flaschel, Pietro Carrara, and Laura De~Lorenzis.
\newblock A review on data-driven constitutive laws for solids.
\newblock {\em Archives of Computational Methods in Engineering}, pages 1--43, 2024.

\bibitem{Linden2023}
Lennart Linden, Dominik~K. Klein, Karl~A. Kalina, J{\"{o}}rg Brummund, Oliver Weeger, and Markus K{\"{a}}stner.
\newblock {Neural networks meet hyperelasticity: A guide to enforcing physics}.
\newblock {\em Journal of the Mechanics and Physics of Solids}, 179:105363, 2023.

\bibitem{tacc2023data}
Vahidullah Ta{\c{c}}, Manuel~K. Rausch, Francisco~Sahli Costabal, and Adrian~Buganza Tepole.
\newblock Data-driven anisotropic finite viscoelasticity using neural ordinary differential equations.
\newblock {\em Computer methods in applied mechanics and engineering}, 411:116046, 2023.

\bibitem{as2022mechanics}
Faisal As'~ad, Philip Avery, and Charbel Farhat.
\newblock A mechanics-informed artificial neural network approach in data-driven constitutive modeling.
\newblock {\em International Journal for Numerical Methods in Engineering}, 123(12):2738--2759, 2022.

\bibitem{Linka2021}
Kevin Linka, Markus Hillg{\"{a}}rtner, Kian~P. Abdolazizi, Roland~C. Aydin, Mikhail Itskov, and Christian~J. Cyron.
\newblock {Constitutive artificial neural networks: A fast and general approach to predictive data-driven constitutive modeling by deep learning}.
\newblock {\em Journal of Computational Physics}, 429:110010, 2021.

\bibitem{Abdolazizi2023}
Kian~P. Abdolazizi, Kevin Linka, and Christian~J. Cyron.
\newblock {Viscoelastic constitutive artificial neural networks (vCANNs) – A framework for data-driven anisotropic nonlinear finite viscoelasticity}.
\newblock {\em Journal of Computational Physics}, 499:112704, 2023.

\bibitem{Linka2022}
Kevin Linka, Cristina Cavinato, Jay~D. Humphrey, and Christian~J. Cyron.
\newblock {Predicting and understanding arterial elasticity from key microstructural features by bidirectional deep learning}.
\newblock {\em Acta Biomaterialia}, 147:63--72, 2022.

\bibitem{Flaschel2021a}
Moritz Flaschel, Siddhant Kumar, and Laura {De Lorenzis}.
\newblock {Unsupervised discovery of interpretable hyperelastic constitutive laws}.
\newblock {\em Computer Methods in Applied Mechanics and Engineering}, 381:113852, 2021.

\bibitem{Linka2022a}
Kevin Linka and Ellen Kuhl.
\newblock {A new family of Constitutive Artificial Neural Networks towards automated model discovery}.
\newblock {\em Computer Methods in Applied Mechanics and Engineering}, 403:115731, 2023.

\bibitem{holthusen2024theory}
Hagen Holthusen, Lukas Lamm, Tim Brepols, Stefanie Reese, and Ellen Kuhl.
\newblock Theory and implementation of inelastic constitutive artificial neural networks.
\newblock {\em Computer Methods in Applied Mechanics and Engineering}, 428:117063, 2024.

\bibitem{pierre2023discovering}
Skyler R.~S.t Pierre, Divya Rajasekharan, Ethan~C. Darwin, Kevin Linka, Marc~E. Levenston, and Ellen Kuhl.
\newblock Discovering the mechanics of artificial and real meat.
\newblock {\em Computer Methods in Applied Mechanics and Engineering}, 415:116236, 2023.

\bibitem{McCulloch2024}
Jeremy~A. McCulloch, Skyler~R. {St. Pierre}, Kevin Linka, and Ellen Kuhl.
\newblock {On sparse regression, Lp-regularization, and automated model discovery}.
\newblock {\em International Journal for Numerical Methods in Engineering}, 125(14):e7481, 2024.

\bibitem{holthusen2025automated}
Hagen Holthusen, Tim Brepols, Kevin Linka, and Ellen Kuhl.
\newblock Automated model discovery for tensional homeostasis: Constitutive machine learning in growth and remodeling.
\newblock {\em Computers in Biology and Medicine}, 186:109691, 2025.

\bibitem{wang2019symbolic}
Yiqun Wang, Nicholas Wagner, and James~M Rondinelli.
\newblock Symbolic regression in materials science.
\newblock {\em MRS Communications}, 9(3):793--805, 2019.

\bibitem{Abdusalamov2022a}
Rasul Abdusalamov, Markus Hillg{\"{a}}rtner, and Mikhail Itskov.
\newblock {Automatic generation of interpretable hyperelastic material models by symbolic regression}.
\newblock {\em International Journal for Numerical Methods in Engineering}, 124(9):2093--2104, 2023.

\bibitem{Liu2024a}
Ziming Liu, Yixuan Wang, Sachin Vaidya, Fabian Ruehle, James Halverson, Marin Solja{\v{c}}i{\'{c}}, Thomas~Y. Hou, and Max Tegmark.
\newblock {KAN: Kolmogorov-Arnold Networks}.
\newblock {\em arXiv preprint arXiv:2404.19756}, 2024.

\bibitem{Liu2024}
Ziming Liu, Pingchuan Ma, Yixuan Wang, Wojciech Matusik, and Max Tegmark.
\newblock {KAN 2.0: Kolmogorov-Arnold Networks Meet Science}.
\newblock {\em arXiv preprint arXiv:2408.10205}, 2024.

\bibitem{Holzapfel2000a}
Gerhard~A Holzapfel.
\newblock {\em {Nonlinear solid mechanics: a continuum approach for engineering science}}.
\newblock John Wiley \& Sons Ltd., 2000.

\bibitem{Kearsley1989}
Elliot~A. Kearsley.
\newblock {Note: Strain Invariants Expressed as Average Stretches}.
\newblock {\em Journal of Rheology}, 33(5):757--760, 1989.

\bibitem{Arruda1993}
Ellen~M. Arruda and Mary~C. Boyce.
\newblock {A three-dimensional constitutive model for the large stretch behavior of rubber elastic materials}.
\newblock {\em Journal of the Mechanics and Physics of Solids}, 41(2):389--412, 1993.

\bibitem{Truesdell1984}
Clifford Truesdell.
\newblock {\em {The Elements of Continuum Mechanics}}.
\newblock Springer Berlin Heidelberg, Berlin, Heidelberg, 1984.

\bibitem{Ogden1985}
Ray~W. Ogden.
\newblock {\em {Nonlinear Elastic Deformations}}.
\newblock Dover Publications, 1997.

\bibitem{Xu2015}
Hongyi Xu, Funshing Sin, Yufeng Zhu, and Jernej Barbi{\v{c}}.
\newblock {Nonlinear material design using principal stretches}.
\newblock {\em ACM Transactions on Graphics}, 34(4):1--11, 2015.

\bibitem{Valanis1967}
Kirk~C. Valanis and Robert~F. Landel.
\newblock {The Strain-Energy Function of a Hyperelastic Material in Terms of the Extension Ratios}.
\newblock {\em Journal of Applied Physics}, 38(7):2997--3002, 1967.

\bibitem{Heinrich1997}
Gert Heinrich and Michael Kaliske.
\newblock {Theoretical and numerical formulation of a molecular based constitutive tube-model of rubber elasticity}.
\newblock {\em Computational and Theoretical Polymer Science}, 7(3-4):227--241, 1997.

\bibitem{Shariff2000}
Mohd~H.B.M. Shariff.
\newblock {Strain energy function for filled and unfilled rubberlike material}.
\newblock {\em Rubber Chemistry and Technology}, 73(1):1--18, 2000.

\bibitem{Attard2004}
Mario~M. Attard and Giles~W. Hunt.
\newblock {Hyperelastic constitutive modeling under finite strain}.
\newblock {\em International Journal of Solids and Structures}, 41(18-19):5327--5350, 2004.

\bibitem{Ehret2022}
Alexander~E. Ehret and Alberto Stracuzzi.
\newblock {Variations on Ogden's model: Close and distant relatives}.
\newblock {\em Philosophical Transactions of the Royal Society A: Mathematical, Physical and Engineering Sciences}, 380(2234), 2022.

\bibitem{Kaliske1999}
Michael Kaliske and Gert Heinrich.
\newblock {An extended tube-model for rubber elasticity: Statistical-mechanical theory and finite element implementation}.
\newblock {\em Rubber Chemistry and Technology}, 72(4):602--632, 1999.

\bibitem{Xiang2018}
Yuhai Xiang, Danming Zhong, Peng Wang, Guoyong Mao, Honghui Yu, and Shaoxing Qu.
\newblock {A general constitutive model of soft elastomers}.
\newblock {\em Journal of the Mechanics and Physics of Solids}, 117:110--122, 2018.

\bibitem{Davidson2013}
Jacob~D. Davidson and N.~C. Goulbourne.
\newblock {A nonaffine network model for elastomers undergoing finite deformations}.
\newblock {\em Journal of the Mechanics and Physics of Solids}, 61(8):1784--1797, 2013.

\bibitem{Itskov2015}
Mikhail Itskov.
\newblock {\em {Tensor Algebra and Tensor Analysis for Engineers}}.
\newblock Mathematical Engineering. Springer Cham, 2015.

\bibitem{polo2024monokan}
Alejandro Polo-Molina, David Alfaya, and Jose Portela.
\newblock {MonoKAN: Certified Monotonic Kolmogorov-Arnold Network}.
\newblock {\em arXiv preprint arXiv:2409.11078}, 2024.

\bibitem{Treloar1944}
Leslie~R.G. Treloar.
\newblock {Stress-strain data for vulcanised rubber under various types of deformation}.
\newblock {\em Transactions of the Faraday Society}, 40:59--70, 1944.

\bibitem{Kawabata1981}
Sueo Kawabata, Masatoshi Matsuda, K.~Tei, and Hiromichi Kawai.
\newblock {Experimental Survey of the Strain Energy Density Function of Isoprene Rubber Vulcanizate}.
\newblock {\em Macromolecules}, 14(1):154--162, jan 1981.

\bibitem{Budday2017}
Silvia Budday, Gerhard Sommer, Johannes Haybaeck, Paul Steinmann, Gerhard~A. Holzapfel, and Ellen Kuhl.
\newblock {Rheological characterization of human brain tissue}.
\newblock {\em Acta Biomaterialia}, 60:315--329, 2017.

\bibitem{Liao2020a}
Zisheng Liao, Mokarram Hossain, and Xiaohu Yao.
\newblock {Ecoflex polymer of different Shore hardnesses: Experimental investigations and constitutive modelling}.
\newblock {\em Mechanics of Materials}, 144:103366, 2020.

\bibitem{Steinmann2012}
Paul Steinmann, Mokarram Hossain, and Gunnar Possart.
\newblock {Hyperelastic models for rubber-like materials: Consistent tangent operators and suitability for Treloar's data}.
\newblock {\em Archive of Applied Mechanics}, 82(9):1183--1217, 2012.

\bibitem{Marckmann2006}
Gilles Marckmann and Erwan Verron.
\newblock {Comparison of hyperelastic models for rubber-like materials}.
\newblock {\em Rubber Chemistry and Technology}, 79(5):835--858, 2006.

\bibitem{Amores2020c}
V{\'{i}}ctor~J. Amores, Jos{\'{e}}~M. Ben{\'{i}}tez, and Francisco~J. Mont{\'{a}}ns.
\newblock {Data-driven, structure-based hyperelastic manifolds: A macro-micro-macro approach to reverse-engineer the chain behavior and perform efficient simulations of polymers}.
\newblock {\em Computers and Structures}, 231:106209, 2020.

\bibitem{Khiem2016}
Vu~Ngoc Khi{\^{e}}m and Mikhail Itskov.
\newblock {Analytical network-averaging of the tube model: Rubber elasticity}.
\newblock {\em Journal of the Mechanics and Physics of Solids}, 95:254--269, 2016.

\bibitem{Budday2017b}
Silvia Budday, Gerhard Sommer, Christoph Birkl, Christian Langkammer, Johannes Haybaeck, J.~Kohnert, Melanie Bauer, Friedrich~P. Paulsen, Paul Steinmann, Ellen Kuhl, and Gerhard~A. Holzapfel.
\newblock {Mechanical characterization of human brain tissue}.
\newblock {\em Acta Biomaterialia}, 48:319--340, 2017.

\bibitem{Mihai2017}
L.~Angela Mihai, Silvia Budday, Gerhard~A. Holzapfel, Ellen Kuhl, and Alain Goriely.
\newblock {A family of hyperelastic models for human brain tissue}.
\newblock {\em Journal of the Mechanics and Physics of Solids}, 106:60--79, 2017.

\bibitem{Anssari-Benam2022}
Afshin Anssari-Benam, Michel Destrade, and Giuseppe Saccomandi.
\newblock {Modelling brain tissue elasticity with the Ogden model and an alternative family of constitutive models}.
\newblock {\em Philosophical transactions. Series A, Mathematical, physical, and engineering sciences}, 380(2234):20210325, 2022.

\bibitem{kuhl2024too}
Ellen Kuhl and Alain Goriely.
\newblock {I too I$_2$: A new class of hyperelastic isotropic incompressible models based solely on the second invariant}.
\newblock {\em Journal of the Mechanics and Physics of Solids}, 188:105670, 2024.

\bibitem{Bernardi2017}
Laura Bernardi, Raoul. Hopf, Daniela Sibilio, Aldo Ferrari, Alexander~E. Ehret, and Eduardo Mazza.
\newblock {On the cyclic deformation behavior, fracture properties and cytotoxicity of silicone-based elastomers for biomedical applications}.
\newblock {\em Polymer Testing}, 60:117--123, 2017.

\bibitem{Case2015}
Jennifer~C. Case, Edward~L. White, and Rebecca~K. Kramer.
\newblock {Soft material characterization for robotic applications}.
\newblock {\em Soft Robotics}, 2(2):80--87, 2015.

\bibitem{Jiang2018}
Yuting Jiang, Yang Wang, Yogendra~Kumar Mishra, Rainer Adelung, and Ya~Yang.
\newblock {Stretchable CNTs‐Ecoflex Composite as Variable‐Transmittance Skin for Ultrasensitive Strain Sensing}.
\newblock {\em Advanced Materials Technologies}, 3(12):1800248, 2018.

\bibitem{Mullins1965}
Leonard Mullins and N.~R. Tobin.
\newblock {Stress softening in rubber vulcanizates. Part I. Use of a strain amplification factor to describe the elastic behavior of filler‐reinforced vulcanized rubber}.
\newblock {\em Journal of Applied Polymer Science}, 9(9):2993--3009, 1965.

\bibitem{Linka2018}
Kevin Linka, Markus Hillg{\"{a}}rtner, and Mikhail Itskov.
\newblock {Fatigue of soft fibrous tissues: Multi-scale mechanics and constitutive modeling}.
\newblock {\em Acta Biomaterialia}, 71:398--410, 2018.

\bibitem{Linka2023}
Kevin Linka, Sarah~R. {St. Pierre}, and Ellen Kuhl.
\newblock {Automated model discovery for human brain using Constitutive Artificial Neural Networks}.
\newblock {\em Acta Biomaterialia}, 160:134--151, 2023.

\bibitem{brunton2016discovering}
Steven~L. Brunton, Joshua~L. Proctor, and J.~Nathan Kutz.
\newblock Discovering governing equations from data by sparse identification of nonlinear dynamical systems.
\newblock {\em Proceedings of the national academy of sciences}, 113(15):3932--3937, 2016.

\bibitem{brunton2022data}
Steven~L. Brunton and J.~Nathan Kutz.
\newblock {\em Data-driven science and engineering: Machine learning, dynamical systems, and control}.
\newblock Cambridge University Press, 2022.

\bibitem{hospedales2021meta}
Timothy Hospedales, Antreas Antoniou, Paul Micaelli, and Amos Storkey.
\newblock Meta-learning in neural networks: A survey.
\newblock {\em IEEE transactions on pattern analysis and machine intelligence}, 44(9):5149--5169, 2021.

\bibitem{Kalina2025}
Karl~A. Kalina, J{\"{o}}rg Brummund, WaiChing Sun, and Markus K{\"{a}}stner.
\newblock {Neural networks meet anisotropic hyperelasticity: A framework based on generalized structure tensors and isotropic tensor functions}.
\newblock {\em Computer Methods in Applied Mechanics and Engineering}, 437:117725, 2025.

\bibitem{Franke2023}
Marlon Franke, Dominik~K. Klein, Oliver Weeger, and Peter Betsch.
\newblock {Advanced discretization techniques for hyperelastic physics-augmented neural networks}.
\newblock {\em Computer Methods in Applied Mechanics and Engineering}, 416:116333, 2023.

\bibitem{fritsch1980monotone}
Frederick~N. Fritsch and Ralph~E. Carlson.
\newblock Monotone piecewise cubic interpolation.
\newblock {\em SIAM Journal on Numerical Analysis}, 17(2):238--246, 1980.

\end{thebibliography}
	\bibliographystyle{unsrt}
	
\end{document}